\documentclass[usenatbib]{mn2e}
\usepackage{amsmath}
\usepackage{xspace} 
\usepackage{natbib}
\usepackage{epsfig}
\usepackage{txfonts}
\usepackage{pict2e}
\usepackage{cleveref}
\usepackage{url}
\usepackage[dvipsnames,usenames]{color}

\newcommand{\lsim}{\lesssim}
\newcommand{\gsim}{\gtrsim}

\newcommand{\taudot}{\dot{\tau}}

\newcommand{\xe}{x_{\rm e}}

\newcommand{\id}{{\,\rm d}}

\newcommand{\beq}{\begin{equation}}   %

\newcommand{\eeq}{\end{equation}}   %

\newcommand{\beqa}{\begin{eqnarray}}   %

\newcommand{\eeqa}{\end{eqnarray}}   %

\newcommand{\beal}{\begin{align}}
\newcommand{\enal}{\end{align}}

\newcommand{\bspl}{\begin{split}}

\newcommand{\espl}{\end{split}}

\newcommand{\bsub}{\begin{subequations}}

\newcommand{\esub}{\end{subequations}}

\newcommand{\bmulti}{\begin{multline}}   %

\newcommand{\beqm}{\begin{mathletters}}   %

\newcommand{\eeqm}{\end{mathletters}}   %

\newcommand{\me}{m_{\rm e}}

\newcommand{\sigT}{\sigma_{\rm T}}

\newcommand{\omb}{\omega_{\rm b}}

\newcommand{\omc}{\omega_{\rm c}}

\newcommand{\thetaMC}{\theta_{\rm MC}}
\newcommand{\thetaxMC}{100\,\thetaMC}
\newcommand{\ns}{n_{\rm s}}
\newcommand{\As}{A_{\rm s}}
\newcommand{\logA}{\ln\left(10^{10} A_{\rm s}\right)}
\newcommand{\aEMs}{\alpha_{\rm EM, 0}}
\newcommand{\mes}{m_{\rm e, 0}}
\newcommand{\aEM}{\alpha_{\rm EM}}

\newcommand{\LCDM}{$\Lambda$CDM\xspace}

\newcommand{\ho}{H_0}
\newcommand{\sig}{\sigma_8}

\newcommand{\planck}{\emph{Planck}\xspace}
\newcommand{\fc}{\mathcal{C}}
\newcommand{\cl}{C_\ell}

\newcommand{\dcl}{\partial\cl}
\newcommand{\dclvec}{\partial\vec{C}_\ell}
\newcommand{\dlncl}{\partial\ln\cl}

\newcommand*\dif{\mathop{}\!\mathrm{d}}

\newcommand{\DelL}{\Delta\ln\mathcal{L}}

\newcommand{\parA}{\mathcal{A}}
\newcommand{\dA}{\Delta\parA/\parA}
\newcommand{\delA}{\Delta\parA}

\renewcommand{\xe}{X_{\rm e}}
\newcommand{\Dell}{\mathcal{D}_\ell}
\newcommand{\DellT}{\Dell^{\rm TT}}
\newcommand{\DellE}{\Dell^{\rm EE}}

\title[Varying fundamental constants PCA]
{Varying fundamental constants principal component analysis: additional hints about the Hubble tension}

\author[L. Hart and J. Chluba]{
Luke Hart$^{1}$\thanks{Email: luke.hart@manchester.ac.uk} and Jens Chluba$^{1}$
\\
$^{1}$Jodrell Bank Centre for Astrophysics, Alan Turing Building, University of Manchester, Manchester M13 9PL \\
}

\date{\vspace{-0mm}Accepted  --. Received --.}

\pubyear{2021}

\begin{document}
\label{firstpage}
\pagerange{\pageref{firstpage}--\pageref{lastpage}}
\maketitle
\begin{abstract}
Varying fundamental constants (VFC) [e.g., the fine-structure constant, $\aEM$] can arise in numerous extended cosmologies. Through their effect on the decoupling of baryons and photons during last scattering and reionisation, these models can be directly constrained using measurements of the cosmic microwave background (CMB) temperature and polarization anisotropies. Previous investigations focused mainly on time-independent changes to the values of fundamental constants. Here we generalize to time-dependent variations. Instead of directly studying various VFC parameterizations, we perform a model-independent principal component analysis (PCA), directly using an eigenmode decomposition of the varying constant during recombination. After developing the formalism, we use \planck 2018 data to obtain new VFC limits, showing that three independent VFC modes can be constrained at present. No indications for significant departures from the standard model are found with \planck data. Cosmic variance limited modes are also compared and simple forecasts for The Simons Observatory are carried out, showing that in the future improvements of the current constraints by a factor of $\simeq 3$ can be anticipated.
Our modes focus solely on VFC at redshifts $z\geq 300$. This implies that they do not capture some of the degrees of freedom relating to the reionisation era. This aspect provides important new insights into the possible origin of the Hubble tension, hinting that indeed a combined modification of recombination and reionisation physics could be at work. An extended PCA, covering both recombination and reionisation simultaneously, could shed more light on this question, as we emphasize here.
\end{abstract}

\maketitle

\section{Introduction}\label{sec:introduction}
%----------------------------------------------
For the last few decades, modern cosmology has been dominated by the study and observations of the cosmic microwave background (CMB) anisotropies. The results from \planck, ACT and SPT have transformed the way we look at the microwave sky and cosmology \citep{Planck2015params, Planck2018params, actpol_polresults, sptpol_results}. These experiments have followed the fine work of their predecessors \emph{COBE} and \emph{WMAP} \citep{COBE4yr,wmap9results} and the many ground and balloon-based experiments \citep[e.g.,][]{Netterfield2002, VSA2003, CBI03}. 
Currently, attention is turning to larger ground-based telescopes such as AdvancedACTPol \citep{AdvancedACTPol},
POLARBEAR \citep{polarbear_results}, The Simons Observatory \citep{SOWP2018}
and
CMB-Stage-VI \citep{CMBS42016,CMBS4WP}, which will give us further insight into the CMB anisotropies, with unparalleled precision for the polarisation power spectra and spanning a vast range of angular scales.

Beyond the now well-established \LCDM model, the immense experimental progress also enabled us to probe new physics. This includes neutrino physics through tests of the neutrino masses and relativistic degrees of freedom \citep{PlanckNeutrino,BattyeNeutrinos,Abazajian2015,Planck2018params}. In addition, we have been able to consider a variety of models linked to \emph{dark matter annihilation} and \emph{decay} \citep{Chen2004,Padmanabhan2005,Galli2009, Slatyer2009, Huetsi2009, Chluba2010a,Finkbeiner2012,Slatyer2016,Chen2021} and \emph{primordial magnetic fields} \citep{Sethi2005,Shaw2010PMF,Kunze2014,Chluba2015PMF,Paoletti2019,Jedamzik2019}. 
CMB anisotropies can furthermore be used to constrain more complex dark energy theories, including \emph{k-essence}, \emph{early} and \emph{interacting dark energy} \citep{Silvestri2009,DiValentino2017,Poulin2018,Pace2019,Lin2020}. Many of these extensions have been proposed to alleviate the \emph{Hubble constant tension} that is currently dominating discussions in the field of cosmology \citep{Poulin2019,DiValentino2019, Knox2020, Nils2021}.

One of the interesting extensions to the standard cosmological model is \emph{varying fundamental constants} (VFC). Whilst fundamental constants are thought to be just that --- constant, there are numerous theories that motivate changes to these parameters at early and late times. Several exhaustive reviews have discussed the mechanisms and motivations for such variations \citep{Uzan2003,Uzan2011,Martins2017review}. 
Two compelling parameters that affect electromagnetism in the early (and late) Universe are the fine structure constant $\aEM$ and the effective electron mass $\me$\footnote{Strictly speaking we allow the effective electron mass to vary so that more formally, we are varying the electron-proton mass ratio $\mu$, which is non-dimensional \citep[see][for a clearer motivation]{Uzan2011}.}. These fundamental constants can change across cosmic history through modifications to the electromagnetic Lagrangian and the introduction of additional scalar fields or particles \citep{Bekenstein1982,Sandvik2001,Mota2004,Barrow2013}. 

Many previous studies have looked at constraining the variations to $\aEM$ using astrophysical probes such as \emph{quasar absorption spectra} \citep{Bonifacio2014, Kotus2017, Murphy2017, Wilczynska2020VFC}, \emph{thermonuclear supernovae} \citep{Negrelli2018}, \emph{white dwarfs} \citep{Hu2020VFC}, \emph{supermassive black holes} \citep{Hees2020VFC} and the \emph{Magellanic Clouds} \citep{Levshakov2019}. More recently, studies have used the detailed structure of CO clouds to constrain the electron-proton mass ratio during the epoch of reionisation ($z\sim6$) \citep{Levshakov2020_me}. 
These works all show that at late times both $\aEM$ and $\me$ cannot depart by more that $\simeq 0.001-0.01\%$ from their standard lab values.

Given the clear connection between the decoupling of photons and the atomic processes during recombination, several groups have furthermore studied the changes in the CMB anisotropies arising from VFC \citep{Kaplinghat1999,Battye2001,Avelino2001,Scoccola2009,Menegoni2009, Menegoni2012,Planck2015var_alp}. 
These probe VFC mainly at recombination, complementing the aforementioned late-time constraints and limiting possible departures from the standard values to $\lesssim 0.1\%$ at $z\simeq 10^3$ \citep[see][for most recent constraints]{Hart2020a}.
These previous studies all focused on simple constant (i.e., time-independent) departures of $\aEM$ and $\me$ from their standard values. However, this picture ought to be unphysical and does not follow the motivations given by the aforementioned theoretical frameworks.
A more general treatment is therefore desirable.

In \citet{Hart2017}, the detailed effects of changes to $\aEM$ and $\me$ on the ionisation history were explored using the recombination code {\tt CosmoRec} \citep{Chluba2010b,Shaw2011}. 
In addition, \citet{Hart2017} considered a phenomenological time-dependence to the VFC using a power-law around pivot redshift $z=1100$, showing explicitly that more than just one model-parameter can be meaningfully constrained using \planck data. 
However, rather than propagating a phenomenological variation of fundamental constants, we can also use information about the recombination era (i.e., from the CMB anisotropies) to constrain the most likely time-dependent variations of the constants using a dimensional reduction technique known as \emph{principal component analysis} (PCA). This kind of analysis has been frequently used in cosmology \citep[see][for various examples]{Mortonson2008,Ishida2011,Finkbeiner2012,Farhang2011, Farhang2013, Dai2018,Campeti2019,Sharma2020}, but so far was not applied to VFC. 

In \citet[henceforth PCA20]{Hart2020b}, we developed our own PCA implementation code in C++ known as {\tt FEARec++} as a means to constrain the strongest principal components in the free electron fraction, $\xe$, as a function of redshift. There we created extensively orthogonal modes optimized specifically for the \planck 2015 likelihood, extending and improving on the pioneering works of \citet{Farhang2011, Farhang2013}. In PCA20, we also introduced a new parameter constraint apparatus coined the \emph{direct projection method}, which allows one to obtain constraints on explicit model parameters without the need to run the full analysis. 

In this paper, we revisit the formalism from PCA20 and directly apply it to the VFC modelling we developed for {\tt CosmoRec} in \citet{Hart2017}. The basic formalism includes the generation of Gaussian basis functions and the propagated responses to both the opacity and the CMB power spectra (Sect.~\ref{sec:howtoPCA}). In Sect.~\ref{sec:cvl}, we first generate the eigenmodes for a cosmic-variance-limited (CVL) experimental setup and investigate the structure and propagation from these variations in $\aEM$ and $\me$ to the CMB anisotropies. The \emph{direct likelihood} method from PCA20 is utilised in Sect.~\ref{sec:planckModes} to constrain the VFC principal components attainable from the \planck 2018 likelihood using a \emph{selective sampling} module patched onto {\tt CosmoMC} \citep{COSMOMC}. The obtained eigenmodes are included in a detailed MCMC analysis in Sect.~\ref{sec:mcmc}, where we present the marginalised results and contours using \planck 2018 baseline data. 
We find that with \planck data, three VFC modes can be constrained. No indication for significant departures from \LCDM are found (e.g., Tables~\ref{tab:alphaPlanck} and \ref{tab:mePlanck}).

Next, we briefly discuss the implications of the PCA for $\me$ variations on the Hubble tension (Sect.~\ref{sec:troubleshoot}). 
The basic idea was discussed for the \planck 2018 likelihood in \citet[henceforth referred to as VFC20]{Hart2020a}, where it was highlighted that $\me$ could play an important role through its combined effect on recombination and reionisation. 
Finally, in Sect.~\ref{sec:so} we use simulated noise curves from The \emph{Simons Observatory} (SO) forecasts with the analytic PCA method to generate predicted modes for this future CMB ground-based experiment.
Our conclusions are presented in Sect.~\ref{sec:conclusion}.
Several Appendices support our analysis and for completeness also present the latest $\xe$-PCA for \planck 2018, with marginal changes to PCA20, which was based on \planck 2015 data. 

\vspace{-3mm}
\section{Recap of the formalism}\label{sec:howtoPCA}
%----------------------------------------------

In this section, we briefly recapitulate on the PCA method used in PCA20 and how it is carried forward for this work. We also discuss the differences required for the fundamental constant analysis. For our study, the PCA methodology is fully implemented in the software package {\tt FEARec++}. Following PCA20, we generate a complete set of basis functions, $\phi_i(z)$, over a large redshift space $z_i\in\{300,2000\}$. These functions can be any given continuous shape, even periodic \citep[as shown in][]{Farhang2011}. The functions used in this work are Gaussians centred on $z_i$. It is important that they maximise \emph{orthogonality}, minimising overlap between neighbouring functions and optimise for \emph{completeness}, where the function space is covered as much as possible. In this analysis, we add these basis functions to the fine-structure constant $\aEM$ and the effective electron mass $\me$, such that
%----------------------------------------------
\begin{equation}
    \fc(z,z_i) = \fc_0\left(1+\frac{\Delta\fc}{\fc_0}\left(z,z_i\right)\right) = \fc_0\left[1+\phi_i(z)\right]
\end{equation}
%----------------------------------------------
where the fundamental constants $\fc\in\{\aEM,\me\}$ are perturbed by the basis function around $z_i$.

Once these functions are added to a recombination code such as {\tt CosmoRec} they induce a response in the CMB temperature and polarisation spectra, $C_\ell$, due to the changes during recombination. The CMB anisotropies can be calculated using a Boltzmann code: in our case {\tt CAMB} \citep{CAMB}. If we measure the relative difference between the \emph{`new'} power spectra with the added basis function and the fidicual power spectra as $\dlncl/\partial{p_i} \equiv 1/\cl \left(\dcl/\partial{p_i}\right)$, we can construct a \emph{Fisher matrix} of these responses by using the fiducial cosmology and a given noise specification as the effective covariance matrix for the experiment. This Fisher machinery can be thought of as an $n-$dimensional signal-to-noise matrix where $p_i$ defines the amplitudes of the Gaussian functions centred on $z_i$. 

\begin{figure}
    \centering
    \includegraphics[width=.97\linewidth]{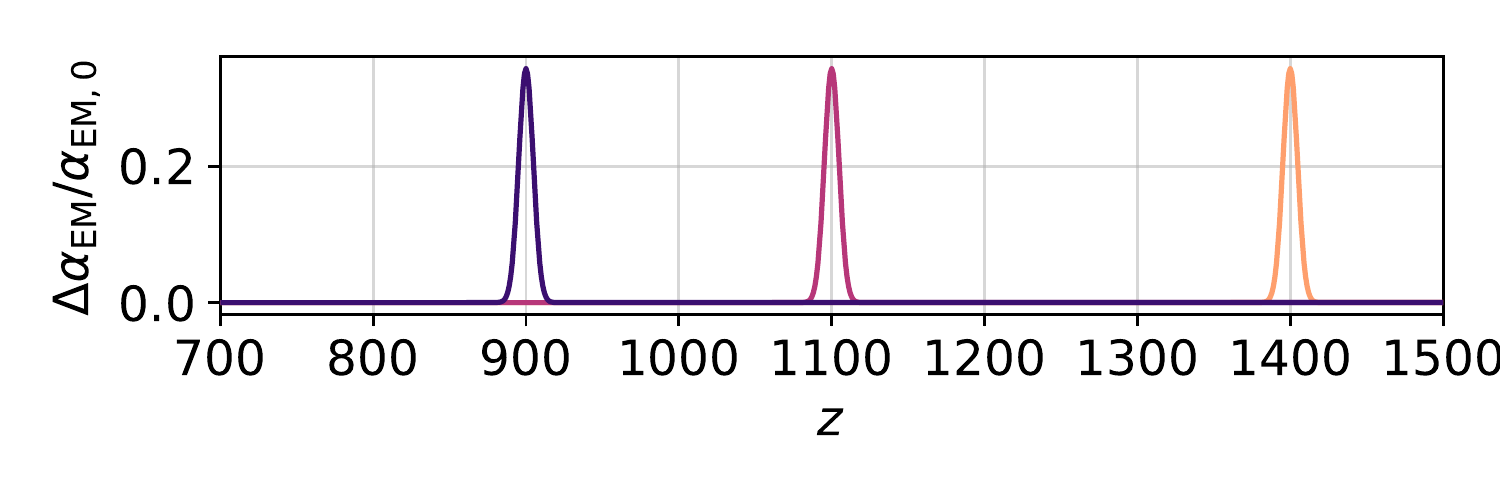}
    \includegraphics[width=\linewidth]{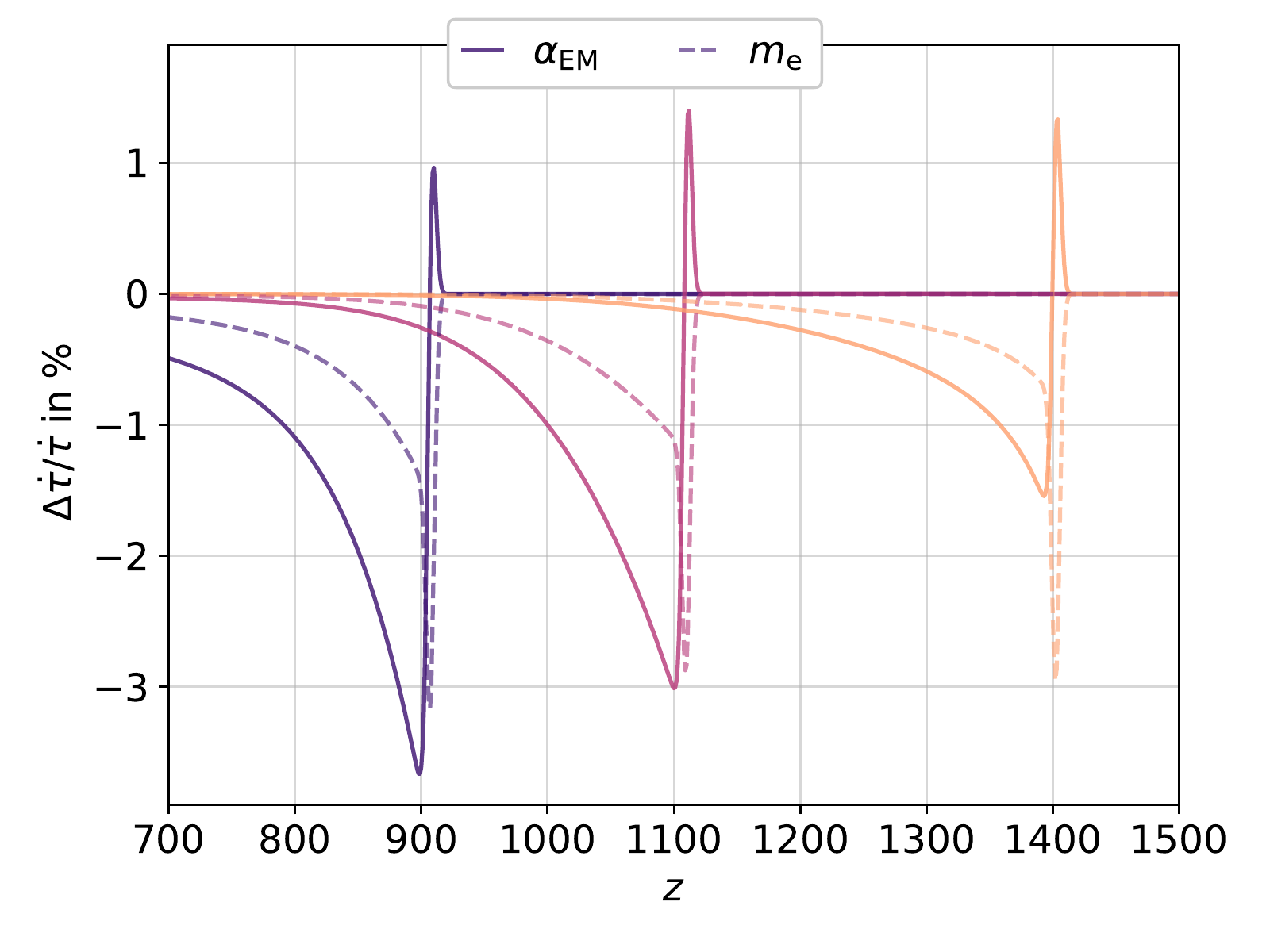}
    \includegraphics[width=\linewidth]{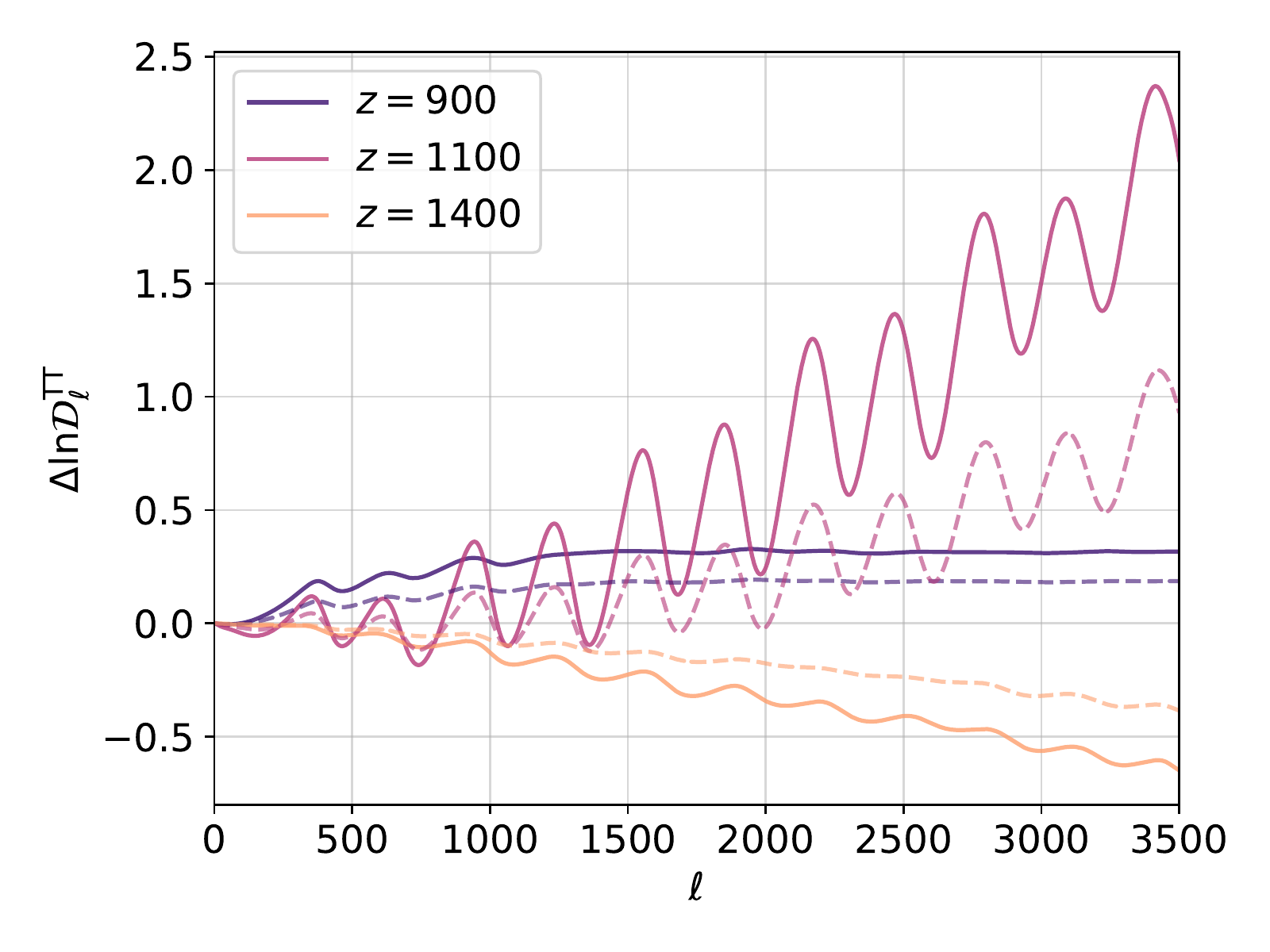}
    \caption{Responses in the weighted free electron fraction $\taudot$ \emph{(central)} and the $\ell$-weighted CMB temperature angular power spectra $\mathcal{D}_\ell$ \emph{(bottom)} for Gaussian basis functions (example for $\aEM$ shown in \emph{top} panel). These are given for $\aEM$ \emph{(solid)} and $\me$ \emph{(dashed)} around the pivot redshifts $z_i  =\{900,1100,1400\}$. All curves are plotted as relative differences against the \LCDM model.}
    \label{fig:responses}
\end{figure}

In Fig.~\ref{fig:responses}, we have shown how the Gaussian changes in $\mathcal{C}$ propogate through the opacity/free electron fraction, and consequentially project onto the CMB power spectra\footnote{Movies of these responses will be made available online at:\\  \url{https://cosmologyluke.github.io}.}. From the Gaussian responses to the opacity, there are sweeping negative variations in $\taudot$ (for $\delta\fc/\fc>0$) that arise from the $\xe$ variations. The superposed peaks on both types of variations (positive peaks for $\aEM$ and negative for $\me$) result from the $\sigT$ changes that affect the visibility functions \citep[see][and Sect.~\ref{sec:sigT}]{Hart2017}. These variations translate into $\delta\Dell$ variations that are most responsive around the redshift of most probable last scattering, $z_*\simeq 1100$. Given that the $\xe$ variations for a given $\Delta\fc$ are largest around this epoch as well, the responses in $\DellT$ and $\DellE$ are hyper-focused around this epoch, with greater diminishes in the tails. 

%----------------------------------------------
%----------------------------------------------

\vspace{-2mm}
\subsection{Fisher matrices}\label{sec:fisher}
%----------------------------------------------
The Fisher matrix can be written as the second derivative of the log-likelihood function, $\ln\mathcal{L}\left(\vec{p}\,|\,{\bf d}, M\right)$ around the maximum likelihood location, where $\vec{p}$ are the parameter values of a given model $M$ and ${\bf d}$ is the data (from an experiment such as \emph{Planck}). However for a simple, CMB-like experiment, we can simplify this using the following equation:
%----------------------------------------------
\begin{equation}
    F_{ij} = \left<\frac{\partial^2\ln\mathcal{L}}{\partial p_i^2}\right> = \sum_{\ell=0}^{\ell_{\rm max}}\frac{\dclvec}{\partial p_i}\cdot{\bf \Sigma}_\ell^{-1}\cdot\frac{\dclvec}{\partial p_j},
\label{eq:Fij}
\end{equation}
%----------------------------------------------
where the CMB power spectra vector is given by,
%----------------------------------------------
\begin{equation}
    \vec{C}_\ell = \left(C_\ell^{TT}, C_\ell^{EE}, C_\ell^{TE} \right)
\end{equation}
%----------------------------------------------
and the covariance matrix for a given multipole $\ell$ is,
%----------------------------------------------
\begin{equation}
    {\bf \Sigma}_\ell = \frac{2}{2\ell+1}
    \begin{bmatrix}
    C^{\rm TT^2} & C^{\rm TE^2} & C^{\rm TT}C^{\rm TE} \\
    C^{\rm TE^2} & C^{\rm EE^2} & C^{\rm TE} C^{\rm EE} \\
    C^{\rm TT}C^{\rm TE} & C^{\rm TE}C^{\rm EE} & \frac{1}{2}\left(C^{\rm TE^2}+C^{\rm TT}C^{\rm EE}\right)
    \end{bmatrix}_\ell.
\end{equation}
%----------------------------------------------
Note that here we have assumed there is no cross-multipole correlations ($\ell\times\ell'$ terms are 0) allowing us to use the summation  in Eq.~\eqref{eq:Fij}. Effects of detector noise have been investigated for changes to recombination in previous works \citep{Farhang2011}. This formalism of the Fisher matrix has been used extensively in the literature \citep{Tegmark1997,Verde2009,Finkbeiner2012}\footnote{It is also important to point out that the derivative of the log-likelihood in Eq.~\eqref{eq:Fij} is over an ensemble average.
This is an important detail we have assumed for our data-driven \emph{direct likelihood} approach.}. 

\subsubsection{Direct likelihood method}\label{sec:direct}
%----------------------------------------------
For the \planck data, the likelihood function is directly sampled along with the same basis functions and then the Fisher matrix is calculated using the \emph{finite difference method} with a second-order stencil. The likelihood is evaluated using {\tt CosmoMC} and the current \planck 2018 likelihood code \citep{COSMOMC,Planck2018like}. This is an effective way of extracting eigenmodes whilst also removing correlations induced by cosmological parameters and nuisance parameters associated with the \planck data\footnote{The object-oriented nature of the {\tt FEARec++} code means this is malleable towards any alternative dataset. One such generality is the addition of nuisance parameters external to \emph{Planck}.}. Details on the implementation of  {\tt FEARec++} and the validation of the direct likelihood method are explained in PCA20. Subsequently, the stability analysis of the \planck likelihood code required for this method, along with a comparison to the 2015 likelihood approach, are included in Appendix~\ref{app:planck}.

\subsubsection{Principal components}\label{sec:PCA}
%----------------------------------------------
To generate principal components, the Fisher matrix is diagonalised and decomposed into its eigenbasis such that,
%----------------------------------------------
\begin{equation}
    F_{ij} = S_{im}\cdot\mathcal{F}_{mn}\cdot S_{nj},
\end{equation}
%----------------------------------------------
where $S_{im}$ is the matrix of eigenvectors of the Fisher matrix and $\mathcal{F}_{ab}$ is a diagonalised matrix of the eigenvalues. These  eigenvectors are recast as eigenfunctions using the basis functions we generated initially. If we create $N$ basis functions initially, this can be written formally as,
%----------------------------------------------
\begin{equation}
    E_m(z) = \sum_{i=1}^N S_{im}\,\phi_i(z).
\end{equation}
%----------------------------------------------
The $E_m(z)$ functions are the principal components we have been wishing to generate and they are ranked by their eigenvalues (i.e., the largest eigenvalue gives the most probable principal component). In reality, we take the amplitude of each of the matrix elements for a given function $E_m$ and then interpolate over this since this is much smoother for the Boltzmann code to process. All the linear algebra stages of this implementation are done by {\tt Eigen3} due to their efficient C++ libraries that have been utilised \citep{EigenCode}. We can use the Kramer-Rao inequality to estimate the error of each mode such that $\sigma_i\gtrsim \sqrt{\left(\mathcal{F}^{-1}\right)_{ii}}$. 

\vspace{-2mm}
\subsection{Using Monte Carlo simulations to constrain the modes}\label{sec:constraints}
%----------------------------------------------
Once the VFC modes have been constructed, both the analytical and direct-method generated eigenmodes can be incorporated onto a Markov Chain Monte Carlo (MCMC) simulation using amplitudes $\mu_i$ such that,
%----------------------------------------------
\begin{equation}
    \mathcal{C}\left(z\right) = \mathcal{C}_0\left(1+\sum_i^M\mu_iE_i(z)\right).
\end{equation}
%----------------------------------------------
Here, $\mu_i$ amplifies the relative strength of a given mode, where lower $i$ correspond to the \emph{more} constrainable components. Equally we can set $M$ as the limit of the modes hierarchy that have enough relevant information depending on a particular condition. The criterion here is the error information defined by the Kramer-Rao inequality, where in this analysis (as in the last analysis), the first three eigenmodes hold the majority of the information ($\simeq 99\%$). Our configuration for MCMC analysis is explained in more detail in Sect.~\ref{sec:mcmc}, with a focus on the \emph{direct projection method} in Sect.~\ref{sec:proj}.

\vspace{-2mm}
\subsection{Changes in the Fisher machinery for VFC}
\label{sec:changes}
%----------------------------------------------
There have been several modifications to the approach from PCA20 to optimise the analysis for time-varying fundamental constants. Firstly, as shown in \citet{Hart2017}, there is a sharp cut-off in the effects to the recombination history, $\xe$ when $z\gsim 1500$. This means that the responses in the CMB radically disappear above this redshift. For this reason, the number of basis functions has been reduced to $N=80$ over a narrower range for the generation of these eigenmodes. The modes have been created up to $z_i = 2000$ for both constants. Even though there were small effects on helium recombination coming from variations in $\aEM$ and $\me$, the larger effects from around the peak of the Thomson visibility function ($z\sim1100$) coupled with the weaker constraining power in the CMB anisotropies from higher-redshift recombination features washes out these variations. Since the higher-order principal components have much larger errors (much smaller eigenvalues), it is unlikely that these redshifts can be constrained with CMB data as part of a principal component analysis.

\subsubsection{Propagating additional contributions from VFC}
\label{sec:changesXe}
%----------------------------------------------
When adding the basis functions to the ionization history, the small perturbations are propagated through to the CMB anisotropies as discussed in previous papers \citep{Farhang2011,Finkbeiner2012,Hart2020b}. However when we include fundamental constants, the effects are not exclusive to the free electron fraction. This was clarified in previous studies of the \planck 2015 data \citep{Planck2015var_alp,Hart2017}. We showed that there is a non-negligible contribution from the rescaling of the Thomson cross section ($\sigT$). As a result, one can reparametrise the fundamental constant variations arising from recombination by using the \emph{opacity}, also known as the differential Thomson optical depth\footnote{In other pieces of literature, $\taudot$ refers to a derivative with respect to conformal time; however we restrict ourselves to redshift in this analysis.}, $\taudot$, where in this study,
%----------------------------------------------
\begin{equation}
    \taudot\equiv\frac{\dif{\tau}}{\dif{z}} = -\frac{N_{\rm H}(z)\,\xe(z)\,\sigT(z)\,c}{H(z)\,(1+z)},
\end{equation}
%----------------------------------------------
where $N_{\rm H}$ is the total hydrogen number density and the Hubble factor, $H(z)$, is independent of the fundamental constants\footnote{Strictly speaking, there are  models where fundamental constant variations affect the background energy density, and by proxy $H(z)$, depending on the underlying mechanism. Here we only discuss phenomenological variations in $\aEM$ and $\me$ arising from recombination. For further discussion of these theories, we point to a recent review in \citet{Martins2017review}.}. Therefore, if we measure the responses in $\taudot$ we will extract the full variation with respect to the fundamental constant basis functions. The opacity variations are illustrated in Fig.~\ref{fig:responses} for both $\aEM$ and $\me$. The spikes in positive or negative directions close to $z_i$ arise directly from the extra $\left[1+\phi_i(z)\right]^2$ term in $\taudot$ that is convolved with the variation arising from the free electron fraction $\xe$. Note that the Thomson cross section depends on the fundamental constants discussed such that $\sigT=\left(\aEM/\aEMs\right)^2\left(\me/\mes\right)^{-2}$.

\subsubsection{Amplitude normalisation for the MCMC code}
\label{sec:amps}
%----------------------------------------------
In PCA20, we discussed the amplitude adjustments required for different redshifts when generating basis functions for the direct likelihood method with {\tt CosmoMC}. Once the $\aEM$ and $\me$ modes have been constructed for an idealised CVL experiment, the diagonal of the Fisher matrix serves as the weighting function for the different redshift bins used in the direct likelihood method. Given the delicate nature of the direct-likelihood method, this was required to insist on numerical stability when generating the eigenmodes. Though these new responses in $\xe$ are non-trivial when a Gaussian is added to the $\aEM$ or $\me$ parameter during recombination, these \emph{weighting template functions} were very similar and helped constrain numerically stable modes such as those presented in Sect.~\ref{sec:planckModes}. 

\subsubsection{Differences in the marginalisation}
\label{sec:marg}
%----------------------------------------------
As in the previous paper, we remove the correlations of the principal components from both cosmological and nuisance parameters by using the identity,
%----------------------------------------------
\begin{equation}
  \left({\bf F}^{-1}\right)_{pp} = \left({\bf F}_{pp}-{\bf F}_{ps}{\bf F}_{ss}^{-1}{\bf F}_{sp}\right)^{-1},
  \label{eq:marg}
\end{equation}
%----------------------------------------------
where ${\bf F}_{pp}$ refers to the sub-matrix of the Fisher matrix pertaining to the \emph{principal components} and ${\bf F}_{ss}$ refers to the sub-matrix pertaining to the standard parameters: cosmological and nuisance. The nuisance parameters are a combination of foregrounds and systematics from the data-processing of the \planck data, however the majority of this machinery remains unchanged between 2015 and 2018. The only difference as far as the simulations are concerned is that the 2018 baseline polarisation data includes no dust-contamination amplitude parameters (referred to as $A^{{\rm dust}EE}_{\mathcal{F}}$ in Table~C1 of PCA20). This is due to the cosmology being insensitive to the dust amplitudes of these particular parameters for $EE$ CMB power spectra \citep[see Sect.~3.3.2 of][for more details]{Planck2018like}. For full transparency, this leads to $N_s = 25$ with 6 less parameters than our previous analysis.

\section{Cosmic variance limited (CVL) experiment}\label{sec:cvl}
%----------------------------------------------

%----------------------------------------------
\begin{figure}
    \centering
    \includegraphics[width=\linewidth]{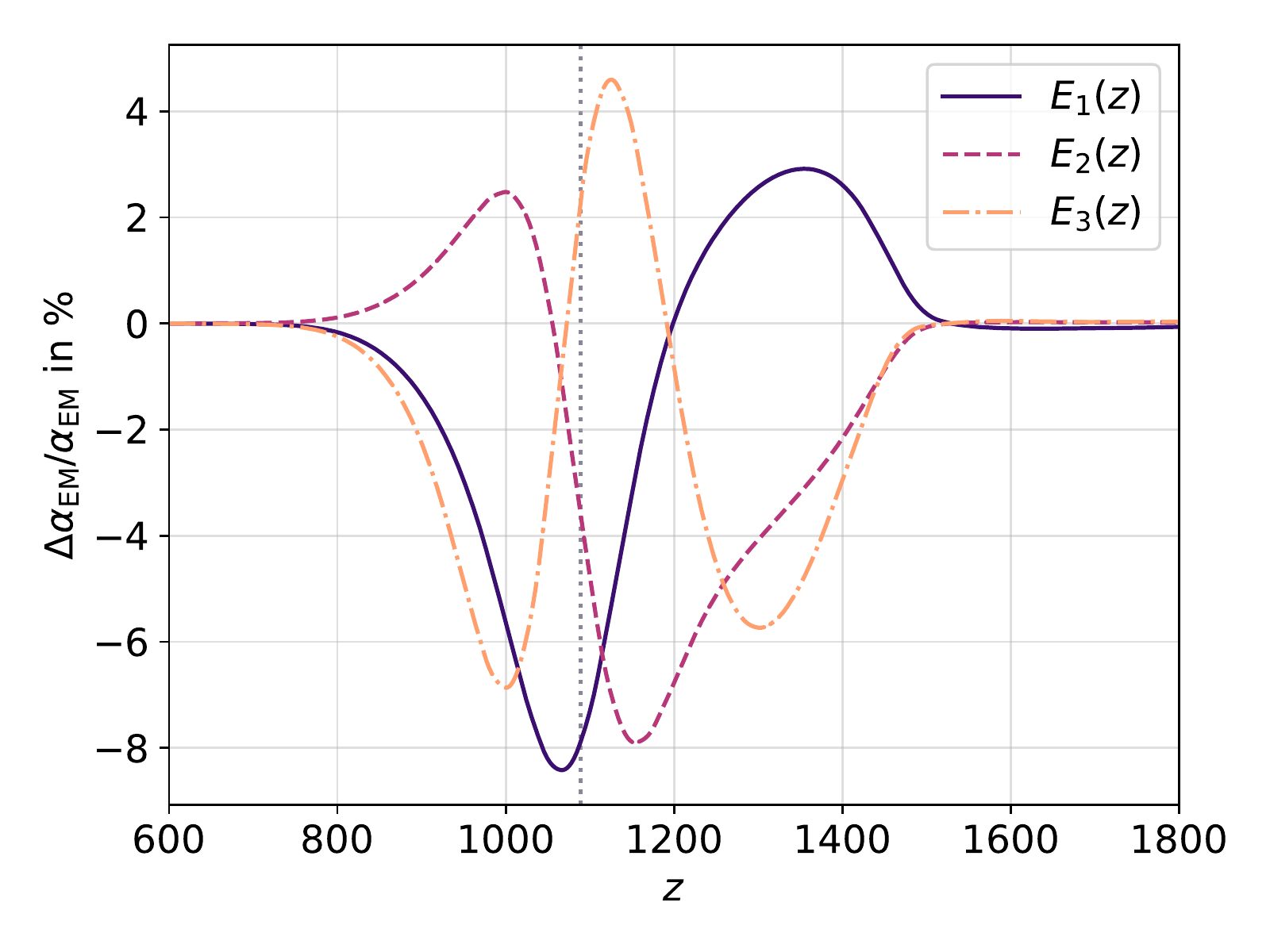}
    \includegraphics[width=\linewidth]{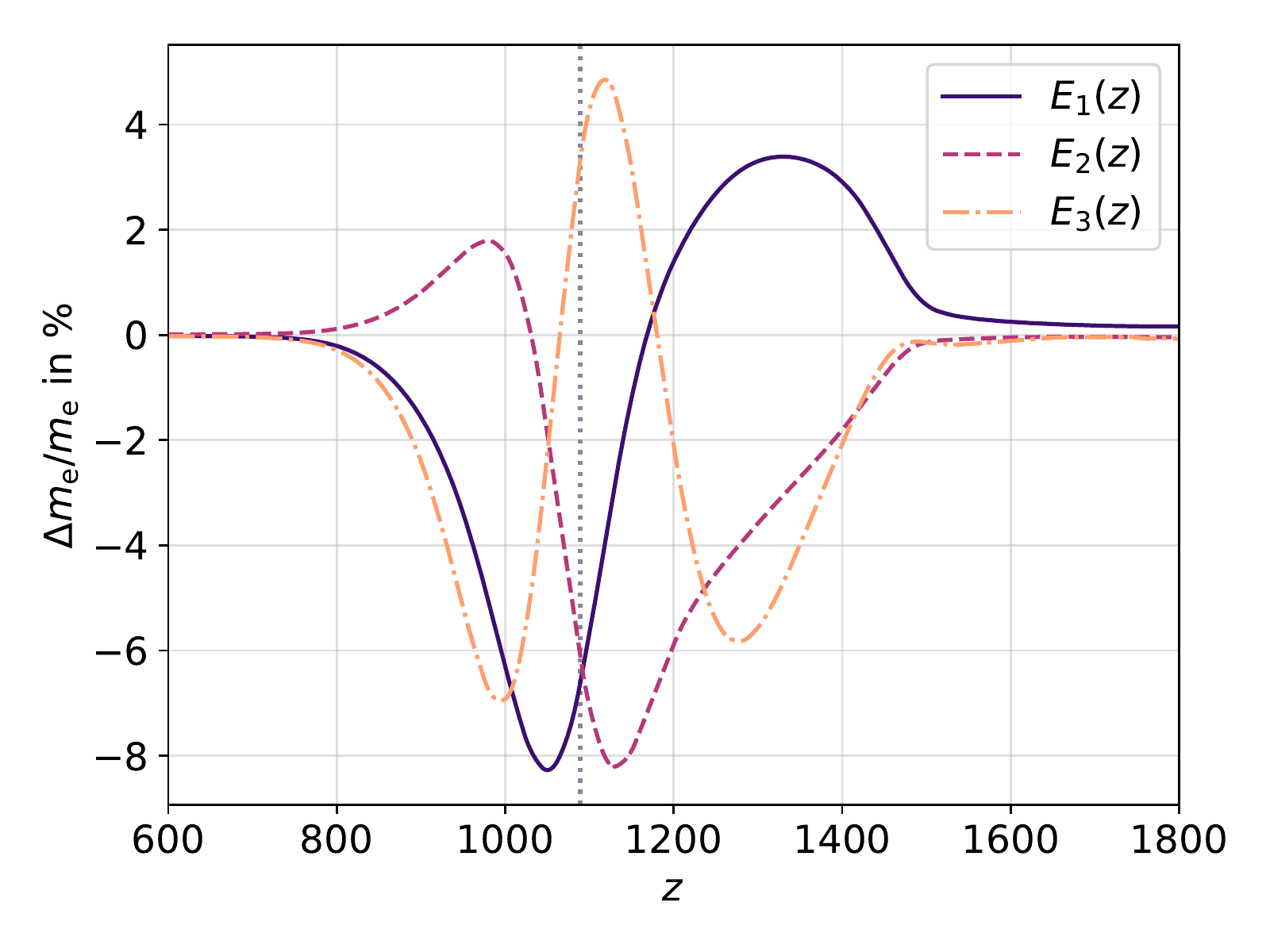}
    \caption{The fine-structure constant \emph{(top)} and electron-mass \emph{(bottom)} eigenmodes for a CVL-experiment with $\ell_{\rm max} = 3500$. Here the redshift associated with the last scattering surface, $z_*=1088$ is shown as a dashed curve. Note here that the resultant modes for $\me$ have been multiplied by -1 to compare symmetry with $\aEM$.}
    \label{fig:modesCVL}
\end{figure}
%----------------------------------------------

%----------------------------------------------
\begin{figure}
    \centering
    \includegraphics[width=\linewidth]{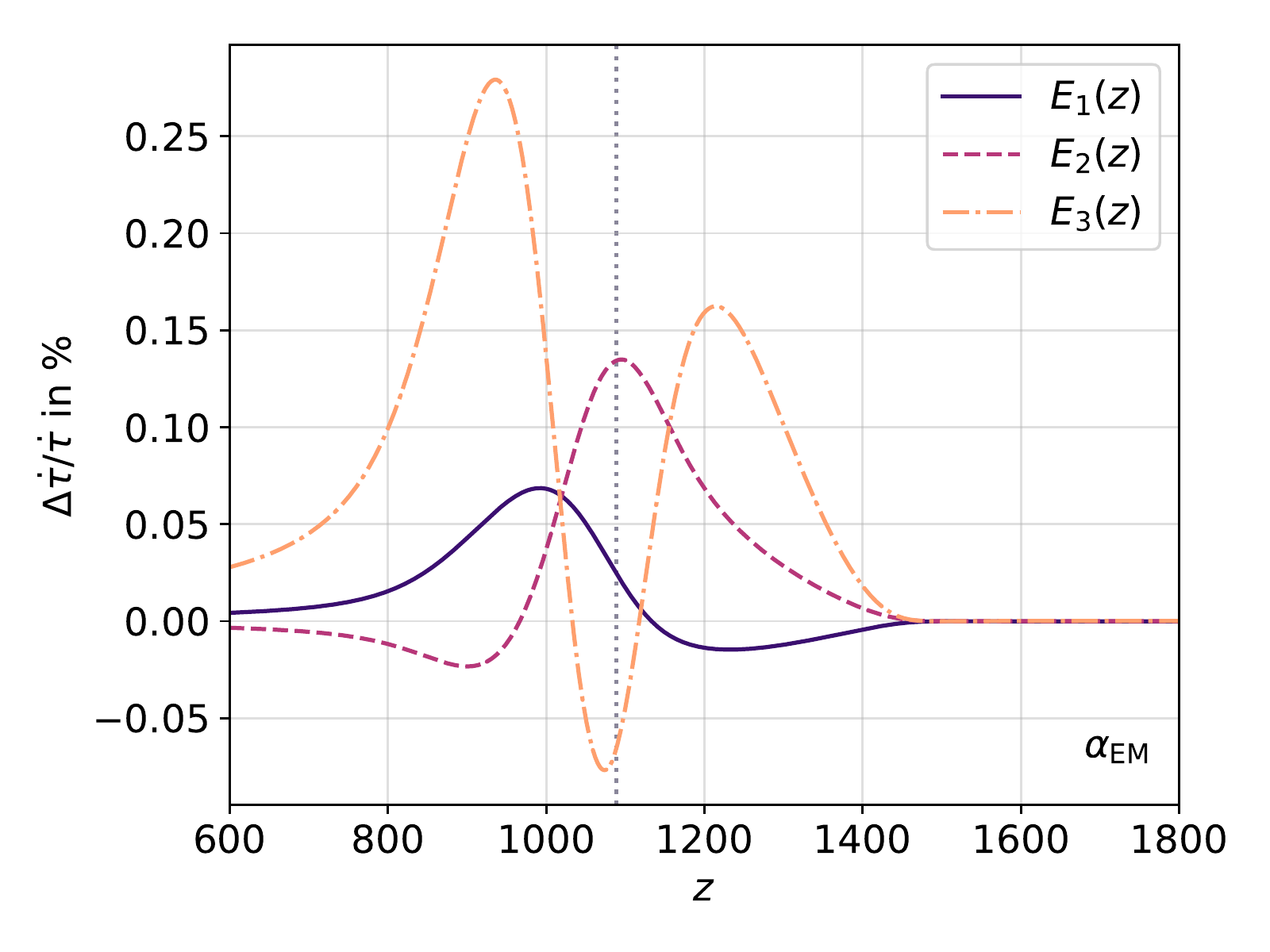}
    \includegraphics[width=\linewidth]{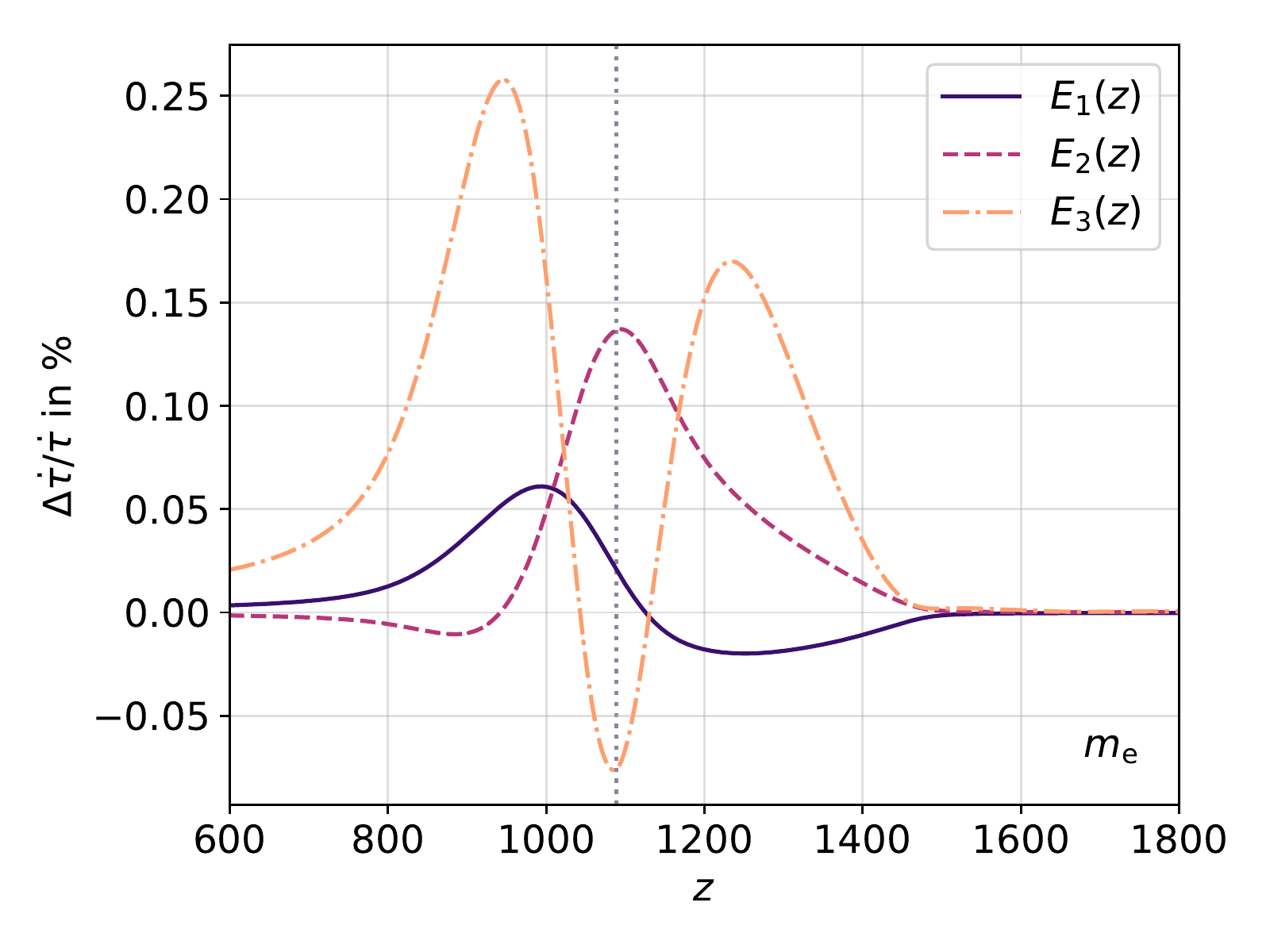}
    \caption{The differential optical depth (opacity) $\taudot$ variations that are caused by the fundamental constant CVL modes from Fig.~\ref{fig:modesCVL}: $\aEM$ \emph{(top)} and $\me$ \emph{(bottom)}. The amplitudes of these eigenmodes are lifted directly from the predicted errors of the Fisher matrix calculation. As in Fig.~\ref{fig:modesCVL}, the $\me$ case has been multiplied by $-1$ for symmetry comparisons.}
    \label{fig:taudotCVL}
\end{figure}
%----------------------------------------------

%----------------------------------------------
\begin{figure}
    \centering
    \includegraphics[width=\linewidth]{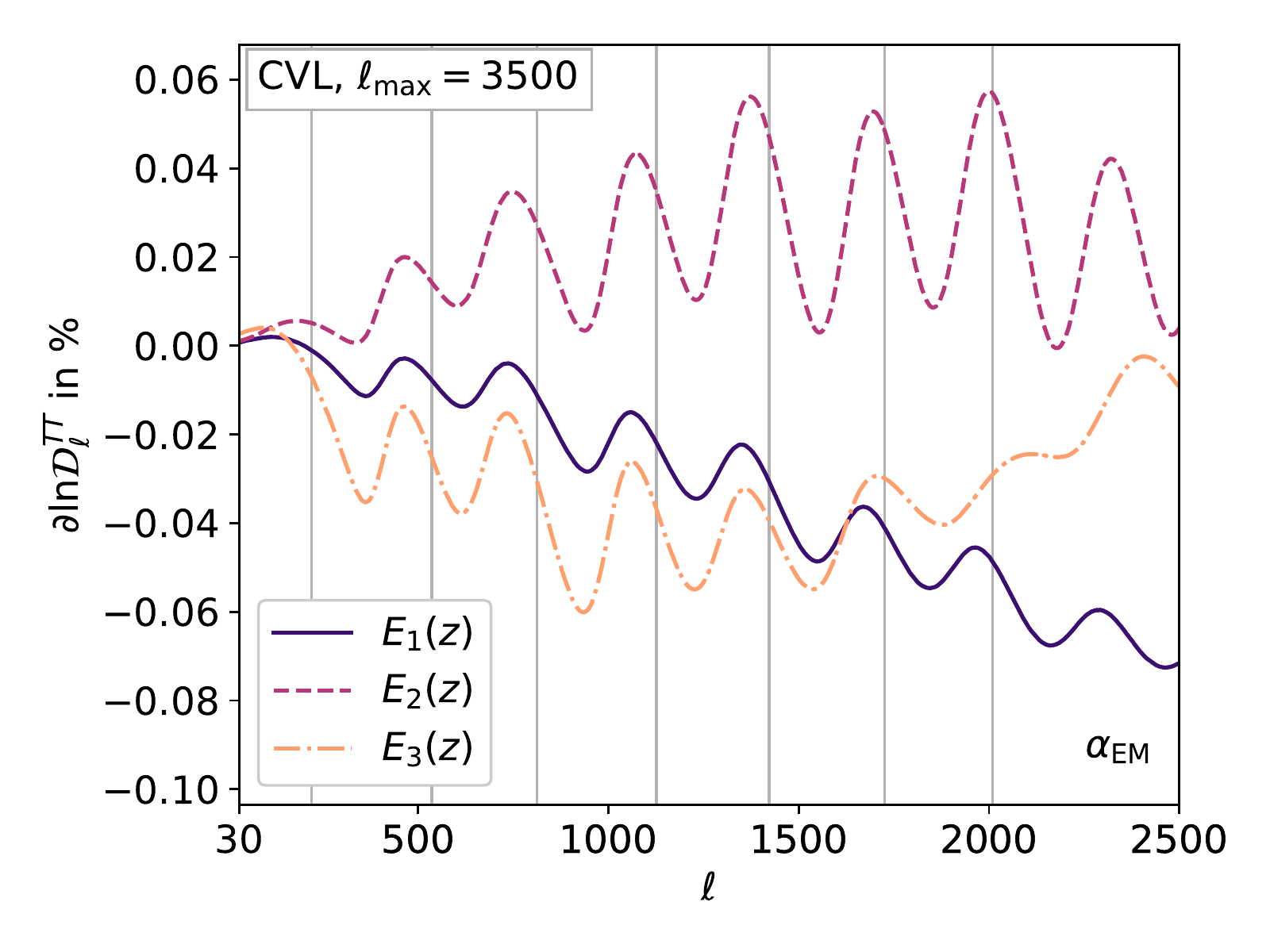}
    \includegraphics[width=\linewidth]{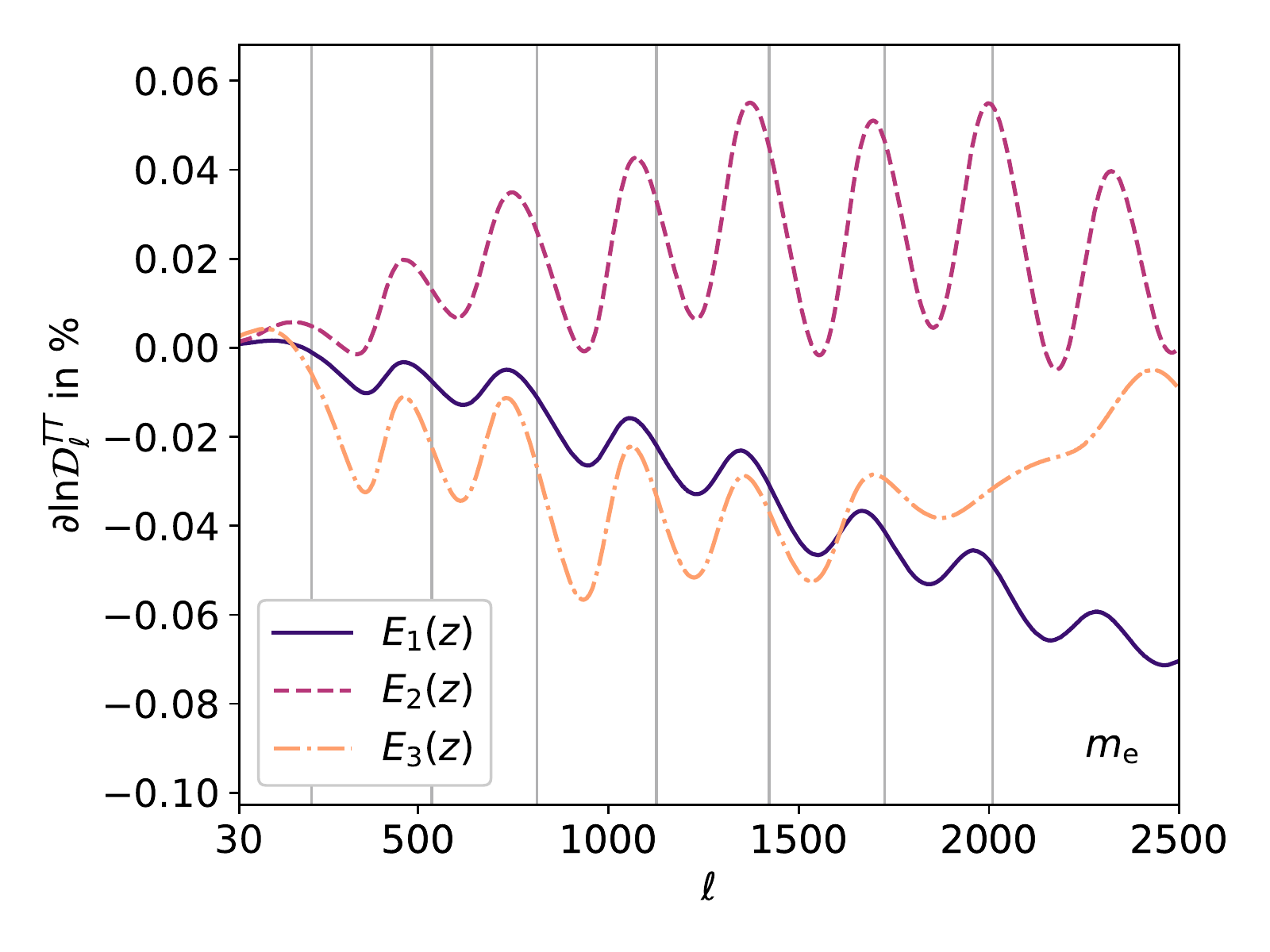}
    \caption{Responses of the CMB temperature angular power spectra according to the fine-structure constant modes \emph{(top)} and electron mass modes \emph{(bottom)} constrained by a CVL-experiment with $\ell_{\rm max} = 3500$ in Fig.~\ref{fig:modesCVL}. These eigenmodes propagate through the Thomson optical depth (Fig.~\ref{fig:taudotCVL}) and then onto the CMB anisotropies. The grey lines correspond to the peaks of the \planck 2018 \LCDM fiducial power spectra.}
    \label{fig:clTTCVL}
\end{figure}
%----------------------------------------------

%----------------------------------------------
\begin{figure}
    \centering
    \includegraphics[width=\linewidth]{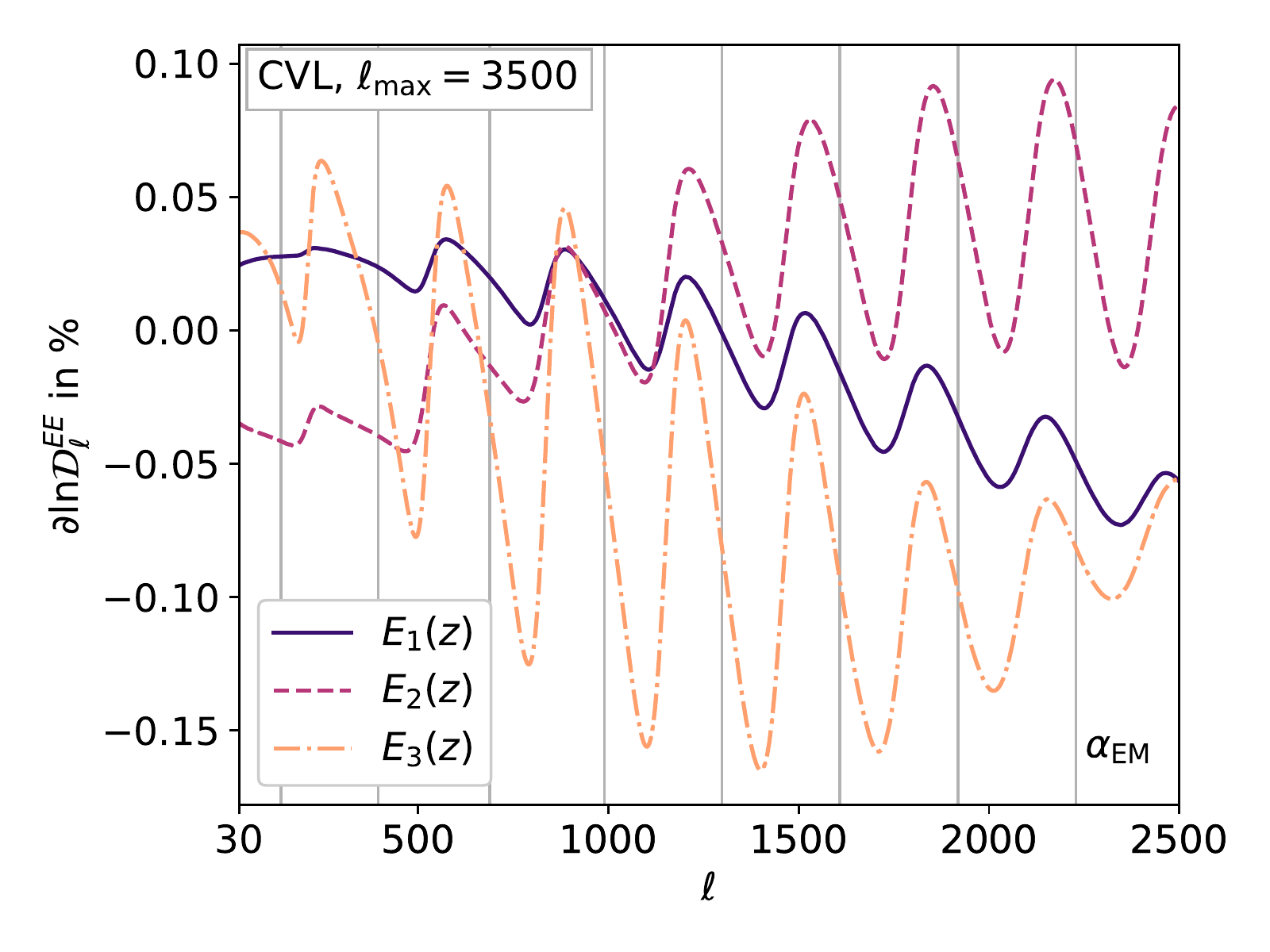}
    \includegraphics[width=\linewidth]{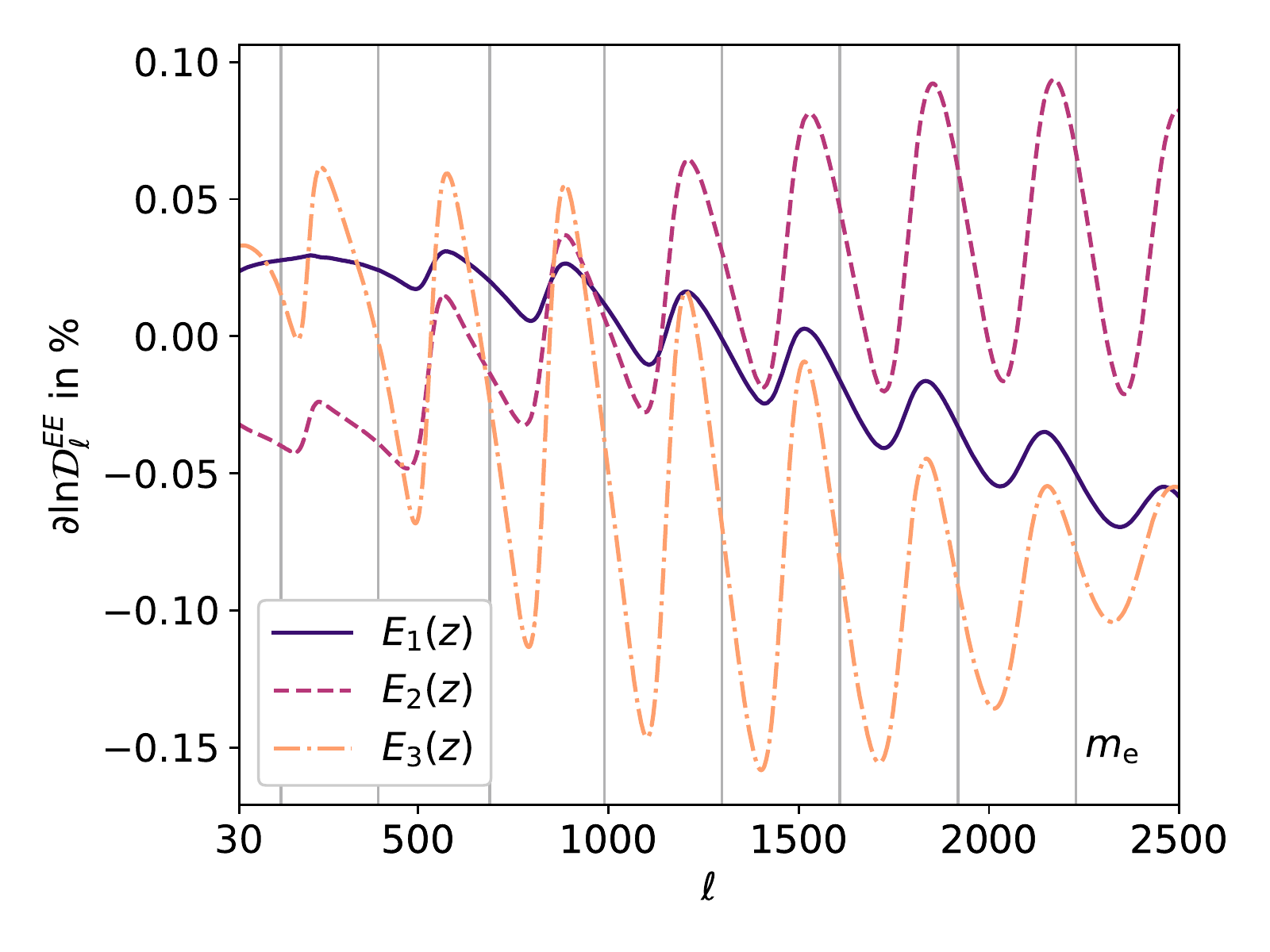}
    \caption{Responses of the CMB $EE$ polarisation angular power spectra according to the fine-structure constant modes \emph{(top)} and electron mass modes \emph{(bottom)} constrained by a CVL-experiment with $\ell_{\rm max} = 3500$ in Fig.~\ref{fig:modesCVL}. The grey lines correspond to the peaks of the \planck 2018 \LCDM fiducial polarisation power spectra.}
    \label{fig:clEECVL}
\end{figure}
%----------------------------------------------

As mentioned in Sect.~\ref{sec:howtoPCA}, the simplest configuration for a PCA with the CMB anisotropies simulates a CVL-like experiment. The covariance (effective noise) of the Fisher matrix for this setup is made up solely from the fiducial CMB $C_\ell$s. Here we present the results for the eigenanalysis. 
For this section, we have flipped the sign of the $\me$ eigenmodes so they can be more directly compared to the $\aEM$ variations. This has also been propagated to the responses in the CMB power spectrum, $C_{\ell}$. We ask the reader to bear this in mind when the full parameter constraints are shown in Sect.~\ref{sec:constraints}. 

The modes for $\aEM$ and $\me$ are shown in Fig.~\ref{fig:modesCVL}. In this figure, they have been normalised as explained in Sect.~\ref{sec:PCA} and the most likely redshift for a photon to decouple, $z_*=1088$, is indicated by a vertical dotted line.

In our previous paper, we found that the first eigenmodes in the hierarchy are most sensitive around the FWHM of the Thomson visibility function ($970<z<1170$). The relative changes of the $\aEM$ and $\me$ modes in that window are incredibly similar. One key difference is the higher redshift behaviour for both modes at $z>1500$, leaking into the neutral helium recombination era. In this epoch, the fine structure constant modes sharply fall to $\Delta\aEM\simeq0$ however the $\me$ components tail off parallel to the origin for these higher redshifts. This will be discussed more in Sect.~\ref{sec:sigT}. In the second eigenmode, $E_2(z)$, we can see from Fig.~\ref{fig:modesCVL} that the wiggly shape is more pronounced for the $\aEM$ modes at $z\simeq1300$. 

The modes are suspiciously similar when first compared to the independent changes to $\xe$ arising from variations in $\aEM$ and $\me$. The $\me$ and $\aEM$ variations affect the $\xe$ fraction in distinctly different ways, particularly at lower redshifts, $z<500$. However, the most constrainable eigenmodes in the hierarchy are all centered around the recombination redshift, $z_*$. At this redshift, the variations become almost indistinguishable, save for their relative magnitudes (encoded in their eigenvalues, see Table~\ref{tab:fisherErrors}). 

The propagation of these modes into the opacity (differential optical depth, $\taudot$ as previously mentioned) as a residual $\Delta\taudot/\taudot$ are shown in Fig.~\ref{fig:taudotCVL}. The responses from the first 3 modes are almost identical, mirroring the mode structures in Fig.~\ref{fig:modesCVL}; however, the 3rd opacity residual of $\me$ is slightly shifted to higher redshifts. For both constants ($\aEM$ and $\me$), the opacity arises from the modes with their predicted Fisher errors from Table~\ref{tab:fisherErrors}. Since these are larger for $E_3$, the amplitude is much higher. However, the responses from the CMB are similar in magnitude. 

We have included the impact on both the temperature and $E$-mode polarisation angular power spectra\footnote{Note that in this work, we will interchangeably talk about $C_\ell$ and $\mathcal{D}_\ell$ spectra. Here $\mathcal{D}_\ell \equiv \ell\left(\ell+1\right)C_\ell/\left(2\pi\right)$. The function $\mathcal{D}_\ell$ highlights the smaller scale features of the CMB spectra more effectively.} in Figs.~\ref{fig:clTTCVL} and~\ref{fig:clEECVL}. The responses for both constants in the $TT$ power spectra show the relative changes with the same $\sim\pi/2$ phase change with respect to the CMB acoustic peaks (grey lines in Fig.~\ref{fig:clTTCVL}). The magnitude of these responses is propagated from the same responses in the opacity from Fig.~\ref{fig:taudotCVL}, hence the similar magnitudes in $\partial\ln\Dell$. 
The shift is consistent with a drift to smaller multipoles (larger scales), however the overall downward trend of the residual corresponds to sharper damping of the peaks. This mimics several aspects of the modes discussed in PCA20, notably that the CMB $TT$ spectra responses that emerge when varying $\ns$. By increasing the matter power spectral index, this sharply modifies the Silk damping envelope for the CMB power spectra. This effect is less prominent for the 2nd and 3rd modes where in particular the 2nd mode gives a sinusoidal-like residual in the $TT$ power spectra. This indicates shifting in the CMB acoustic peaks, in phase with the variations from $E_1$. The third mode is a complicated superposition of the two effects where the damping effect becomes less prominent at higher multipoles. Furthermore, the sinusoidal-like pattern of the $E_3$ residual goes gradually out of phase with the first two eigenmodes at higher $\ell>2000$. This reflects similar mode patterns in the CMB temperature spectra from previous PCA analyses \citep{Planck2015params, Hart2020b}. In Fig.~\ref{fig:clEECVL}, there is a similar effect on the polarisation responses, $\partial\ln\mathcal{D}_\ell^{\rm EE}$. For the $EE$ polarisation signal, the responses in the CMB behave similarly for modes $E_1$ to $E_3$, with a larger residual envelope size. This is due to the smaller magnitude of the $EE$ polarisation power spectra. 

\subsection{Effects on the Thomson cross section}\label{sec:sigT}
\begin{figure}
    \centering
    \includegraphics[width=\linewidth]{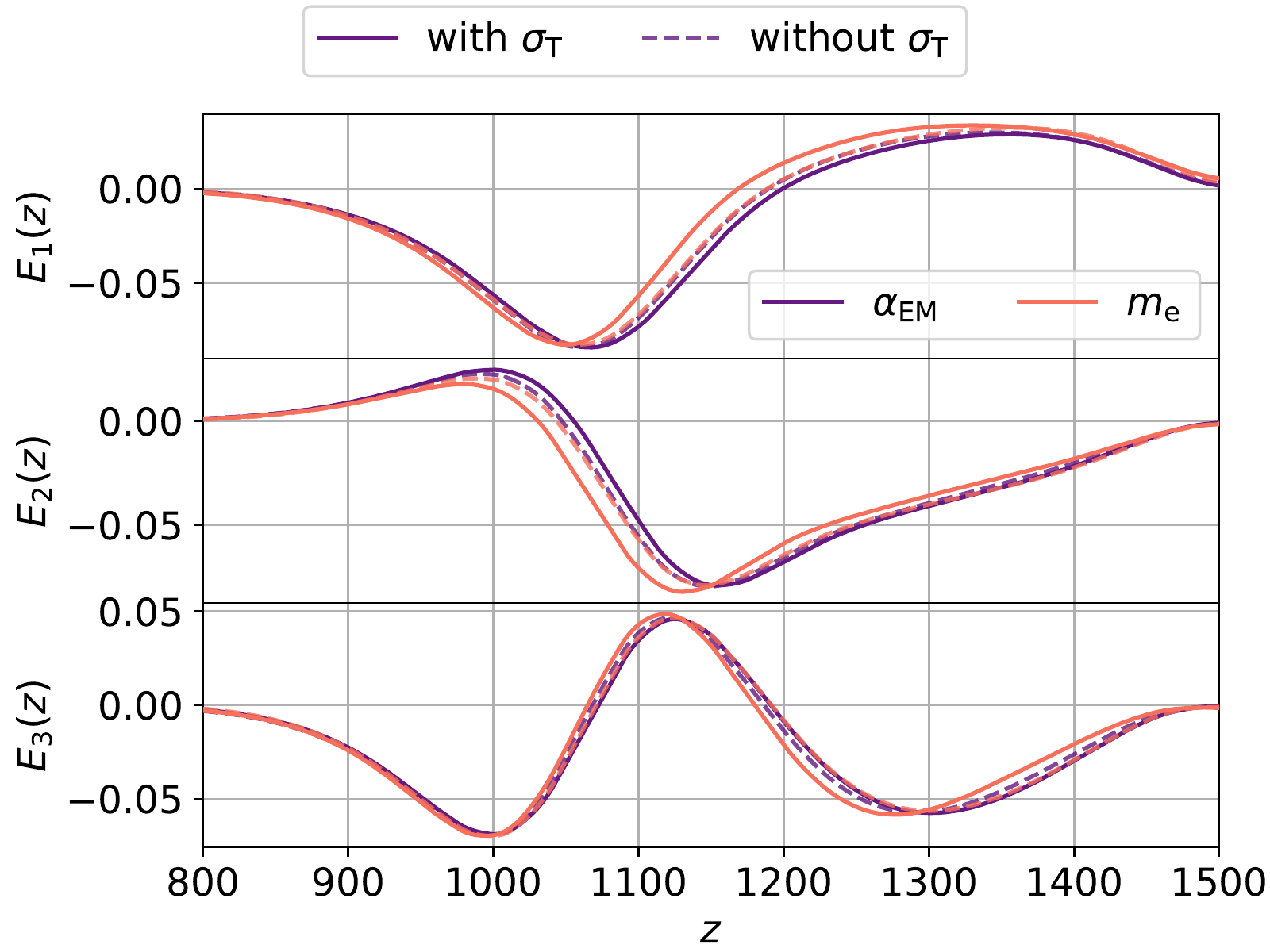}
    \caption{The first three eigenmodes in a CVL experiment both with (\emph{solid}) and without (\emph{dashed}) $\sigT$ changes arising from a varying $\aEM$ and $\me$. The two parameters shown are coloured \emph{purple} and \emph{orange} respectively. Outside of the range shown ($800<z<1500$) the observed differences between the modes are $<0.1\%$.}
    \label{fig:sigT}
\end{figure}

%----------------------------------------------
As mentioned in Sect.~\ref{sec:changesXe}, the Thomson cross section needs to be rescaled when propagating the variations of $\aEM$ and $\me$ to the CMB anisotropies. The effects of including that correction for the eigenmodes, for a CVL-experiment are shown in Fig.~\ref{fig:sigT}. In this figure, we show the comparison when including this correction for the first 3 eigenmodes. When the $\sigT$ rescaling is removed from the analysis, the eigenmodes for $\me$ and $\aEM$ almost entirely overlap. However, when the full correction is included, the first peaked features of $E_1$ at $z\sim1050$ and $z\sim1350$ begin to shift. For $\aEM$ the features slightly drift to higher redshifts, whereas they drift to lower redshifts for $\me$. There is also a peculiar feature at $z\gsim1500$ where the $\me$ mode tails off less sharply. From inspecting the responses in Fig.~\ref{fig:responses}, this comes from the additional negative change to $\taudot$ from the $\sigT$ scaling, prolonging the effects of the basis functions at higher redshifts. However, these high redshift features were also seen in the $\xe$ eigenmodes in PCA20 and they were hindered greatly when real data like \planck was included \citep[see Figs.~3-4 of][for comparison]{Hart2020b}. 

From the Fisher matrix eigenvalues (see Sect.~\ref{sec:howtoPCA}) we see that the predicted errors for both fundamental constant modification examples are different when including variations to $\sigT$. This latter example is what we show in Table.~\ref{tab:fisherErrors}. For $\aEM$ the errors are $\simeq8\%$ larger when $\sigT$ is included. By comparison, the predicted errors for $\me$ are $\simeq20\%$ larger when $\sigT$ is included. Though these modes appear more constrainable, this becomes much harder to disentangle when we generate data-driven eigenmodes and marginalise over the cosmological/nuisance parameters.

% %----------------------------------------------
% \begin{figure}
%     \centering
%     \includegraphics[width=\linewidth]{figures/vfc_cvl_sigT}
%     \caption{The absolute value of the first eigenmode, $|E_1(z)|$, shown for $\aEM$ and $\me$ with and without the rescaling of $\sigT$ outside of the ionisation history as discussed in Sect.~\ref{sec:changesXe}. Here we show the absolute value to highlight the drift of the peak structures in the modes when the $\sigT$ correction is included and the most-probable redshift for last scattering, $z_* = 1088$ is also included.}
%     \label{fig:sigT}
% \end{figure}
% %----------------------------------------------

%----------------------------------------------
%----------------------------------------------

%----------------------------------------------
\begin{table}
    \centering
    \begin{tabular}{l | c | c c c}
    \hline\hline
        Parameter & Error & CVL & \planck 2018 & SO (forecast) \\
        \hline
        $\aEM$ & $\sigma_1$ & 0.00039 & 0.0060 & 0.0015\\
         & $\sigma_2$ & 0.00092 & 0.012 & 0.0040\\
         & $\sigma_3$ & 0.0022 & 0.036 & 0.0079\\
         \hline
         $\me$ & $\sigma_1$ & 0.00076 & 0.0089 & 0.0022\\
          & $\sigma_2$ & 0.0017 & 0.017 & 0.0060\\
          & $\sigma_3$ & 0.0041 & 0.055 & 0.011\\
    \hline\hline
    \end{tabular}
    \caption{Errors calculated from the eigenvalues $\lambda_i$ of the principal components $E_i$ of $\aEM$ and $\me$. These are listed for each of the configurations discussed in Sect.~\ref{sec:cvl}-\ref{sec:so}. Note that these values have been used as the amplitudes for the $\taudot$ responses and the $C_\ell$ responses throughout this work (e.g., Figs.~\ref{fig:taudotCVL} - \ref{fig:clEECVL}).} 
    \label{tab:fisherErrors}
\end{table}
%----------------------------------------------

\section{Eigenmodes constrained with \planck data}\label{sec:planckModes}
%----------------------------------------------
Applying the direct-likelihood method described in Sect.~\ref{sec:howtoPCA} and Appendix~\ref{app:planck}, we can use the likelihood function from the \planck dataset to find the most constrainable eigenmodes. As an additional test for the method with this particular dataset, we also re-constructed the $\xe$ modes for the \planck 2018 dataset since PCA20 was limited to \planck 2015. The full comparison and details of these modes presented in Appendix~\ref{app:xe18} show that the modes have not varied significantly between datasets. This means the step-size choices and stability confirmations in Appendix~\ref{app:planck} coupled with the consistent results indicate that the direct likelihood method has been optimally configured for the following analysis\footnote{These eigenmodes are numerically stable and converged yet they are not 100\% optimised (as we discussed in PCA20). However the noisiness in the \planck likelihood function limits the precision of constraints. One could improve these limits by modifying the likelihood function sampling method with future studies.}.

\begin{figure}
    \centering
    \includegraphics[width=\linewidth]{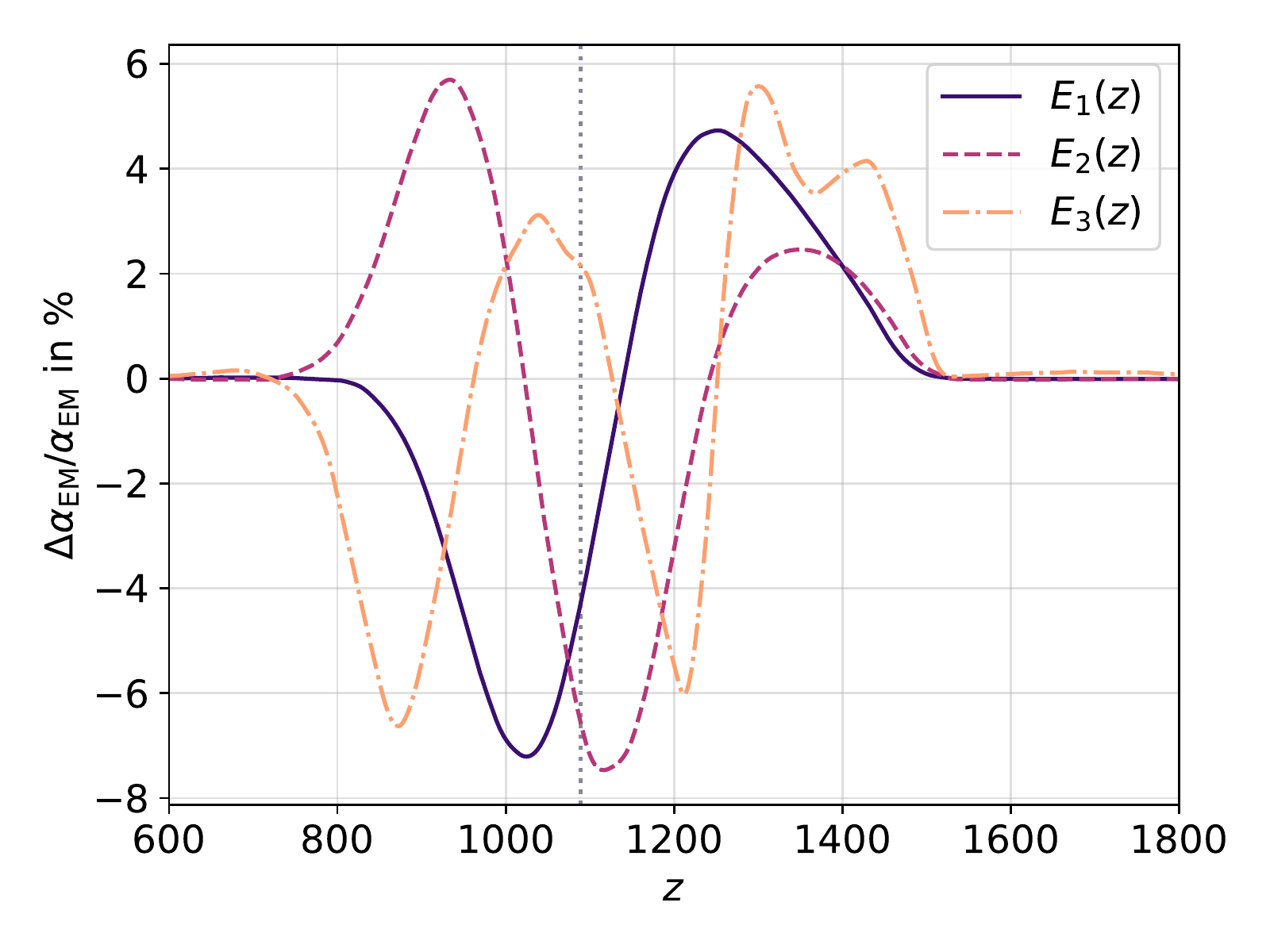}
    \includegraphics[width=\linewidth]{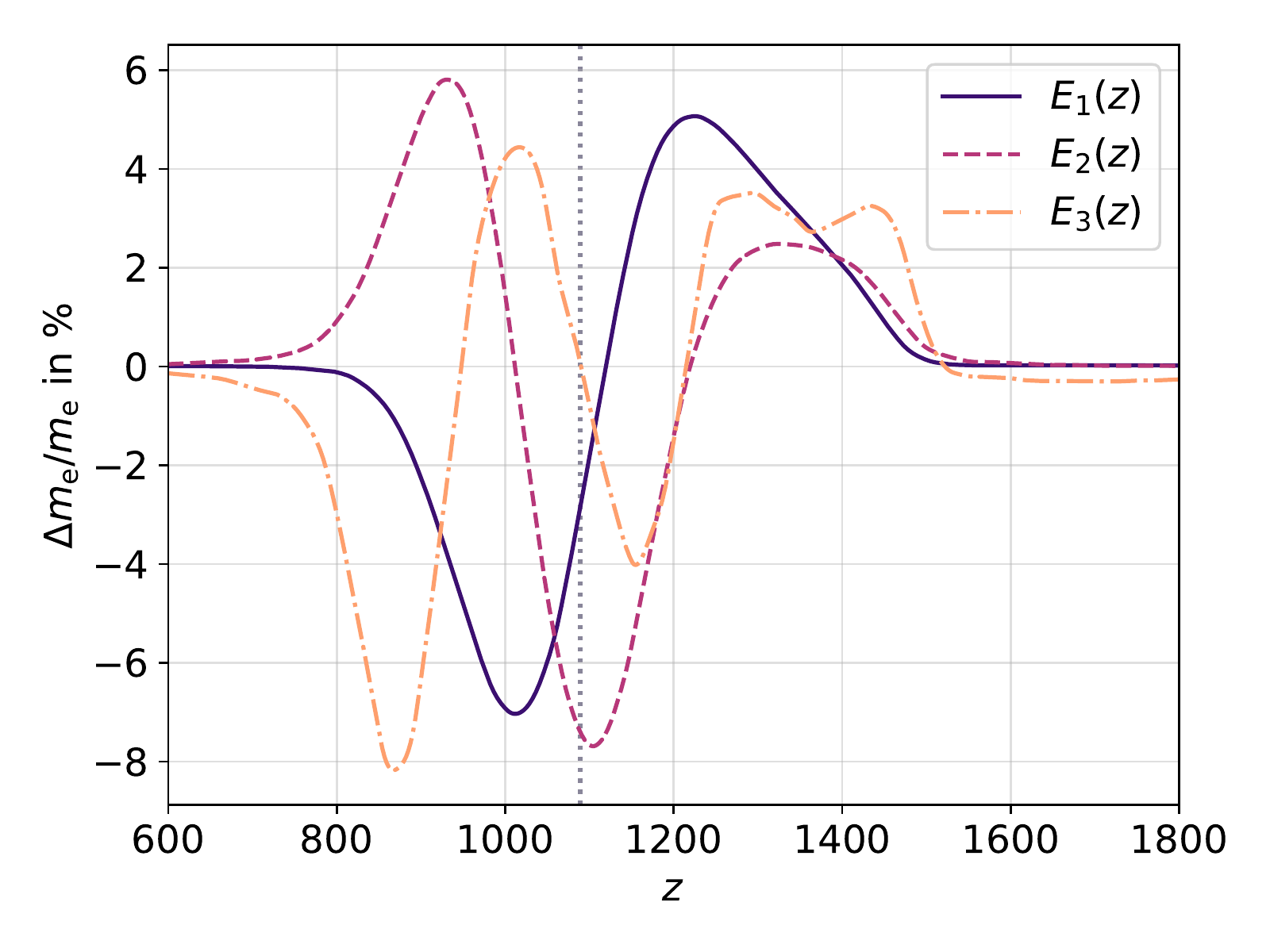}
    \caption{The first three principal components for $\aEM$ \emph{(top)} and $\me$ \emph{(bottom)} constrained with the \planck 2018 data. The eigenmodes are all normalised as previous modes such that $\int|E_i^2(z)|\dif z = 1$. As in Fig.~\ref{fig:modesCVL} and~\ref{fig:taudotCVL}, the maxima of the Thomson visibility function for a \LCDM cosmology with \planck data has been included.}
    \label{fig:modesPlanck}
\end{figure}

In this section, we will discuss the direct-likelihood constrained eigenmodes of $\aEM$ and $\me$ and the resultant responses on the CMB power spectrum. For illustration purposes, we have multiplied the second eigenmode $E_2$ for $\me$ by $-1$ to compare and contrast the similar structure to $\aEM$. This has propagated to the $\mathcal{D}_\ell$ responses in Figs.~\ref{fig:clTTPlanck}-\ref{fig:clEEPlanck} also. As with the CVL case in Sect.~\ref{sec:cvl}, this flipping has not been applied to the modes going into the MCMC solver. The constrained eigenmodes are shown in Fig.~\ref{fig:modesPlanck}. The predicted errors from the eigensolver of these modes are shown in the second column of Table~\ref{tab:fisherErrors}.

The fine-structure constant components are shown in the top panel of Fig.~\ref{fig:modesPlanck}. Much like in previous studies of $\xe$ components, the introduction of sourcing direct data introduces unique features to the \planck modes compared to a simple CVL case such as those in Fig.~\ref{fig:modesCVL}. For the most constrained eigenmode, $E_1$, the features of the mode (i.e., dip and trough) have shifted to lower redshifts. The higher redshift peak at $z\sim1250$ is considerably sharper than in the CVL case. In both the second and third eigenmodes, the number of features in each mode has increased. The second mode for the \planck modes in Fig.~\ref{fig:modesPlanck} is more reminiscent of the third mode in the CVL case (Fig.~\ref{fig:modesCVL}). Notably the kinks in $E_2$ we mentioned in Sect.~\ref{sec:cvl} have been removed, where they have been replaced by another peak at $z\sim1350$. We know from PCA20 that these features, where they are asymmetric with peaks around $z_*$, creates large degeneracies in $H_0$ (or $\thetaMC$). These are not present in $E_2$ for the direct-method eigenmodes in Fig.~\ref{fig:modesPlanck}, therefore the degeneracies with the expansion should be removed via marginalisation. In the case of $E_3$, similar to PCA20, there is higher order fine structure at $z\sim1250-1500$ which seems to arise from the marginalisation step of generating these modes.

The effective electron mass modes (also in Fig.~\ref{fig:modesPlanck}) exhibit a very similar behaviour as the $\aEM$ modes, when created with the direct method. However the departures from the CVL modes are not identical to the $\aEM$ modes. The first \planck eigenmode for $\me$ does not have the shift in peaks that is apparent in the CVL case between the two fundamental constants. Instead the differences predominantly manifest in the third eigenmode, $E_3$. 
The peaks at $1200<z<1500$ on $E_3$ in Fig.~\ref{fig:modesPlanck} are dampened for the $\me$ case. There also is a non-zero floor in $E_3$ for $z>1500$. The same floor is seen in $E_3$ at low redshift, $z<600$. Both of these features could point to more information locked in the fourth eigenmode $E_4$, however the predicted errors on these components are still very high and therefore, this may need a more rigorous analysis in the future, potentially when an improved likelihood approach is introduced. 

%----------------------------------------------
%----------------------------------------------

\subsection{Differences in the CMB power spectrum responses}\label{sec:planckDiff}
\begin{figure}
    \centering
    \includegraphics[width=\linewidth]{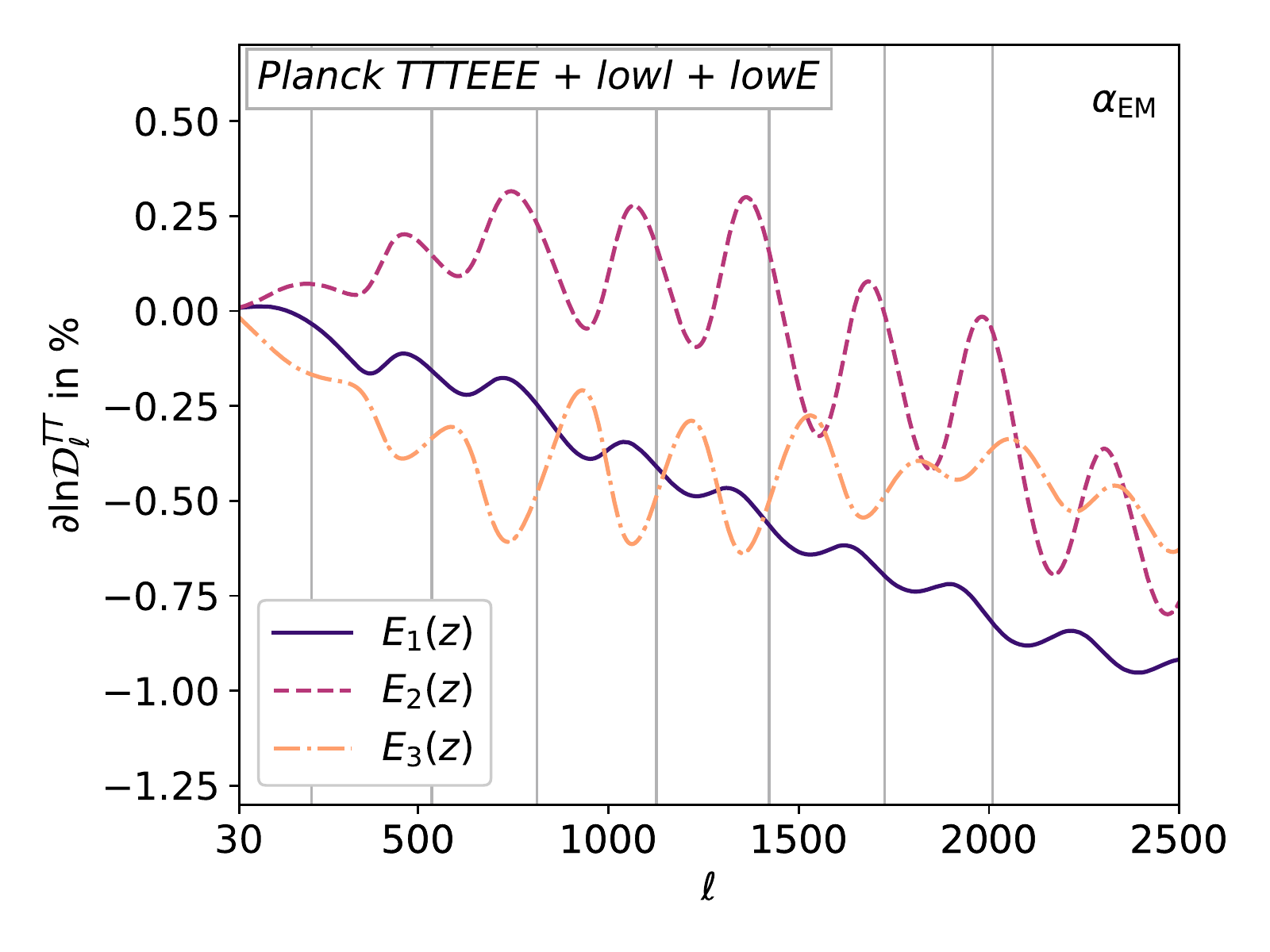}
    \includegraphics[width=\linewidth]{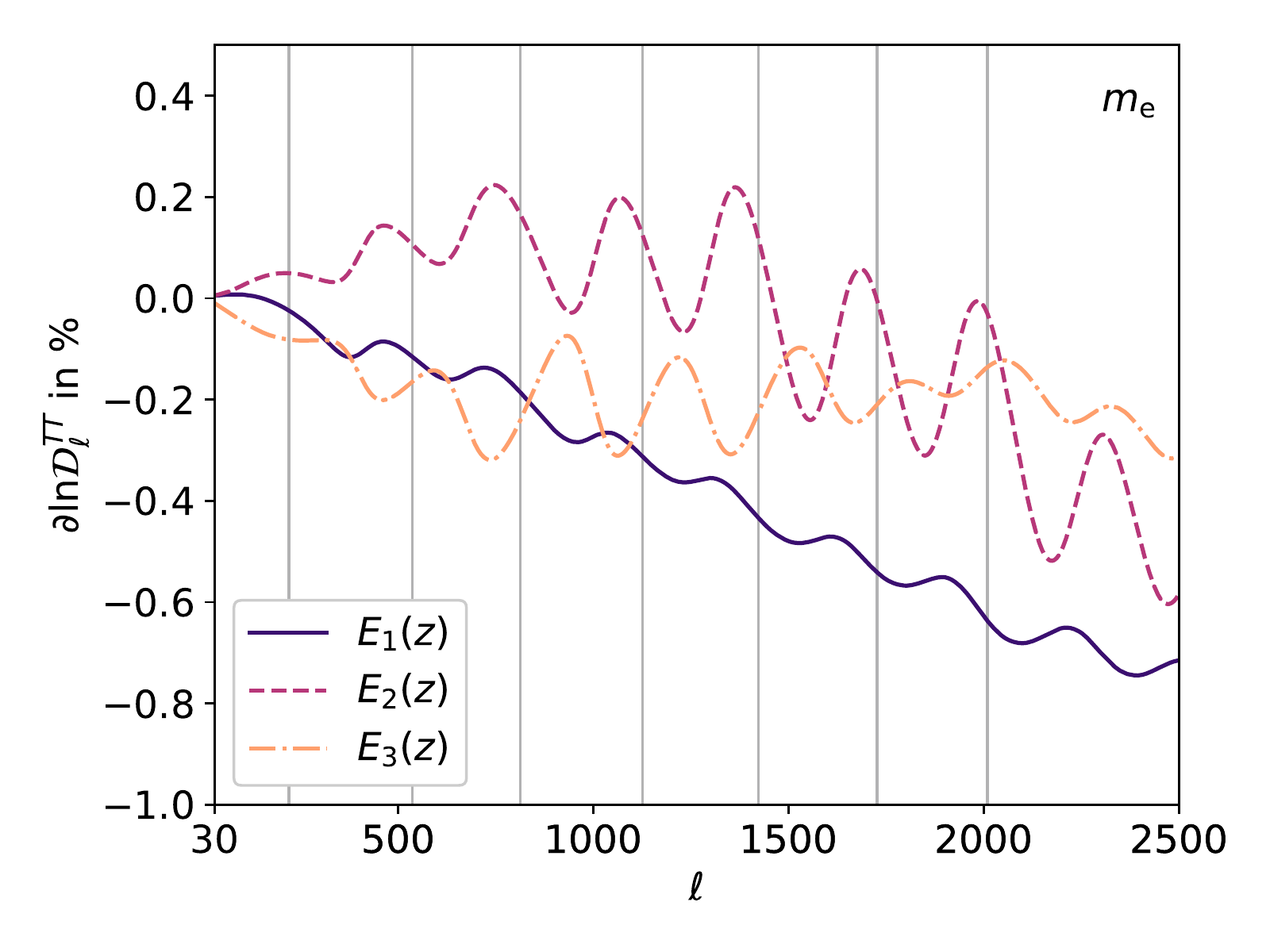}
    \caption{The responses in the CMB power spectra, similar to Fig.~\ref{fig:clTTCVL}, however arising from the \planck converged modes in Fig.~\ref{fig:modesPlanck}. Once again, the grey vertical lines are peaks of the CMB spectra in fiducial \LCDM cosmology with \planck 2018 parameters. All components have once again been multiplied by the predicted Fisher errors shown in Table~\ref{tab:fisherErrors}.} 
    \label{fig:clTTPlanck}
\end{figure}

\begin{figure}
    \centering
    \includegraphics[width=\linewidth]{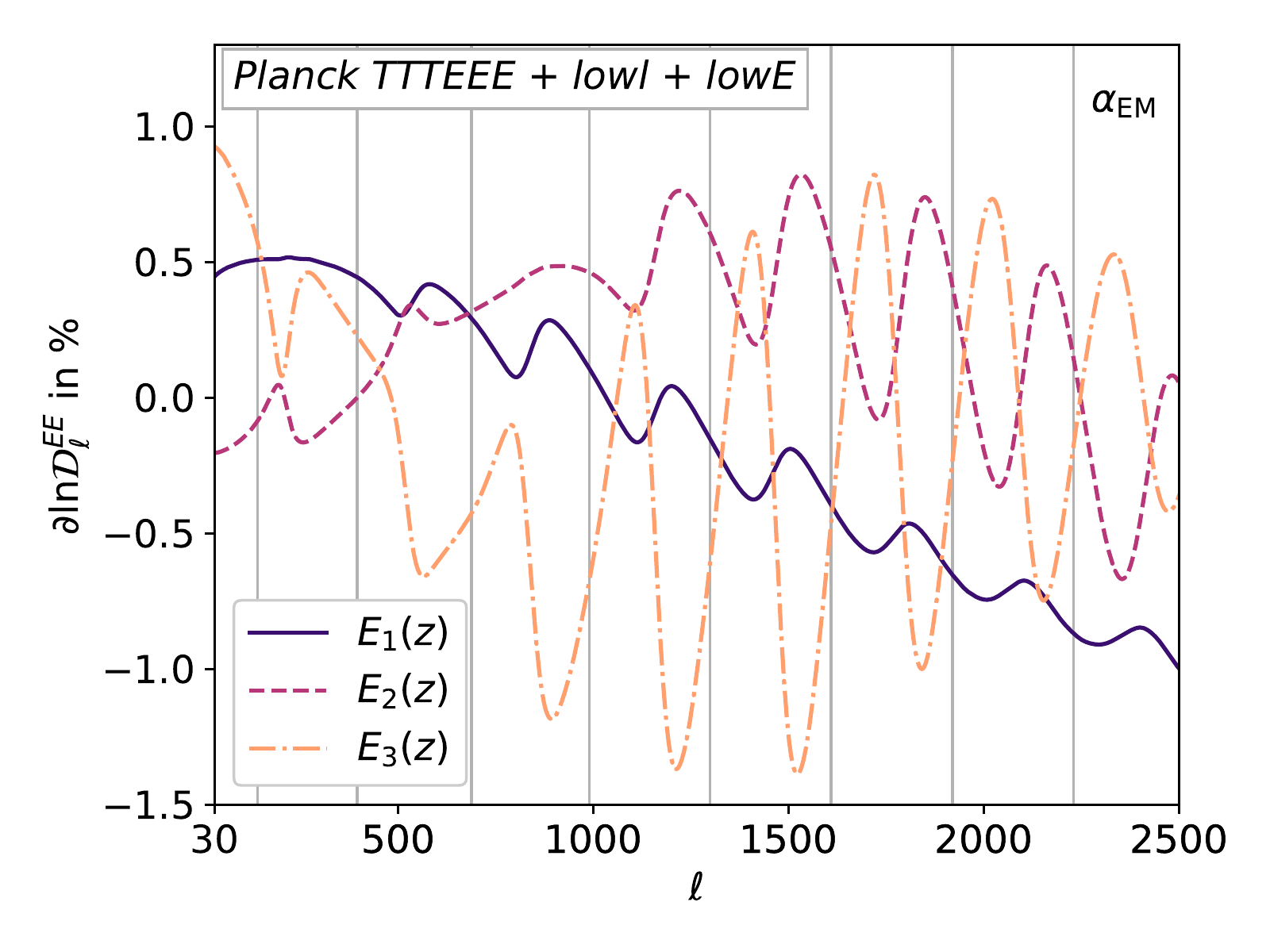}
    \includegraphics[width=\linewidth]{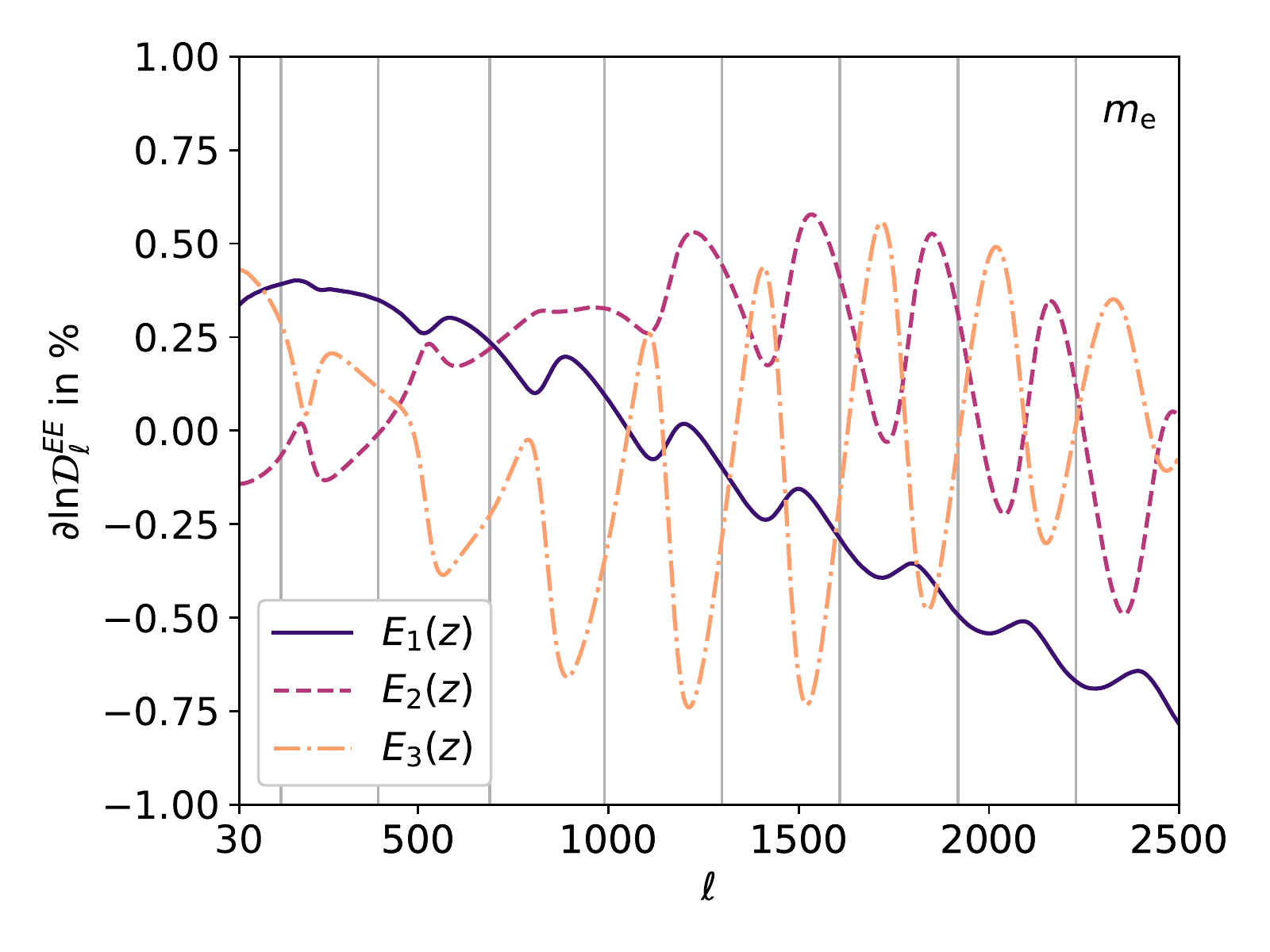}
    \caption{The responses in the CMB power spectra arising from the \planck converged modes in Fig.~\ref{fig:modesPlanck}. Once again, the grey vertical lines are peaks of the CMB spectra in fiducial \LCDM cosmology with \planck 2018 parameters. All components have once again been multiplied by the predicted Fisher errors shown in Table~\ref{tab:fisherErrors}.}
    \label{fig:clEEPlanck}
\end{figure}

%----------------------------------------------
In Fig.~\ref{fig:clTTPlanck}, we show the $\DellT$ responses according to the first 3 eigenmodes in $\aEM$ and $\me$. For consistency and completeness, we also present how these eigenmodes affect the $E$-mode polarisation power spectra encoded in the $\DellE$ variations. These are shown in Fig.~\ref{fig:clEEPlanck}.
Similar to the responses for the CVL modes in Figs.~\ref{fig:clTTCVL}-\ref{fig:clEECVL}, the acoustic peaks of each CMB power spectra (assuming \LCDM) have also been included as grey lines.
The first $\aEM$ and $\me$ mode give similar responses in $\DellT$ to their CVL counterparts as shown in Fig.~\ref{fig:clTTPlanck} however the oscillatory behaviour (which is associated with a slight shift in the peak positions) is far smaller for the \planck eigenmodes. This pattern emerges in the polarisation responses from Fig.~\ref{fig:clEEPlanck} as well.

The second component, $E_2$, starts to show differences between the CVL and \planck cases for both $\aEM$ and $\me$. Instead of creating an average residual of $\partial\ln\Dell>0$, the responses starts to shift downwards. This extra `damping' could be a result of the additional bump to $E_2$ in the \planck modes, where the accelerated recombination has not only knocked the response out of phase, but also moved the $\partial\ln\DellT$ at $\ell = 2500$ from $\sim0.02\%$ to $-\sim0.75\%$. The change in magnitude is related to the Fisher errors from Table~\ref{tab:fisherErrors} being propagated through the $C_\ell$ calculation. The third mode $E_3$ has a very unique impact on the residual $\Dell$ power spectrum due to the majority of the expansion rate ($\thetaMC$) degeneracies being removed at marginalisation. Furthermore, the changes to $E_3$ lead to both changes in the temperature (Fig.~\ref{fig:clTTPlanck}) and polarisation (Fig.~\ref{fig:clEEPlanck}) where the peaks and troughs of the responses are now anti-aligned with the second and first modes. 

Whilst the second eigenmodes for Fig.~\ref{fig:clTTPlanck} are similar for $\aEM$ and $\me$, the third mode $E_3$ is shifted higher for the effective electron mass compared to the downward effect seen in $\aEM$. The most likely reason for this is the large degeneracy seen between $\me$ and the expansion rate parameters (i.e, $\thetaMC$ or $\ho$). As shown in VFC20, there is a mild degeneracy between $\aEM$ and $\ho$ but a far larger geometric degeneracy line between $\me$ and $\ho$. This arises from a small extra tilting in the residual $\Dell$ for $\me$. To remove these degeneracies, $\me$ requires a larger degree of marginalisation (the relevant correlation coefficients in the Fisher matrix for the perturbations and $\thetaMC$ will be larger) and therefore the CMB responses shown for $\me$ in Fig.~\ref{fig:clTTPlanck} are more damped. 
Comparable effects can be seen in the polarisation spectral residuals in Fig.~\ref{fig:clEEPlanck} for $\me$, however these changes are much more subtle. 

The only key difference in the polarisation power spectra, for both $\aEM$ and $\me$ is the breakdown of the periodic residuals in $E_2$ for $\ell<1000$. Here the repetitive wavy pattern has been replaced by a more complex response. In \citet{Hart2017}, the shape of the VFC effects on Silk damping and the location of the horizon $\theta_*$ were explained in detail. However, the effects on the $EE$ polarisation spectra have not been explored in more detail. In the second eigenmode, the $\DellT$ and $\DellE$ responses (shown in Fig.~\ref{fig:clTTPlanck} and~\ref{fig:clEEPlanck}) for both $\aEM$ and $\me$ has broad similarities with the constant variations discussed in our previous papers \citep[and VFC20]{Hart2017}. The changes in the CMB anisotropy power spectra are degenerate with variations expected with a change in the horizon size $\thetaMC$. If those oscillatory variations at $\ell<1000$ were removed, that degeneracy may be removed from the marginalised modes. This is clear for the $\DellE$ variations for $\aEM$ and $\me$ in Fig.~\ref{fig:clEEPlanck} but; this degeneracy may account for the drop in the damping variations for $E_2$ in the $\DellT$ variations at $\ell>1500$. 

%----------------------------------------------
%----------------------------------------------

%----------------------------------------------
%----------------------------------------------

\section{Constraining eigenmode amplitudes using Markov Chain Monte Carlo}\label{sec:mcmc}
%----------------------------------------------

In this section, we present the MCMC results for the $\aEM$ and $\me$ components. With the modes for the two experimental configurations, we can constrain the amplitudes of the eigenmodes $\mu_i$ as explained in Sect.~\ref{sec:constraints}. For the CVL case, the third error starts to contain all but a negligible contribution to the information ($99\%$) and for the \planck case, this is smaller ($97\%$); but, the $i>3$ modes in the hierarchy are numerically unstable using the direct likelihood method. This follows from a similar problem found in PCA20, however, from these justifications we shall restrict ourselves to the first 3 eigenmodes. 

The amplitudes $\mu_i$ are added as free parameters into {\tt CAMB} and {\tt CosmoMC}, where the latter is used to sample over parameter space with the former as the theoretical model for calculating the resultant likelihood. 
% WRITE BAO+SN STUFF HERE
For the recombination-era $\xe$ eigenmodes, we have already concluded that the addition of lensing or BAO likelihood information makes little difference to the marginalised results in PCA20. When the BAO data is added, the error values for the amplitudes $\mu_i$ have negligible differences ($\sigma_\mu\lsim2\%$) and the largest drift in amplitude is $\mu_2$ which shifts by $\simeq0.2\sigma$. The only other drifts occur in $\omc$ and $\ns$ which are consistent with the \LCDM variations found in the \planck 2018 results \citep{Planck2018params}. Therefore, we will focus on the addition of the \planck 2018 baseline likelihood.

In sampling for the Markov chains, the standard \planck priors were used and the same Gelman-Rubin convergence metric that was used in PCA20 where $\mathcal{R}-1\leq0.01$. The parameters varied as part of the MCMC are the standard 6 parameters:  $\left\{\omb,\omc,\thetaxMC,\tau,\ns,\logA\right\}$. The nuisance parameters that were varied in the construction of the \planck modes (Sect.~\ref{sec:planckModes}) will also be sampled over using the \emph{fast-slow} algorithm in the \planck likelihood \citep{Lewis2013}.

%----------------------------------------------
\begin{table*}
\begin{tabular} { l  c c c c}
\hline
Parameter & \planck 2018 TTTEEE + low-$\ell$ & + 1 CVL $\alpha_{\rm EM}$ mode & + 2 CVL $\alpha_{\rm EM}$ modes & + 3 CVL $\alpha_{\rm EM}$ modes\\
\hline\hline
$\omega_b  $ &  $0.02237\pm 0.00015  $ &  $0.02234\pm 0.00019  $ &  $0.02233\pm 0.00018  $ &  $0.02223\pm 0.00022  $\\
$\omega_c  $ &  $0.1199\pm 0.0012  $ &  $0.1202\pm 0.0014  $ &  $0.1203\pm 0.0015  $ &  $0.1194\pm 0.0018  $\\
$100\theta_{MC}  $ &  $1.04088\pm 0.00031  $ &  $1.04089\pm 0.00043  $ &  $1.04097\pm 0.00091  $ &  $1.0370^{+0.0038}_{-0.0051}  $\\
$\tau  $ &  $0.0542\pm 0.0074  $ &  $0.0544^{+0.0073}_{-0.0082}  $ &  $0.0539\pm 0.0080  $ &  $0.0542\pm 0.0080  $\\
${\rm{ln}}(10^{10} A_s)  $ &  $3.044\pm 0.014  $ &  $3.045\pm 0.016  $ &  $3.044\pm 0.016  $ &  $3.044\pm 0.017  $\\
$n_s  $ &  $0.9649\pm 0.0041  $ &  $0.9642\pm 0.0059  $ &  $0.9641\pm 0.0060  $ &  $0.9670\pm 0.0065  $\\
\hline
$\mu_1 \;(\aEM) $ & $--$   &  $-0.0008\pm 0.0074  $ &  $-0.0014\pm 0.0096  $ &  $0.017^{+0.025}_{-0.019}  $\\
$\mu_2 \;(\aEM) $ & $--$   & $--$   &  $-0.002\pm 0.014  $ &  $0.039^{+0.053}_{-0.041}  $\\
$\mu_3 \;(\aEM) $ & $--$   & $--$   & $--$   &  $0.062^{+0.075}_{-0.060}  $\\
\hline
$H_0  $ &  $67.36\pm 0.54  $ &  $67.27\pm 0.62  $ &  $67.24\pm 0.61  $ &  $66.2^{+1.2}_{-1.5}  $\\
$\sigma_8  $ &  $0.8107\pm 0.0059  $ &  $0.8116\pm 0.0076  $ &  $0.8120\pm 0.0084  $ &  $0.806\pm 0.010  $\\
\hline\hline
\end{tabular}
\caption{Marginalised results at the $68\%$ confidence level for the CVL $\aEM$ modes in Fig.~\ref{fig:modesCVL}. This is combined with the \planck 2018 baseline dataset \citep{Planck2018params} and shown against the \LCDM standard case. The comparison of all the standard \LCDM parameters along with two derived parameters, $H_0$ and $\sigma_8$, are shown with the $\mu_i$ amplitudes. The Gelman-Rubin convergence metric for all the chains that generated these results satisfy $\mathcal{R}-1 < 0.01$.}
\label{tab:alphaCVL}
\end{table*}
%----------------------------------------------

%----------------------------------------------
\begin{figure}
    \centering
    \includegraphics[width=\linewidth]{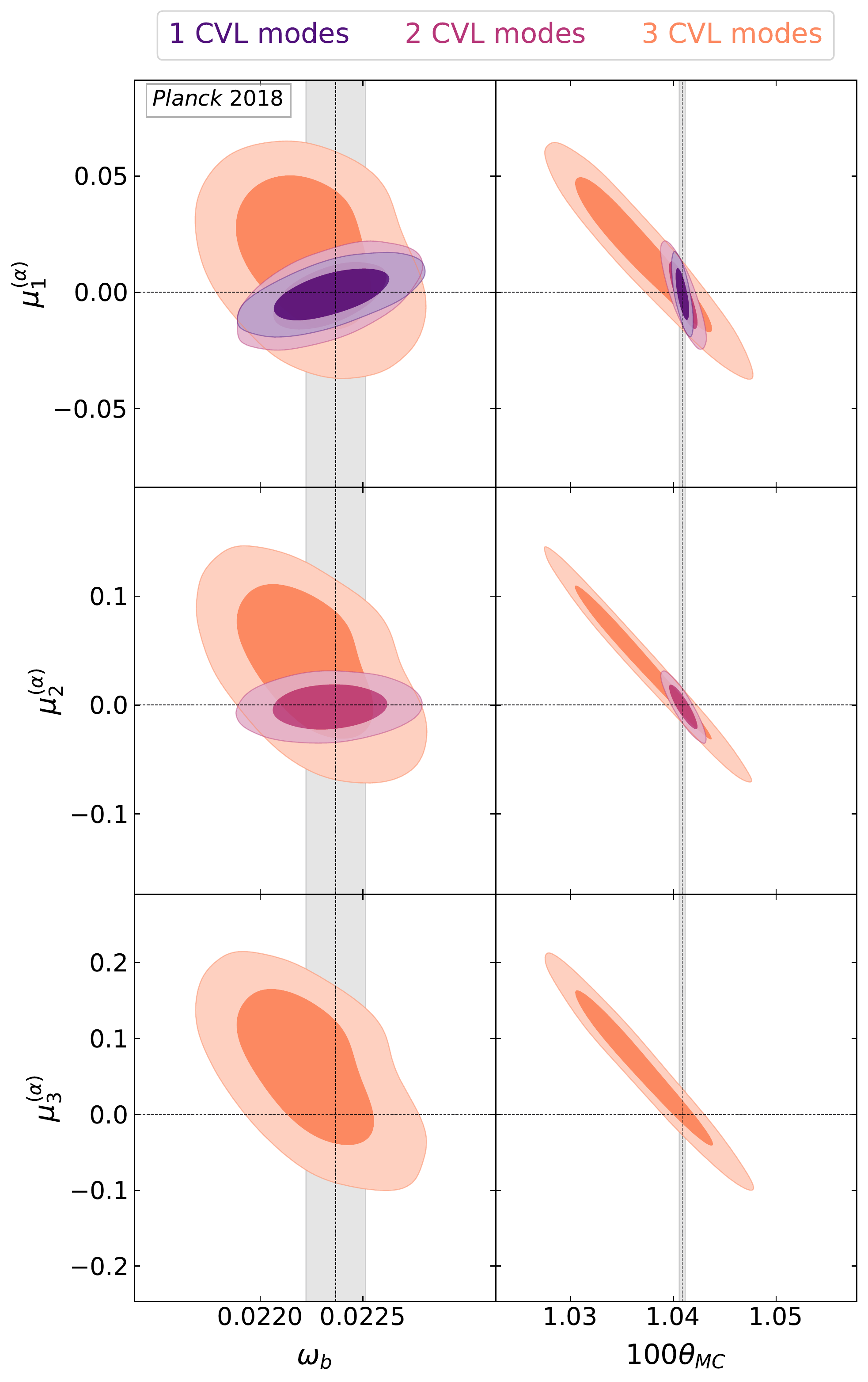}
    \caption{Posterior distribution contours from varying $\aEM$ modes and the most correlated standard \LCDM parameters ($\omb$ and $\thetaMC$) with \planck 2018 data. The amplitudes of the $\aEM$ principal components are categorised by $\mu_i^{(\alpha)}$. Bands of the standard errors coming from \planck TTTEEE + low-$\ell$ 2018 data are shown as well.}
    \label{fig:alphaCVLContours}
\end{figure}
%----------------------------------------------

%----------------------------------------------
\begin{table*}
    \begin{tabular} { l c c c c}
        \hline\hline
        Parameter & Planck 2018 TTTEEE + low-$\ell$ & + 1 CVL $m_{\rm e}$ mode & + 2 CVL $m_{\rm e}$ modes & + 3 CVL $m_{\rm e}$ modes\\
        \hline
        $\omega_b  $ &  $0.02237\pm 0.00015  $ &  $0.02234\pm 0.00019  $ &  $0.02235\pm 0.00019  $ &  $0.02220\pm 0.00022  $\\
        $\omega_c  $ &  $0.1199\pm 0.0012  $ &  $0.1202\pm 0.0014  $ &  $0.1202\pm 0.0015  $ &  $0.1195\pm 0.0016  $\\
        $100\theta_{MC}  $ &  $1.04088\pm 0.00031  $ &  $1.04088\pm 0.00040  $ &  $1.04096\pm 0.00091  $ &  $1.0374^{+0.0026}_{-0.0040}  $\\
        $\tau  $ &  $0.0542\pm 0.0074  $ &  $0.0543\pm 0.0079  $ &  $0.0542\pm 0.0080  $ &  $0.0539\pm 0.0079  $\\
        ${\rm{ln}}(10^{10} A_s)  $ &  $3.044\pm 0.014  $ &  $3.044\pm 0.016  $ &  $3.044\pm 0.016  $ &  $3.043\pm 0.017  $\\
        $n_s  $ &  $0.9649\pm 0.0041  $ &  $0.9642\pm 0.0059  $ &  $0.9644\pm 0.0059  $ &  $0.9654\pm 0.0060  $\\
        \hline
        $\mu_1 \;(\me)  $ & $--$   &  $0.001\pm 0.014  $ &  $0.002\pm 0.018  $ &  $-0.024^{+0.025}_{-0.033}  $\\
        $\mu_2 \;(\me)  $ & $--$   & $--$   &  $0.003\pm 0.027  $ &  $-0.069^{+0.055}_{-0.082}  $\\
        $\mu_3 \;(\me)  $ & $--$   & $--$   & $--$   &  $-0.107^{+0.078}_{-0.11}  $\\
        \hline
        $H_0  $ &  $67.36\pm 0.54  $ &  $67.26\pm 0.61  $ &  $67.29\pm 0.62  $ &  $66.2^{+1.0}_{-1.2}  $\\
        $\sigma_8  $ &  $0.8107\pm 0.0059  $ &  $0.8116\pm 0.0076  $ &  $0.8118\pm 0.0084  $ &  $0.8060\pm 0.0098  $\\
        \hline\hline
    \end{tabular}
\caption{Marginalised results at the $68\%$ confidence level for the CVL $\me$ modes in Fig.~\ref{fig:modesCVL}. This is combined with the \planck 2018 baseline likelihood. The standard 6 cosmological parameters are shown with $\ho$ and $\sig$ as well as the eigenmode amplitude parameters, $\mu_i$. The Gelman-Rubin convergence metric for all the chains that generated these results satisfy $\mathcal{R}-1 < 0.01$.}  
\label{tab:meCVL}
\end{table*}
%----------------------------------------------

%----------------------------------------------
\begin{figure}
    \centering
    \includegraphics[width=\linewidth]{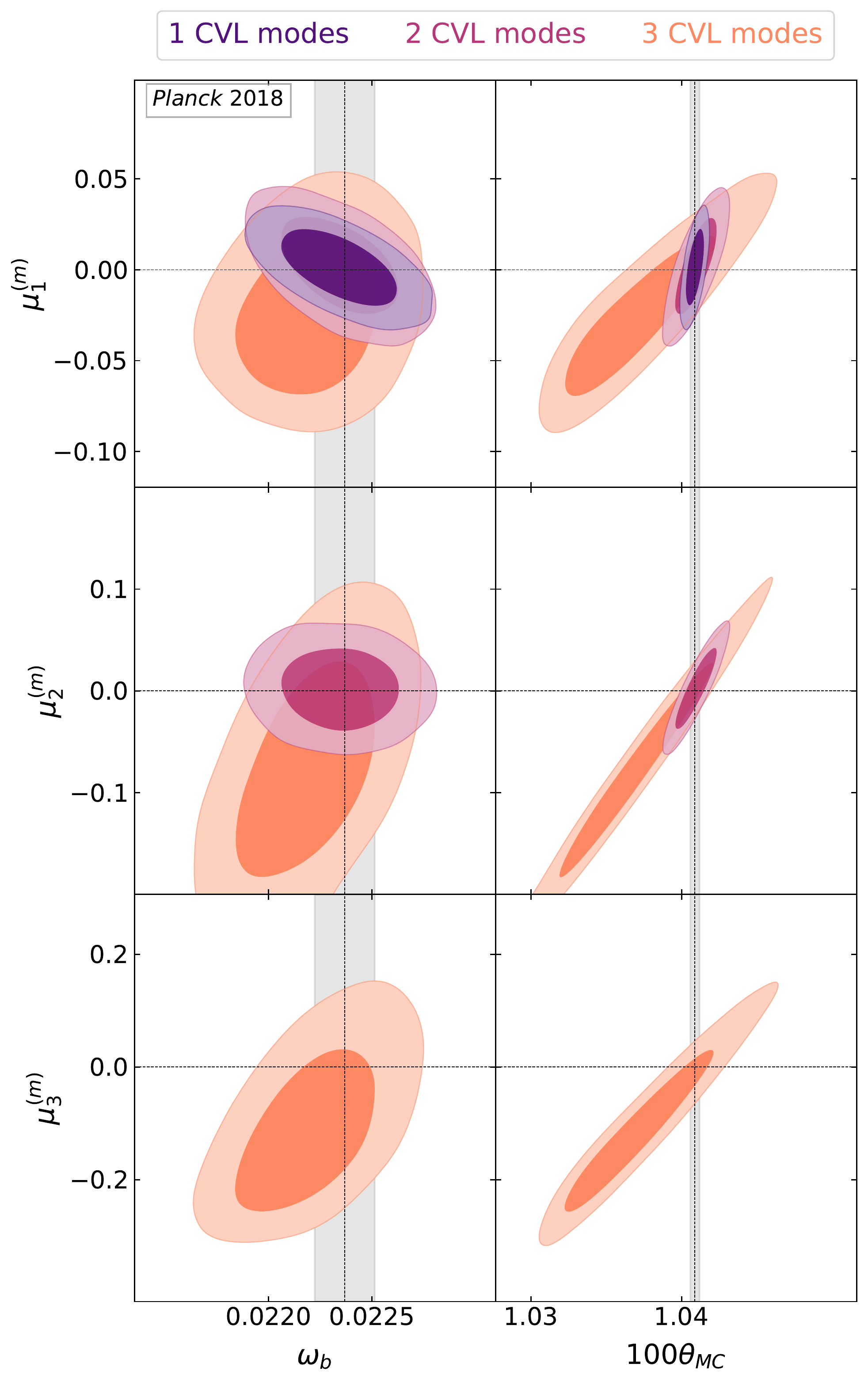}
    \caption{Posterior distribution contours from varying $\me$ modes and the most correlated standard \LCDM parameters ($\omb$ and $\thetaMC$) with \planck 2018 data. The amplitudes of the $\aEM$ principal components are categorised by $\mu_i^{(m)}$. Bands of the standard errors coming from \planck TTTEEE + low-$\ell$ 2018 data are shown as well.}
    \label{fig:meCVLContours}
\end{figure}
%----------------------------------------------

%----------------------------------------------
\begin{figure}
    \centering
    \includegraphics[width=\linewidth]{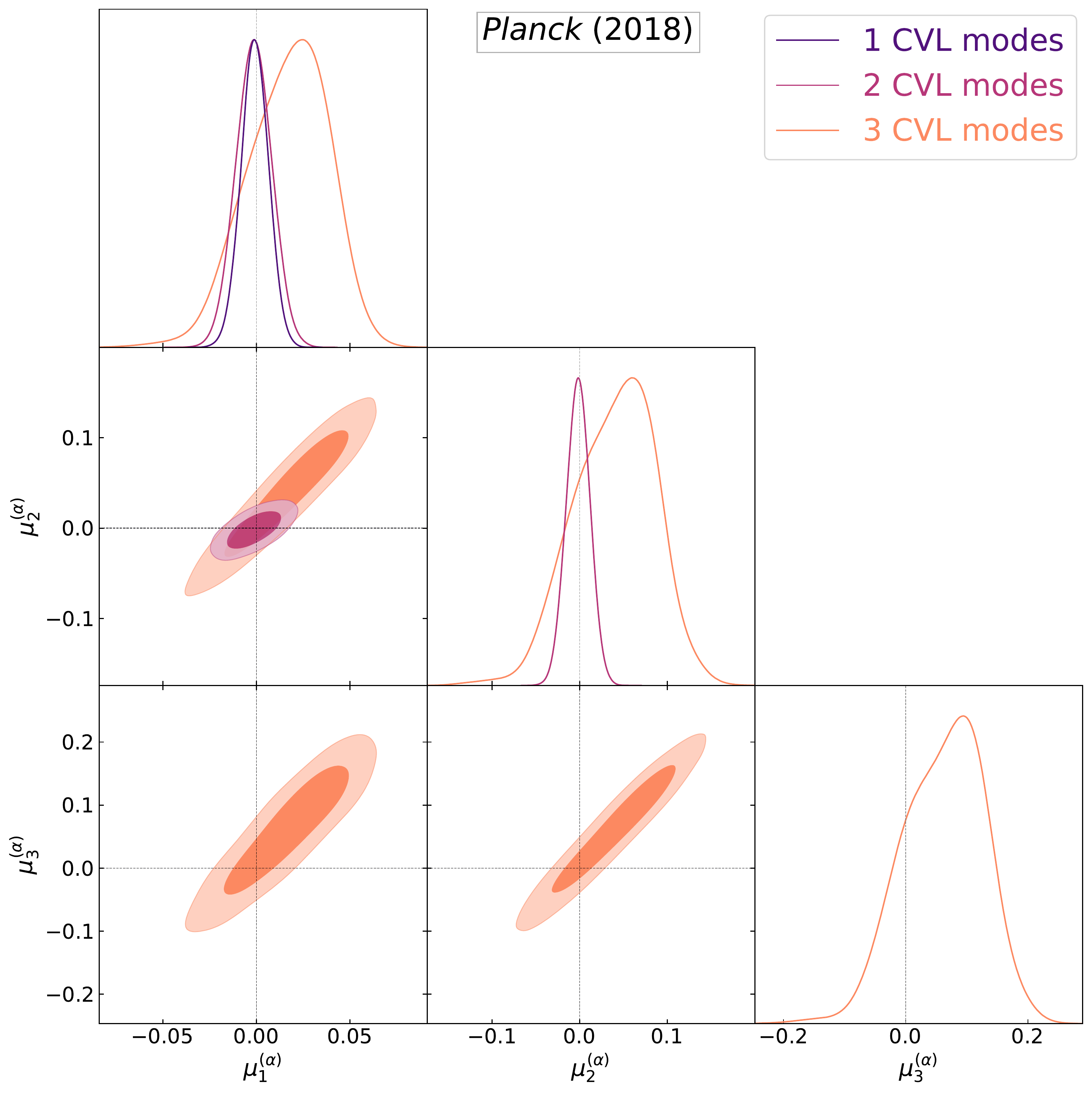}
    \caption{Posterior contours showing the cross-correlations of the first 3 most constrainable components for $\aEM$ defined by $\mu_i^{(\alpha)}$. Here the same \planck TTTEEE+low-$\ell$ baseline data was used as in Fig.~\ref{fig:alphaCVLContours}.}
    \label{fig:alphaCVLMu}
\end{figure}
%----------------------------------------------

%----------------------------------------------
\begin{figure}
    \centering
    \includegraphics[width=\linewidth]{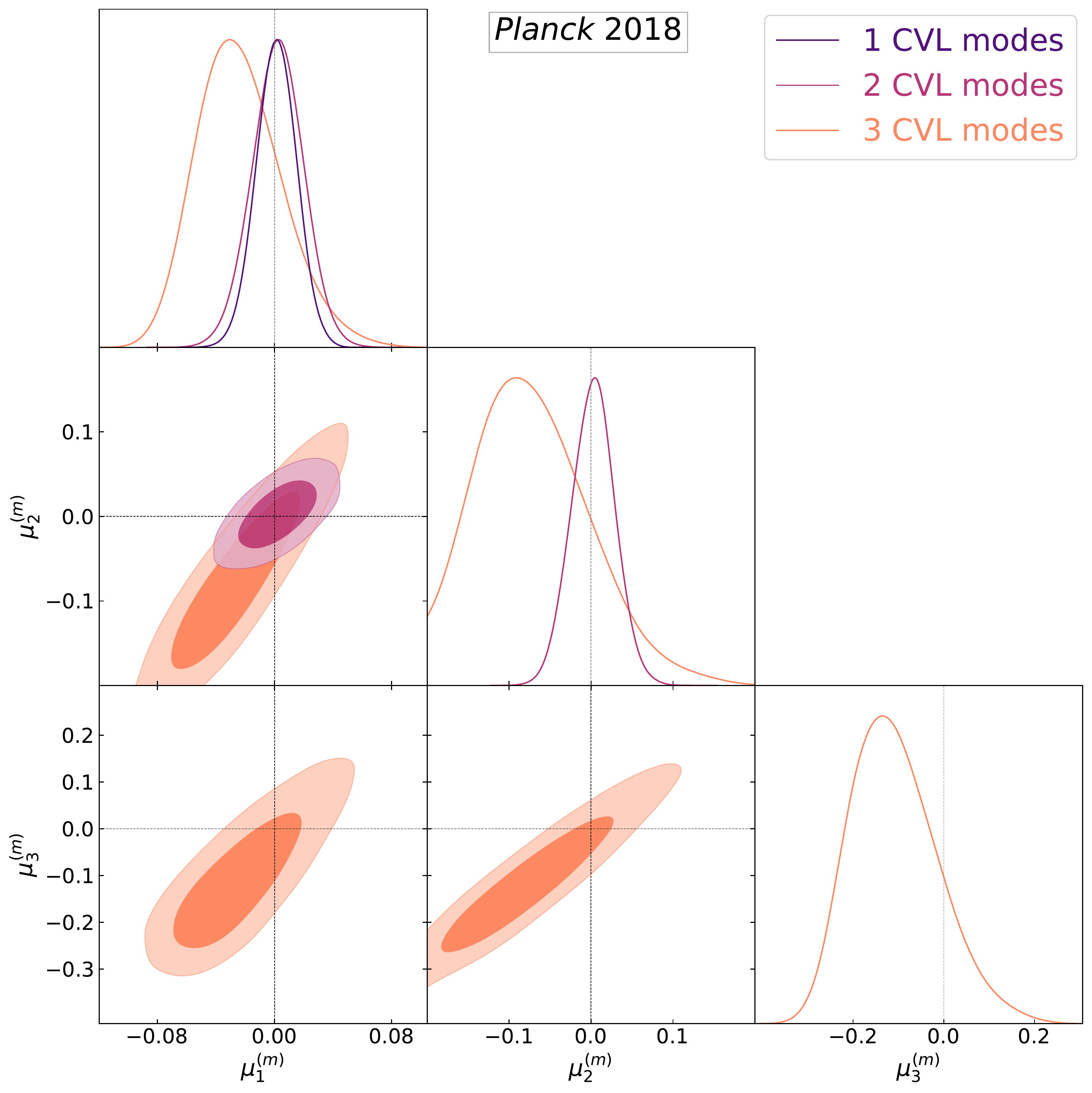}
    \caption{Contours for $\me$ modes with the same data source as before. The degeneracies between each of the first 3 eigenmodes are highlighted here.}
    \label{fig:meCVLMu}
\end{figure}
 %----------------------------------------------

% %----------------------------------------------
% \begin{figure}
%     \centering
%     \includegraphics[width=\linewidth]{figures/muThetaCVL.pdf}
%     \caption{Comparisons of the contours between the third eigenmode amplitude $\mu_3$ and the last scattering surface scale $\theta$ for both $\aEM$ and $\me$ PCAs. Here the \planck 2018 baseline \LCDM limit is shown in the x-band. The contours point in different directions due to the eigenmodes having negative symmetry coming out of the eigensolver as explained in Fig.~\ref{fig:modesCVL}.}
%     \label{fig:muThetaCVL}
% \end{figure}
% %----------------------------------------------

\subsection{Cosmic-variance-limited modes}\label{sec:contoursCVL}
%----------------------------------------------
Initially, we added the CVL modes into {\tt CosmoRec} and {\tt CosmoMC}  to constrain their amplitudes $\mu_i$ alongside the baseline \planck parameters. Since these modes are not optimized using the full data covariance matrix (as we have  done in Sect.~\ref{sec:contoursPlanck}), one expects significant correlations with standard parameters. 
In Table~\ref{tab:alphaCVL}, we present the marginalised results for the $\aEM$ with a CVL setup. The results indeed show that introducing the first eigenmode created a substantial degeneracy with $\ns$ and $\omb$. This is due to the error increase in $\ns$ by $\sim44\%$ and error increase in $\omb$ by $\sim25\%$. When analysing the chains, we calculate the  correlations $\rho\,(\omb,\mu_1^{(\aEM)}) = 0.59$ and $\rho\,(\ns,\mu_1^{(\aEM)}) = 0.67$. The physical origins for this correlation is the tilted spectra in $\DellT$ from Fig.~\ref{fig:clTTCVL} and $\DellE$ from Fig.~\ref{fig:clEECVL}, reminiscent of the tilted residuals from a varied $\ns$. Since $\ns$ and $\omb$ are correlated by $\sim50\%$, this explains the joint correlations and is reflected by the results in Table~\ref{tab:alphaCVL}. 

In Fig.~\ref{fig:alphaCVLContours}, the posterior contours for $\aEM$ also show this correlation for $\omb$ with 1 mode (\emph{purple}) very well. The oscillatory nature of these same residuals lead to a shift in the position of the sound horizon and so this mode has a correlation with $\thetaMC$ (more formally $\theta_*$) where $\rho(\thetaMC,\mu_1^{(\aEM)}) = -0.69$. Cross correlations lead to degeneracies between $\mu_1$ and the derived parameters  $\sig$ and $\ho$. When the second and third modes are added, the degeneracies are predominantly related to $\thetaMC$. This is shown in Table~\ref{tab:alphaCVL} as the error on $\thetaMC$ increases by a factor of two when compared to the baseline \planck case once 3 modes are added. This is reinforced by the shapes of the posterior contours in Fig.~\ref{fig:alphaCVLContours} between $\thetaMC$ and $\mu_i$. A sharp degeneracy line between $\mu_2$, $\mu_3$ and $\thetaMC$ ( $\rho(\thetaMC,\mu_3^{(\aEM)}) = -0.99$, $\rho(\thetaMC,\mu_2^{(\aEM)}) = -0.98$ ) leads to huge jumps in all the parameter errors that have medium-large degeneracies with $\thetaMC$ (i.e., $\omb$, $\omc$, $\ns$). Throughout all this analysis the marginalised values and errors of $\tau$ and $\As$ are unaffected, which is consistent since the $\partial\DellT$ and $\partial\DellE$ spectra shown in Fig.~\ref{fig:clTTCVL}-\ref{fig:clEECVL} do not resemble overall amplitude shifts (where the residual of $\Dell$ would be a flat, non-zero response\footnote{Changes to the CMB spectra in $\tau$ and $\As$ do leave oscillation-like relics but they are far smaller-scale structure than the overall amplification of the power spectra.}). The large degeneracies present for $\aEM$ leave the non-orthogonalities tarnished post-MCMC sampling. This means that whilst the eigenmodes are heavily orthogonal ($>99.9\%$) with each other, they accrue degeneracies through the assorted cross-correlations previously mentioned. In Fig.~\ref{fig:alphaCVLMu}, these correlations between the amplitude parameters are clearly shown and become most apparent when $\mu_3$ is added to the simulation.

Similarly, the marginalised constraints for the first 3 $\me$ mode amplitudes being added to the \planck baseline analysis are shown in Table~\ref{tab:meCVL}. The first difference between the two cases from this analysis is that the errors for $\me$ are twice as large as those for $\aEM$ (i.e., $\sigma_\mu^{\me}\sim2\sigma_\mu^{\aEM}$). This is fairly consistent for the relative change in magnitudes between $\aEM$ and $\me$ variations explored in \citet{Hart2017}, especially since the PCA is focussed around redshifts more associated exclusively with hydrogen and helium recombination ($300<z<3000$). 
The opposite signs of the marginalised values in Table~\ref{tab:meCVL} compared to Table~\ref{tab:alphaCVL} are related to the flipped symmetry of the outputted modes from the eigensolver\footnote{Note this flipping does not affect the orthonormalisation and therefore, does not affect the results, simply the sign of the mean amplitude value $\bar{\mu_i}$.}, as mentioned in Sect.~\ref{sec:cvl}. 

Aside from the normalised errors on the modes, the standard parameter values and their marginalised errors are consistently similar to the results for $\aEM$. One peculiar difference is that the sharpness of the $\thetaMC$ contour for $\aEM$ is larger than $\me$ ($\approx 35\%$ higher). For $\me$ specifically, once again the electron mass is correlated with the horizon size such that $\rho(\thetaMC,\mu_3^{(\me)}) = 0.96$ From our previous analyses, this is inconsistent, but this appears to be related to the degeneracies introduced by the first 2 modes. Additional marginalisation and generation of eigenmodes with the appropriate data (as discussed with the direct likelihood method in Sect.~\ref{sec:direct}, with modes shown in Sect.~\ref{sec:planckModes}) reduce these strong correlations. In Fig.~\ref{fig:meCVLMu}, there are similar contours as in Fig.~\ref{fig:alphaCVLMu} for $\aEM$; however the contours are shifted into the opposite quadrant due to the flipping of the eigenmodes. Note that the contours broaden out as $\mu_1$ and $\mu_3$ deviate further from $\mu_i = 0$ (\LCDM case) due to all 3 modes being consistently correlated with $\thetaMC$\footnote{Similar behaviour happens with $\aEM$ in Fig.~\ref{fig:alphaCVLMu} but the effect is much more subtle and reversed.}\\

%----------------------------------------------
\begin{figure}
    \centering
    \includegraphics[width=\linewidth]{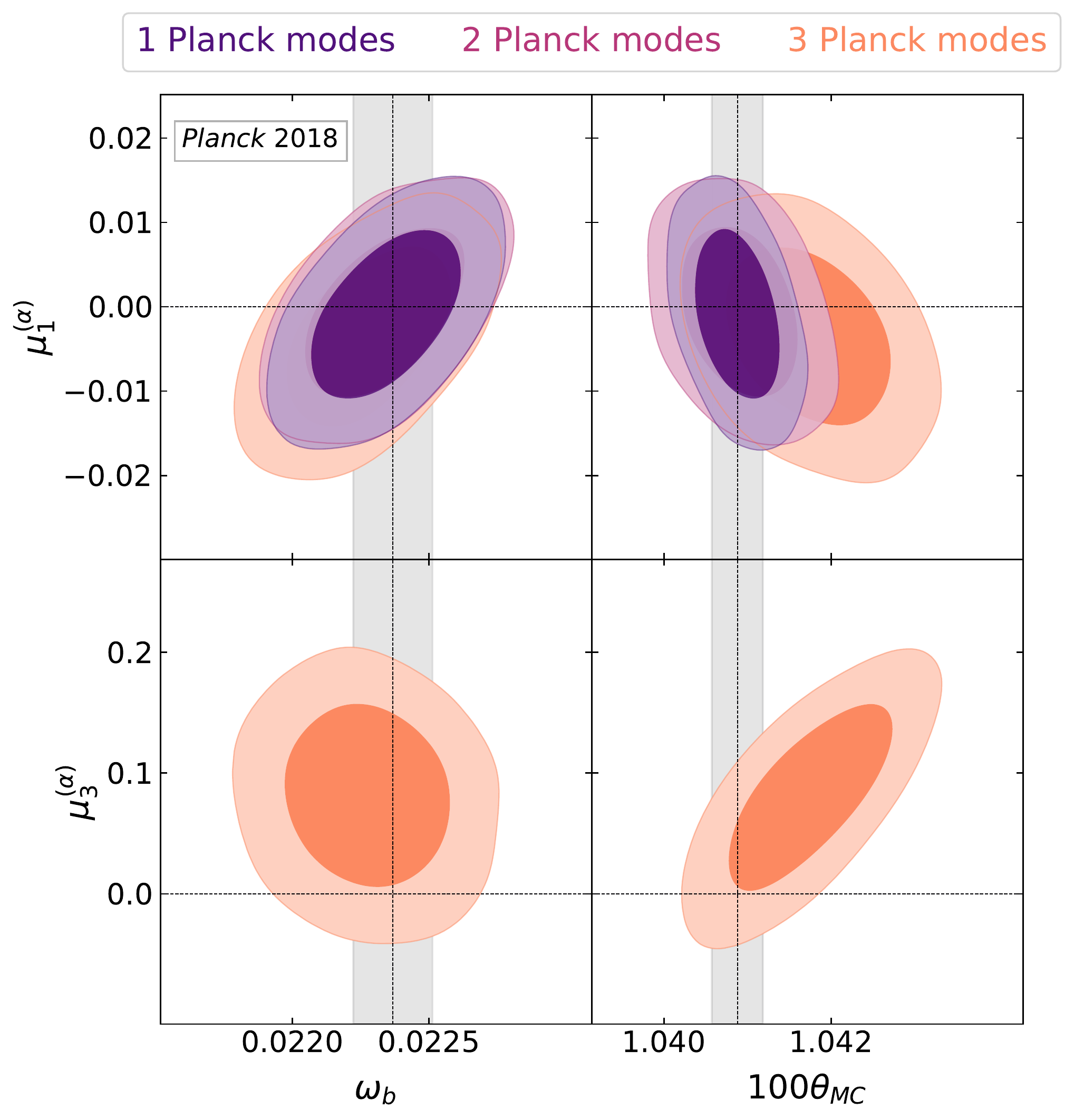}
    \caption{Most correlated likelihood contours from the $\aEM$ \planck modes shown in Fig.~\ref{fig:modesPlanck}. This is the same correlations as in Fig.~\ref{fig:alphaCVLContours} except here we remove the $\mu_2^{(\alpha)}$ contour row because the degeneracies for this parameter are more derived from $\mu_1$ and $\mu_3$. As with all the contour plots for comparing \LCDM parameters, the standard cosmology \planck results are represented by the dark bands.}
    \label{fig:alphaPlanckContours}
\end{figure}
%----------------------------------------------

%----------------------------------------------
\begin{figure}
    \centering
    \includegraphics[width=\linewidth]{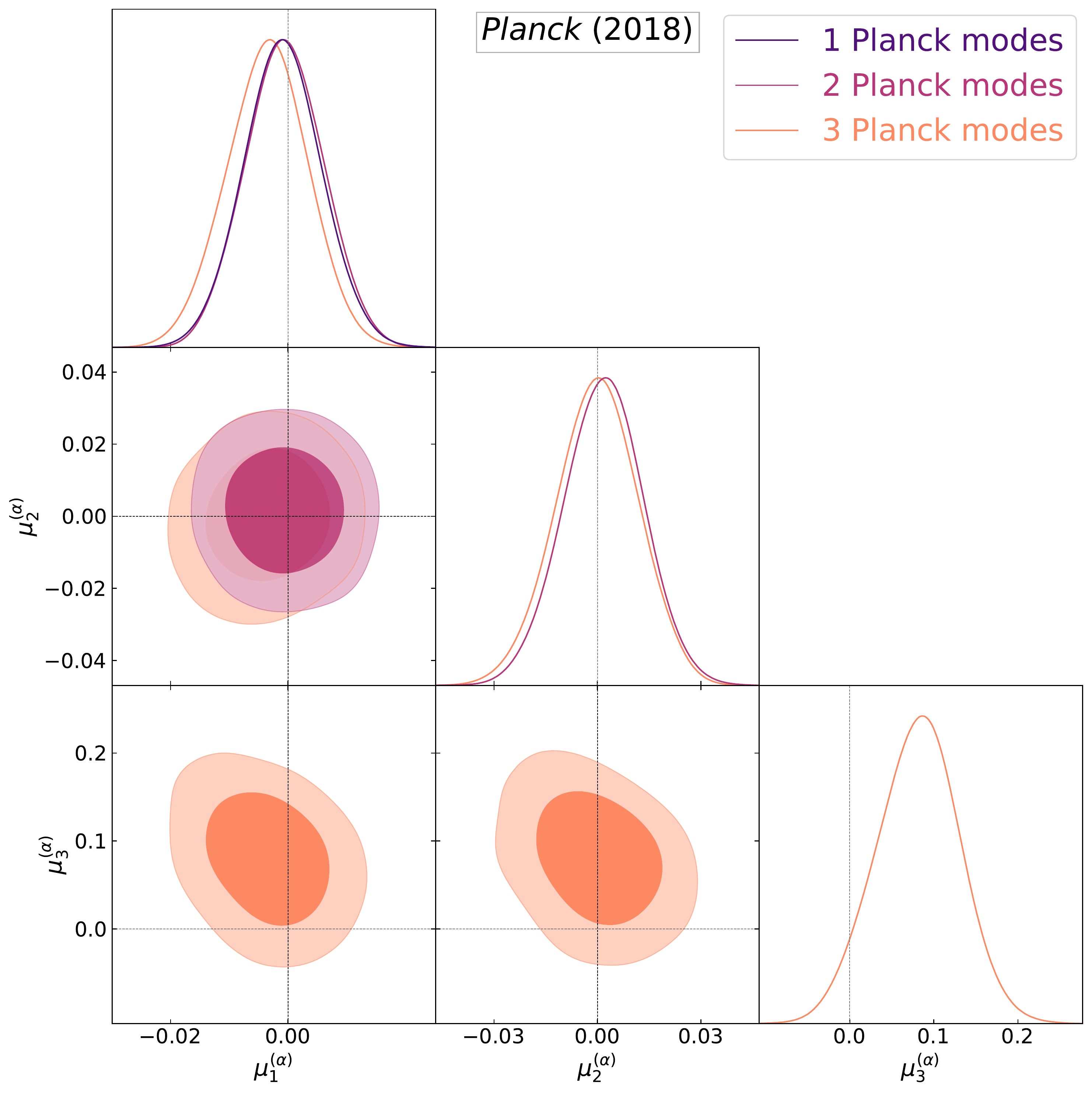}
    \caption{Correlations between the $\mu_i$ amplitude parameters with the \planck likelihood generated $\aEM$ eigenmodes. This plot is comparable to Fig.~\ref{fig:alphaCVLMu} except the modes are generated with the direct likelihood method from Sect.~\ref{sec:direct} instead. The contours are much smaller and close to circular because the modes have been marginalised (see Sect.~\ref{sec:marg}).}
    \label{fig:alphaPlanckMu}
\end{figure}
%----------------------------------------------

\subsection{Planck-data generated modes}\label{sec:contoursPlanck} 
%----------------------------------------------
Following the analysis with the CVL modes, we carried out a similar approach with the \planck direct-likelihood method. The marginalised values of the standard parameters, eigenmode amplitudes and the derived parameters $\ho$ and $\sig$ are shown in Table~\ref{tab:alphaPlanck} for $\aEM$, mirroring the previous analysis. The degeneracy between $\omb$ and $\mu_1$ has slightly reduced to a correlation $\rho(\omb,\mu_1) = 0.50$. 
%----------------------------------------------
\begin{table*}
    \centering
    \begin{tabular} { l  c c c c}
        \hline\hline
        Parameter & \planck 2018 TTTEEE + low-$\ell$ & + 1 \planck $\alpha_{\rm EM}$ mode & + 2 \planck $\alpha_{\rm EM}$ modes & + 3 \planck $\alpha_{\rm EM}$ modes\\
        \hline
        $\omega_b  $ &  $0.02237\pm 0.00015  $ &  $0.02234\pm 0.00018  $ &  $0.02234\pm 0.00019  $ &  $0.02227\pm 0.00020  $\\
        $\omega_c  $ &  $0.1199\pm 0.0012  $ &  $0.1201\pm 0.0014  $ &  $0.1202\pm 0.0015  $ &  $0.1202\pm 0.0016  $\\
        $100\theta_{MC}  $ &  $1.04088\pm 0.00031  $ &  $1.04087\pm 0.00034  $ &  $1.04091\pm 0.00046  $ &  $1.04173\pm 0.00063  $\\
        $\tau  $ &  $0.0542\pm 0.0074  $ &  $0.0541\pm 0.0079  $ &  $0.0538\pm 0.0078  $ &  $0.0535\pm 0.0077  $\\
        ${\rm{ln}}(10^{10} A_s)  $ &  $3.044\pm 0.014  $ &  $3.044\pm 0.016  $ &  $3.044\pm 0.016  $ &  $3.037\pm 0.017  $\\
        $n_s  $ &  $0.9649\pm 0.0041  $ &  $0.9642\pm 0.0060  $ &  $0.9643\pm 0.0060  $ &  $0.9599\pm 0.0065  $\\
        \hline
        $\mu_1 \;(\aEM) $ & $--$   &  $-0.0009\pm 0.0066  $ &  $-0.0006\pm 0.0066  $ &  $-0.0035\pm 0.0069  $\\
        $\mu_2 \;(\aEM) $ & $--$   & $--$   &  $0.002\pm 0.012  $ &  $0.001\pm 0.012  $\\
        $\mu_3 \;(\aEM) $ & $--$   & $--$   & $--$   &  $0.081\pm 0.049  $\\
        \hline
        $H_0  $ &  $67.36\pm 0.54  $ &  $67.28\pm 0.63  $ &  $67.26\pm 0.64  $ &  $67.50\pm 0.68  $\\
        $\sigma_8  $ &  $0.8107\pm 0.0059  $ &  $0.8112\pm 0.0078  $ &  $0.8116\pm 0.0082  $ &  $0.8084\pm 0.0086  $\\
        \hline\hline
    \end{tabular}
    \caption{Marginalised results at the $68\%$ confidence level for the  $\aEM$ modes in Fig.~\ref{fig:modesPlanck} generated with \planck data using the direct likelihood method. This is combined with the \planck 2018 baseline dataset \citep{Planck2018params} and shown against the \LCDM standard case. The comparison of all the standard \LCDM parameters along with two derived parameters, $H_0$ and $\sigma_8$, are shown with the $\mu_i$ amplitudes. The Gelman-Rubin convergence metric for all the chains that generated these results satisfy $\mathcal{R}-1 < 0.01$.}
    \label{tab:alphaPlanck}
\end{table*}
%----------------------------------------------
Notably, the degeneracies between the parameters are no longer affected by the added number of amplitudes. The marginalisation step introduced when creating the \planck modes reduces the standard parameter dependencies. Consequently, the inter-mode orthogonality is relatively preserved. The degeneracy between $\mu_3$ and $\thetaMC$ has not been totally removed, leaving some spurious correlations. This also translates into a $\approx 25\%$ increase in the error to $\ho$, given that the matter density parameters have changed very little with these \planck modes. However, the removal of $\thetaMC$ correlations from $\aEM$ variations in general is much more difficult for marginalisation considering that a broadband, top-hat variation in $\aEM$ will sharply correlate with $\thetaMC$ \citep[see][for more details]{Hart2017}. This is not as rigorously decorrelated compared to PCA20; however it is still heavily improved since the error changes in $\thetaMC$ are increased by a factor of 2 when 3 modes are included. The first two errors are incredibly consistent with the \planck $\aEM$ forecasted errors shown in Table.~\ref{tab:fisherErrors}; however the larger $\thetaMC$ contour leads to the $\mu_3$ error being $36\%$ higher than the Fisher prediction. Though the modes are strongly decorrelated, one can see the influence of $\thetaMC$ degeneracy lines by the $\mu_i\times\mu_j$ correlation contours shown in Fig.~\ref{fig:alphaPlanckMu}. 

%----------------------------------------------
\begin{table*}
    \centering
    \begin{tabular} {l c c c c}
        \hline\hline
        Parameter & \planck 2018 TTTEEE + low-$\ell$ & + 1 \planck $m_{\rm e}$ mode & + 2 \planck$m_{\rm e}$ modes & + 3 \planck$m_{\rm e}$ modes\\
        \hline
        $\omega_b  $ &  $0.02237\pm 0.00015  $ &  $0.02235\pm 0.00018  $ &  $0.02233\pm 0.00019  $ &  $0.02226\pm 0.00020  $\\
        $\omega_c  $ &  $0.1199\pm 0.0012  $ &  $0.1201\pm 0.0014  $ &  $0.1203\pm 0.0015  $ &  $0.1199\pm 0.0015  $\\
        $100\theta_{MC}  $ &  $1.04088\pm 0.00031  $ &  $1.04086\pm 0.00032  $ &  $1.04089\pm 0.00039  $ &  $1.04040\pm 0.00056  $\\
        $\tau  $ &  $0.0542\pm 0.0074  $ &  $0.0541\pm 0.0078  $ &  $0.0542\pm 0.0079  $ &  $0.0535\pm 0.0080  $\\
        ${\rm{ln}}(10^{10} A_s)  $ &  $3.044\pm 0.014  $ &  $3.044\pm 0.016  $ &  $3.045\pm 0.016  $ &  $3.038\pm 0.017  $\\
        $n_s  $ &  $0.9649\pm 0.0041  $ &  $0.9642\pm 0.0057  $ &  $0.9643\pm 0.0057  $ &  $0.9619\pm 0.0061  $\\
        \hline
        $\mu_1 \;(\me)  $ & $--$   &  $-0.001\pm 0.012  $ &  $-0.001\pm 0.012  $ &  $-0.003\pm 0.013  $\\
        $\mu_2 \;(\me)  $ & $--$   & $--$   &  $0.004\pm 0.023  $ &  $0.001\pm 0.023  $\\
        $\mu_3 \;(\me)  $ & $--$   & $--$   & $--$   &  $-0.116\pm 0.092  $\\
        \hline
        $H_0  $ &  $67.36\pm 0.54  $ &  $67.27\pm 0.61  $ &  $67.22\pm 0.64  $ &  $67.12\pm 0.63  $\\
        $\sigma_8  $ &  $0.8107\pm 0.0059  $ &  $0.8113\pm 0.0077  $ &  $0.8121\pm 0.0081  $ &  $0.8075\pm 0.0089  $\\
        \hline\hline
    \end{tabular}
    \caption{Marginalised results at the $68\%$ confidence level for the  $\me$ modes in Fig.~\ref{fig:modesPlanck} generated with \planck data using the direct likelihood method. This is combined with the \planck 2018 baseline dataset \citep{Planck2018params}. The comparison of all the standard \LCDM parameters along with two derived parameters, $H_0$ and $\sigma_8$, are shown with the $\mu_i$ amplitudes. The Gelman-Rubin metric for all the chains that generated these results satisfy $\mathcal{R}-1 < 0.01$.}
    \label{tab:mePlanck}
\end{table*}
%----------------------------------------------
The reduction in inter-correlated degeneracies can be clearly seen in Fig.~\ref{fig:alphaPlanckContours}, where the $\thetaMC$ vs. $\mu_3$ contour is far smaller than the case in Fig.~\ref{fig:alphaCVLContours} for the suboptimal CVL modes. The decorrelation between $\mu_2$ and the other standard parameters is evident from the lack of change in the contours, when the second mode is added. This is corroborated when examining the column of Table.~\ref{tab:alphaPlanck} where 2 modes have been added. The comparison of correlations between $\mu_3$ and $\thetaMC$ for both the CVL and \planck modes is shown in Fig.~\ref{fig:muThetaCompare}. 
The reduction in the error on $\mu_3$ for both fundamental constants has induced a $\sim1-1.5\sigma$ departure from \LCDM for $E_3$ (see Table~\ref{tab:alphaPlanck} and Table~\ref{tab:mePlanck}). 
This is a small deviation, however, it further points to the proposition that constant variations of $\aEM$ and $\me$ do not tell the full story. Physically-motivated models of VFC with more oscillatory behaviour could prove more detectable in future studies. For both $\aEM$ and $\me$, the wider CVL contours show huge improvements when constrained with a marginalisation step since the errors have shrunk by more than a factor of 5. 

%----------------------------------------------
\begin{figure}
    \centering
    \includegraphics[width=\linewidth]{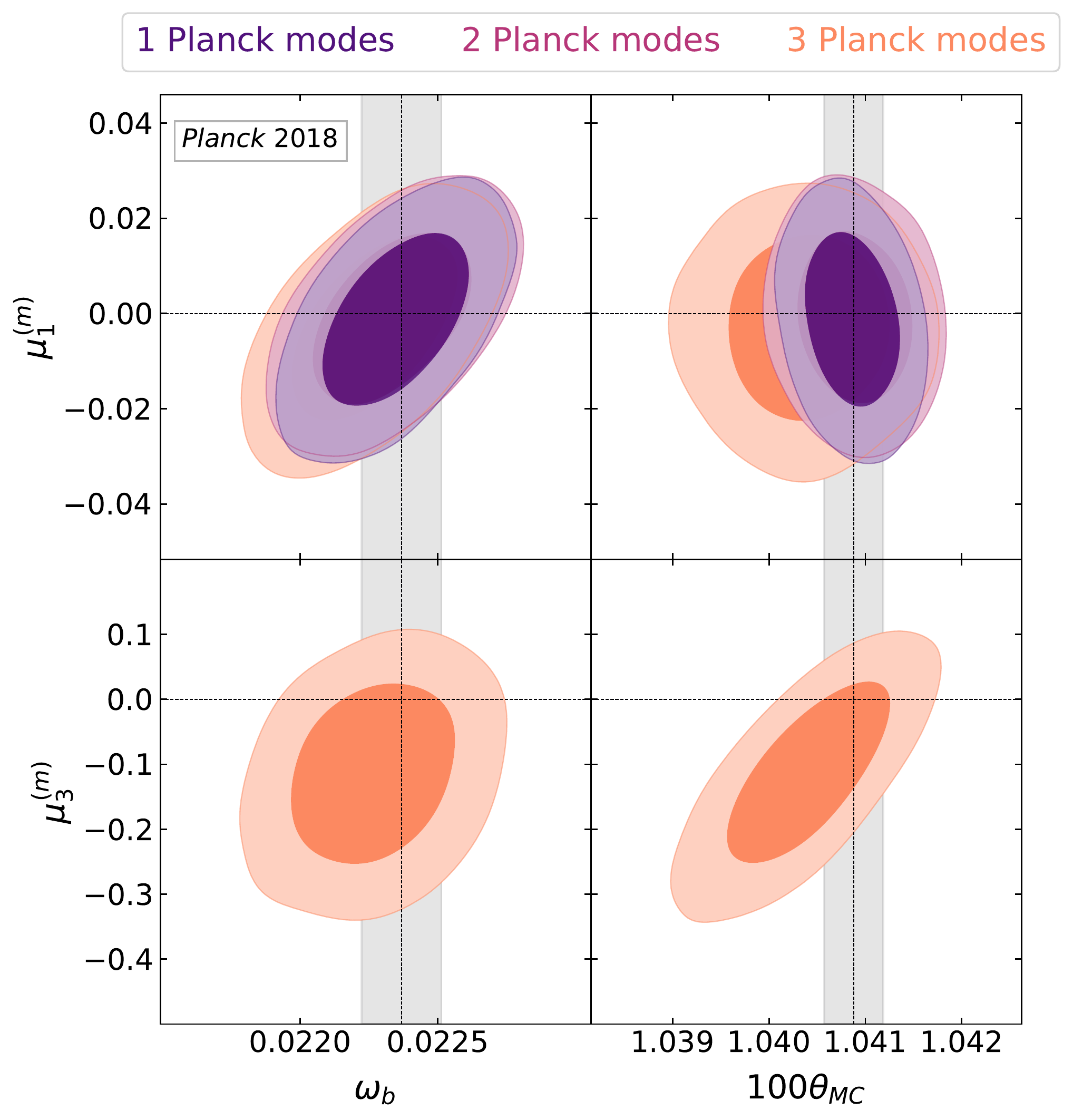}
    \caption{Most correlated likelihood contours from the $\me$ \planck modes shown in Fig.~\ref{fig:modesPlanck}. This is the same correlations as in Fig.~\ref{fig:meCVLContours} except here we remove the $\mu_2^{(\me)}$ contour row because the degeneracies for this parameter are mainly derived from $\mu_1$ and $\mu_3$. Dark bands represent the \LCDM baseline errors.}
    \label{fig:mePlanckContours}
\end{figure}
%----------------------------------------------

In Table.~\ref{tab:mePlanck}, we present the marginalised results for the $\me$ \planck modes previously shown in Fig.~\ref{fig:modesPlanck}. As in the CVL case, the $\me$ results are very similar to those for $\aEM$, however, the errors are slightly larger than the eigensolver predicts (see Table~\ref{tab:fisherErrors}). The correlations between standard parameters and eigenmode amplitudes are also fairly consistent as for $\aEM$. For example, when 3 modes are included, the fine structure correlations, $\rho(\omb,\mu_1^{(\alpha)}) = 0.51$; however the electron mass correlations, $\rho(\omb,\mu_1^{(m)}) = 0.54$. Interestingly, the errors on the standard parameters are modified by a smaller degree in the case of added $\me$ modes as shown by Table~\ref{tab:mePlanck}. Referring to Table~\ref{tab:alphaPlanck} and Table~\ref{tab:mePlanck}, $\sigma(\ho) = 0.68$ when we add $\aEM$ modes whereas the same parameter error when adding $\me$ modes is $\sigma(\ho) =  0.63$. Though these are very small changes, one can see the subtle differences in the contour deformities shown in Fig.~\ref{fig:mePlanckContours}. The electron mass mode amplitudes $\mu_1$ and $\mu_2$ seem thoroughly decorrelated; however the third mode has the same problem with $\thetaMC$ which prevents full decorrelation. More crucially though, the error contours are much narrower than the CVL case thanks to the marginalisation step as illustrated in Fig.~\ref{fig:muThetaCompare}. 

%----------------------------------------------
\begin{figure}
    \centering
    \includegraphics[width=\linewidth]{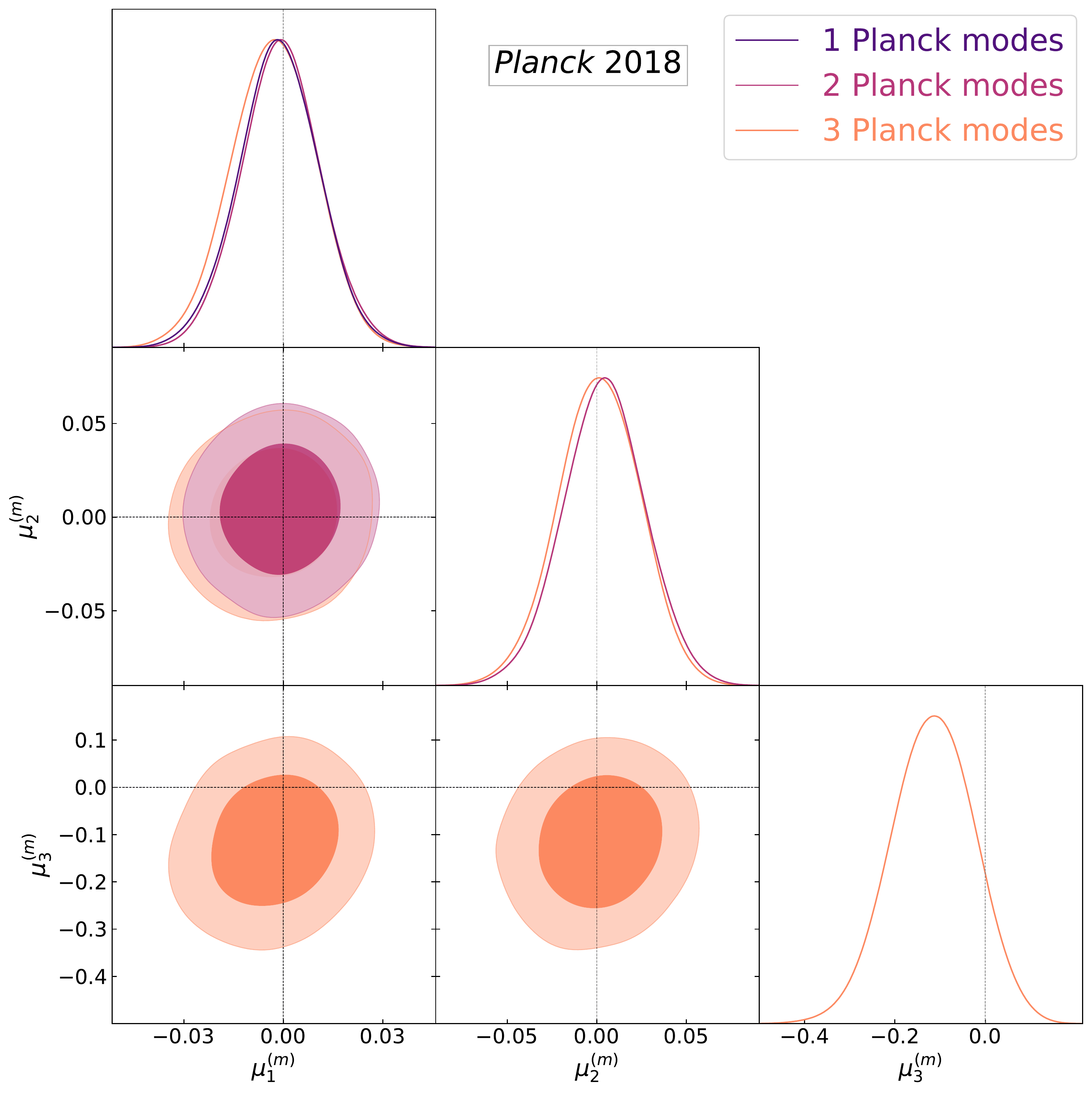}
    \caption{Correlations between the $\mu_i$ amplitude parameters with the \planck likelihood generated $\me$ eigenmodes. Contours are generated from amplitudes using the marginalised eigenmodes as with $\aEM$.}
    \label{fig:mePlanckMu}
\end{figure}
%----------------------------------------------

One point of contention for $\me$, as with the CVL case, is the size of the error bars. Specifically the fact that the $\mu_i$ eigenmode amplitudes are so neatly multiplicative factors of the $\aEM$ modes. This is most clearly shown by the comparable \planck contours between $\mu_3$ and the horizon size $\thetaMC$. In \citet{Hart2017}, the error bars for $\me$ as a constant variation blow up due a degeneracy with $\thetaMC$ (already discussed in Sect.~\ref{sec:planckDiff}); however, in \citet{Hart2017} we showed that the majority of the anomaly relies on the rescaling of the Thomson visibility function. Yet, there is also an interplay between early and late redshifts (pre- and during recombination) which cannot be accounted for if the variations $\fc(z)$ dissipate before later times (i.e., reionisation). 

We will discuss this in more detail in Sect.~\ref{sec:troubleshoot}, however, for now we want to draw the reader's attention to the lack of this geometric degeneracy which is reflected in the contours in Fig.~\ref{fig:mePlanckContours} and~\ref{fig:muThetaCompare}. The change in the horizon scale error from $\sigma(\thetaMC) = 0.00031$ in the \planck baseline case to $\sigma(\thetaMC) = 0.00063$ when 3 modes are added, is far smaller than the $\thetaMC$ error jump expected for constant variations of $\me$. Comparing to the results in VFC20, the error on the horizon size, $\sigma(\thetaMC) = 0.0003\rightarrow0.036$ growing 2 orders of magnitude when including the electron mass variations. This indicates that the $\me$ eigenmodes lack important contributions from $z<300$, which in VFC20 opened the geometric degeneracy line that alleviated the Hubble tension. 

%----------------------------------------------
\begin{figure}
    \centering
    \includegraphics[width=\linewidth]{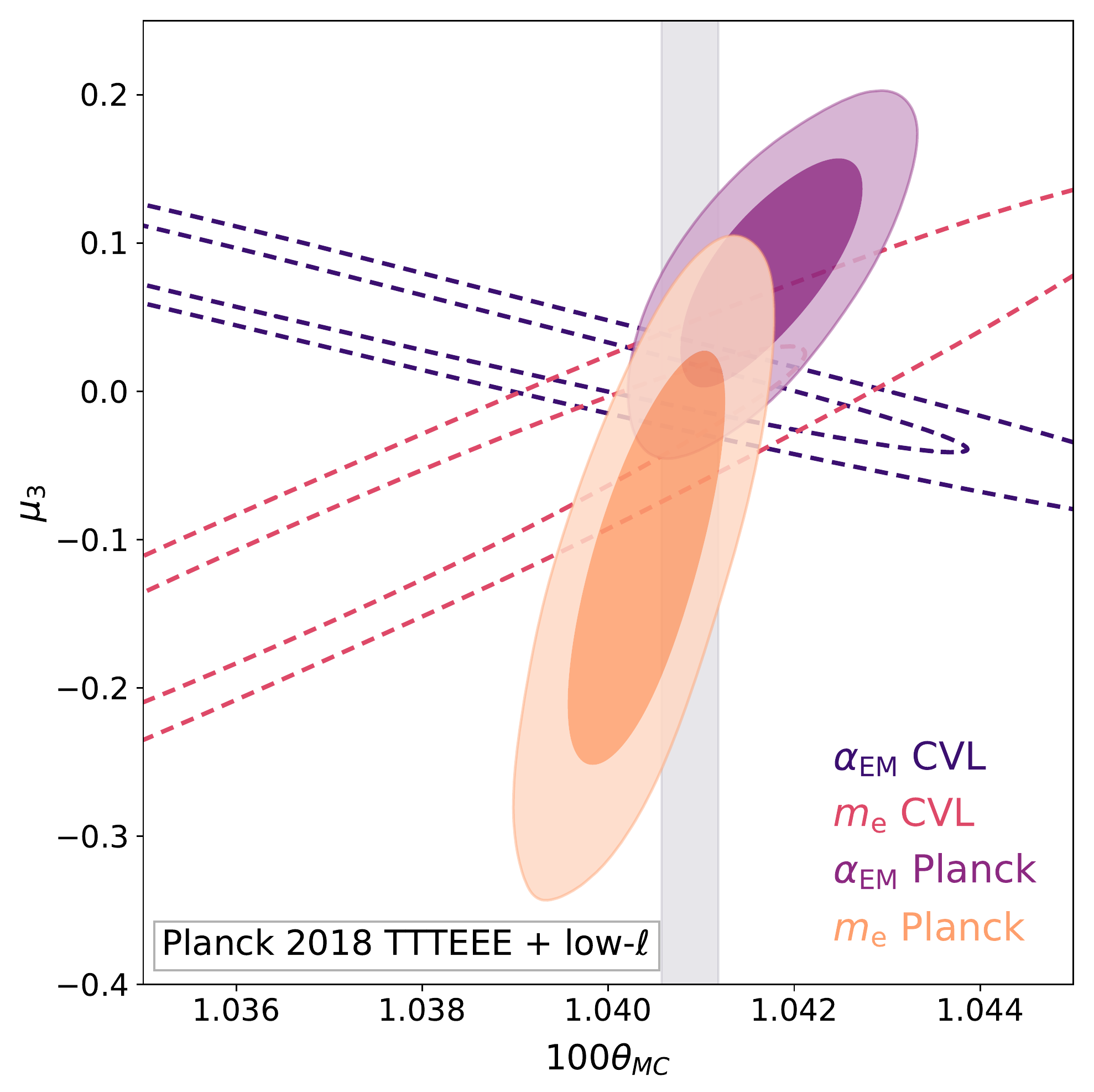}
    \caption{Posterior contour for $\mu_3$ vs. $\thetaMC$ when 3 mode amplitudes are added into the MCMC sampling. Here we compare $\aEM$ (\emph{darker}) with the $\me$ (\emph{lighter}) modes generated with the \planck likelihood (\emph{solid}), against the wider CVL-like mode contours from Sect.~\ref{sec:cvl} (\emph{dashed}).}
    \label{fig:muThetaCompare}
\end{figure}
%----------------------------------------------

\subsection{Direct projections for $\aEM$ and $\me$}\label{sec:proj}
%----------------------------------------------
For eigenmodes that are sufficiently decorrelated, we can recast the variations $\Delta\fc/\fc(z)$ onto a small deviation from the fiducial cosmology and attain excellent, first-order estimates for the parameter values and their errors before jumping onto computationally expensive MCMCs (for certain cosmological problems). The main methodology of the projections formalism has been explained in detail in PCA20; however, we will briefly elucidate some of the key aspects. For the $\xe$ eigenmodes, this approach has already been successfully applied in CMB spectral distortion analysis \citep{Bolliet2020}.

Firstly, we can create a generic variation in the fundamental constants $\fc$ as a function of eigenmodes constrained in the analytic or direct-likelihood method such that,
%----------------------------------------------
\begin{equation}
    \frac{\Delta\fc}{\fc}\left(z\right) = \sum_i\rho_i\,E_i(z)\,, \qquad
    \rho_i = \int\frac{\Delta\fc}{\fc}(z)\cdot E_i(z) \,\id z,
\end{equation}
%----------------------------------------------
where once again, $\rho_i$ is the projection of the fundamental constant eigenmodes onto the given model that one is trying to constrain. If we assume that we are in the perturbative regime that the relative change in the fundamental constant is proportional to the relative change in the model amplitude, (i.e., $\Delta\fc/\fc\propto \dA$ where $\parA$ is the magnitude of a certain model variation\footnote{See the full derivation and motivation for this method in PCA20.}). Since we can suppose the $\Delta\ln\fc$ is proportional to the relative change in the parameter, the projection is now multiplied by the new parameter change $\Delta\parA$ and weighted by the original change $\Delta\parA_0$.

For illustration, the various projections of the eigenmodes with the constant variation and power law models are shown in Table.~\ref{tab:rho}. For constant variations, the fine structure constant modes strongly are strongly projected onto the first two modes, with a slightly weaker contribution from the third mode. In contrast, the $\me$ modes projected predominantly onto the second eigenmode, roughly double the projection onto the third mode. Furthermore, there is negligible projection onto $\mu_1$. For power-law time-dependence, the projections onto the $\aEM$ and $\me$ modes are very similar with the strongest projections onto $E_1$ and $E_3$; however, both modes have much smaller projections onto the second mode, with negative symmetry ($\rho_2\left(p;\aEM\right) = -0.77$, $\rho_2\left(p;\me\right) = 0.64$).

Apply this projection as a $\chi$-squared residual with the $\mu_i$ amplitudes using the MCMC covariance matrices such that, 
%----------------------------------------------
\begin{equation}\label{eq:chi2}
    \chi^2=\left(\delA\,\rho_i-\mu_i\right)^{\rm T}\Sigma_{ij}^{-1}\left(\delA\,\rho_j-\mu_j\right).
\end{equation}
%----------------------------------------------
By solving for the minima of this fit one can find the best-fit value allowed, the accuracy of which is determined by the strength of the marginalisation when generating the modes. If the goodness-of-fit is treated like a likelihood such that the value in Eq.~\eqref{eq:chi2} is transformed by $\mathcal{L} = \exp\left(-\chi^2/2\right)$, the $68\%$ and $95\%$ percentile errors can be found for the given parameter change $\parA$ as well. 
\begin{table}
     \centering
     \begin{tabular}{l|c|c|c|c|c}
       \hline\hline
        Model & Parameter ($\mathcal{A}$) & $\Delta\parA$ & ${\rho}_1$ & ${\rho}_2$ & ${\rho}_3$ \\
        \hline
        Constant & $\aEM/\aEMs$ & $0.01$ & $-2.01$ & $2.00$ & $1.37$ \\[0.5mm]
        Power law & $p$ & $0.001$ & $2.58$ & $-0.77$ & $3.65$ \\[0.5mm]
        \hline\hline
        Model & Parameter ($\mathcal{A}$) & $\Delta\parA$ & ${\rho}_1$ & ${\rho}_2$ & ${\rho}_3$ \\
        \hline
        Constant & $\me/\mes$ & $0.01$ & $0.13$ & $-3.65$ & $-1.63$ \\[0.5mm]
        Power law & $p$ & $0.001$ & $2.80$ & $0.64$ & $2.71$ \\[0.5mm]
        \hline
        \hline 
     \end{tabular}
     \caption{Projections $\rho_i$ of fundamental constant $\fc$ changes onto the \planck eigenmodes alongside the parameter step size $\Delta\parA$ used. Each value ${\rho}_i$ measures how strongly the physical variations from the constant and power-law models project onto our \planck modes in Fig.~\ref{fig:modesPlanck}}
     \label{tab:rho}
 \end{table}
In Table~\ref{tab:projectionResults}, the projection results for $\aEM$ and $\me$ are compared against the simple constant relation and the phenomenological power law from our previous work. It is important to point out that the MCMC parameter values are actually garnered from a best-fit algorithm, since that is a clearer indication from the minimisation of the $\chi^2$. From the models given, the constant $\aEM$ results constrained by the projections method are exceptionally close to the MCMC sampled value. This is also the case for both the phenomenological power law cases where the difference is $\sim0.25\sigma$ for the $\aEM$ modes and $\lsim0.1\sigma$ for the $\me$ modes. The power-law variations were even tested with an added curvature term where $p\rightarrow p+\beta\ln\left[(1+z)/1100\right]$; however, the results were compatible to $0.003\sigma$. Though the curvature term has higher physical consistency ($\fc(z\rightarrow0)\rightarrow0)$, it has a very small impact around the Thomson visibility function where the recombination constraints are most sensitive. All these model projections were far closer to PCA20 results due to the basic functional form these variations for $\fc$ take compared to the free electron fraction, $\xe$ and complicated parameter dependencies.

The key difference is the constant $\me$ projection. As documented in VFC20, the electron mass exposes a huge degeneracy line with $\ho$. This leads to the $\me$ MCMC error being much higher than $\aEM$ (as shown in Table~\ref{tab:projectionResults}). However, here the projection error is an order of magnitude smaller and the central value of $\me/\mes$ is far closer to unity. This suggests that something is amiss with the projection method for $\me$, as we discuss now.

%----------------------------------------------
\begin{table}
    \centering
    \begin{tabular*}{\linewidth}{@{\extracolsep{\fill}} l c c c}
        \hline\hline
        \multicolumn{4}{c}{Fine structure constant variations ($\aEM$)}\\
        \hline
        Model & $\aEM(z)$ & MCMC & Projections \\[0.5mm]
        \hline
        Constant
        & $\aEM/\aEMs$ & $1.0010\pm0.0024$ & $1.0012\pm0.0029$ \\[0.5mm]
        \vspace{0.5mm}
        Power law & $\left(\frac{1+z}{1100}\right)^{\,p}$ & $-0.0002\pm0.0024$ & $0.0004\pm0.0024$\\[0.5mm]
        \hline\hline
        \multicolumn{4}{c}{Effective electron mass variations ($\me$)} \\
        \hline
        Model & $\me(z)$ & MCMC & Projections \\[0.5mm]
        \hline
        Constant & $\me/\mes$ & $0.844\pm0.059$ & $0.9995\pm0.0062$\\[0.5mm]
        \vspace{0.5mm}
        Power law 
         & $\left(\frac{1+z}{1100}\right)^{\,p}$ & $-0.0006\pm0.0042$ & $-0.0009\pm0.0045$ \\[0.5mm]
        \hline\hline
       
    \end{tabular*}
    \caption{Projection results using the first 3 eigenmodes for $\aEM$ and $\me$. The constant and power law models have been compared against the MCMC results from {\tt CosmoMC} constrained with \planck in \citet{Hart2017}. The values from the MCMC are the best fit values along with the marginalised errors (since the projection module finds the best fit point in $\parA$).}
    \label{tab:projectionResults}
\end{table}
%---------------------------------------------- 

\subsection{Problems with the $\me$ projection and new hints about the origin of the Hubble tension}
\label{sec:troubleshoot}
%----------------------------------------------
As we have shown in Sect.~\ref{sec:proj}, the direct projection method works quite well for simple models of fundamental constant variations except for the constant variations in $\me$, which seem to be giving much smaller errors than the direct constraints \citep[e.g.,][]{Hart2017}. 
{\it What is going on here?}

As already mentioned in passing, this may be related with how the modes are constructed in our VFC PCA. In contrast to the direct constraints, our modes, operating at $300<z<2000$, do not capture any changes to the Thomson visibility caused by VFC at $z<300$ and during reionisation.
%----------------------------------------------
\begin{figure}
    \centering
    \includegraphics[width=\linewidth]{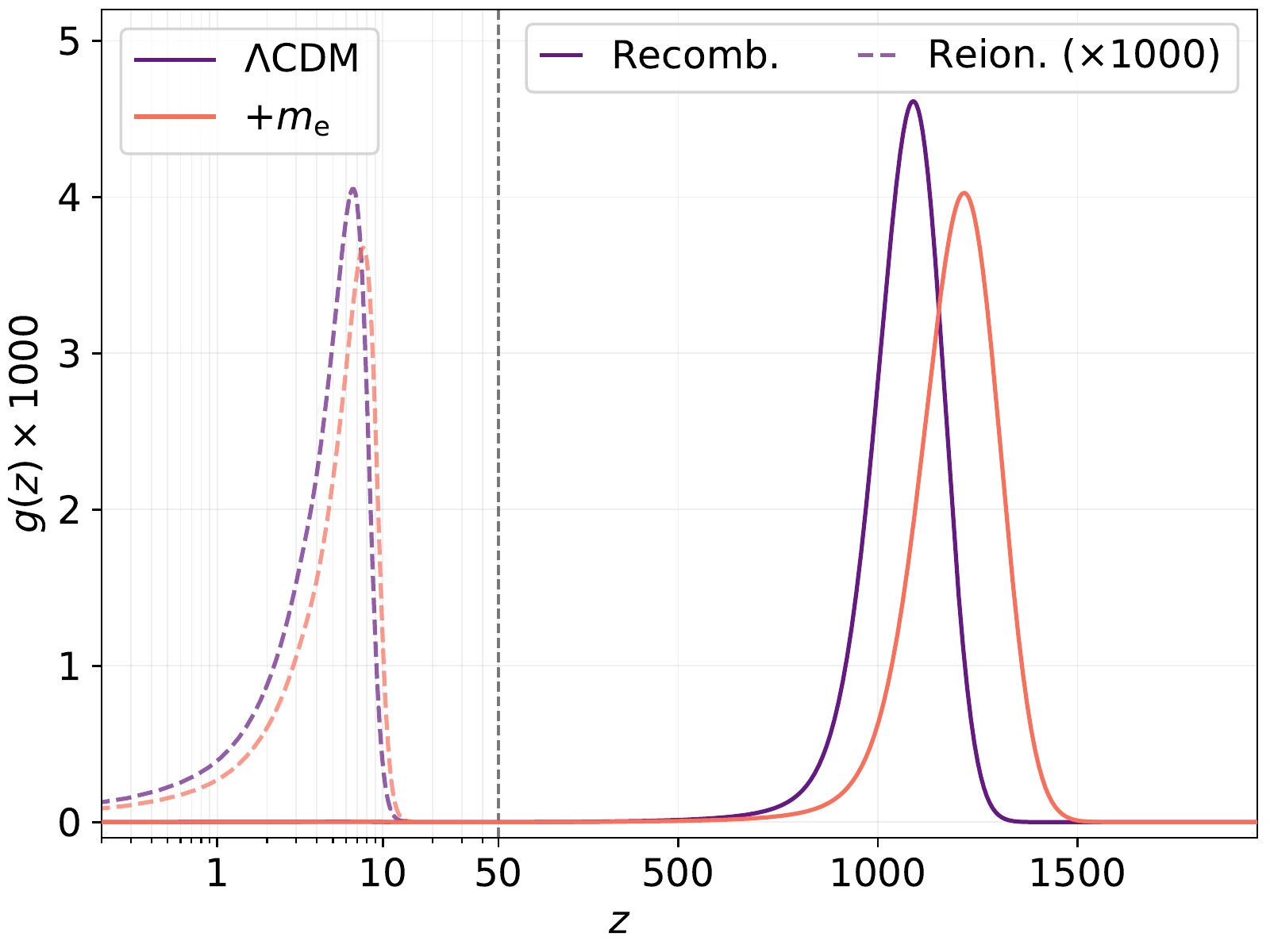}
    \caption{The visibility function $g(z) = \taudot e^{-\tau(z)}$ made from the opacity $\taudot$ discussed in Sect.~\ref{sec:cvl}. The changes from increasing $\me$ by $10\%$ (\emph{orange}) are shown against the \LCDM scenario (\emph{purple}). This includes both the Thomson visibility function at recombination (\emph{solid}) and the residual bump of opacity coming from the reionisation epoch (\emph{dashed}). Reionisation visibility has been multiplied by a factor of $1000$.}
    \label{fig:visiMe}
\end{figure}
%----------------------------------------------
In Fig.~\ref{fig:visiMe}, we present the visibility function variations when we include a constant variation of $\me/\mes = 1.1$. While the right panel focuses on the effect during recombination, the left panel looks at the variations arising in the reionisation era. The latter rely purely on the rescaling of the Thomson cross section which modifies the opacity of electrons during the reionisation epoch, which are not covered by our VFC modes. Note that the visibility from reionisation had to be amplified $\times 1000$ due to the smaller opacity during this era. 

From our previous study, we also know that the geometric degeneracy between $\me$ and $\ho$ lies in the additional $\sigT$ rescaling. Without this rescaling, the errors on $\me$ shrinks by a factor of $\simeq 5$, providing much less freedom along the geometric degeneracy line \citep{Hart2017}.
In VFC20, we further tested the dependence on various likelihood configurations and found no clear data source (e.g., high-$\ell$ likelihood, lensing) that causes the large degeneracies with $\ho$. The exception was a $\sim30\%$ reduction in the tension between $\me$ and $\ho$ which could be accounted by the changes to the $\tau$ value from the new polarisation $EE$ likelihood\footnote{Testing for this was done with {\tt CosmoMC} using the \planck 2015 optical depth prior: $\tau=0.079\pm0.017$.}. The potency of the polarisation likelihood and its proximity to the Hubble tension have also been alluded to in \citet{Addison2021}.

It therefore seems crucial to account for the full time dependence of the electron mass variability as a function of redshift, including later eras such as the dark ages, reionisation and the 21cm regime. This is also corroborated when adding BAO and SN data \citep[using][]{Riess2019} in the MCMC analysis, where one finds a small drift in the parameter values consistent with the likelihood combinations in \LCDM, but negligible changes in the error bars ($\lsim0.01\sigma$). If we recreate the results that are shown in Table.~\ref{tab:projectionResults} using the \planck+ BAO MCMC results instead of the \planck likelihood alone, we find the projection result $\me/\mes = 1.0013\pm0.0060$. This departs slightly from the direct MCMC result when we added BAO in VFC20 ($\me/\mes=1.0078\pm0.0067$); however there are still traces of the geometric degeneracy here, albeit much smaller variations. However, the changes in the projection result when the BAO likelihood is included goes into the right direction and is far closer to the direct projection ($\simeq1\sigma$ deviation).

Our discussion shows that a coordination between the dark ages, reionisation and recombination could be vital for modelling the ionisation history in the future. The link between these epochs and the consequences of a universal ionisation history solution in the atomic physics regime may aid other theories. For example, one of the compelling solutions to the Hubble tension involves a baryon clumping effect that arises from primordial magnetic fields \citep{Jedamzik2020}. However, another study has suggested that small-scale CMB data may contradict this with current Atacama Cosmology Telescope (ACT) data \citep{Thiele2021ACT}. If the baryonic clumping model was refined for a wider range of epochs, small changes during the dark ages and reionisation epoch may restore the consistency problems with the small-scale CMB data. Additional baryon clumping causes an acceleration of recombination at last scattering. Conversely, star formation may be enhanced in denser regions during reionisation, causing an earlier onset and longer duration of reionisation. To leading order, this is consistent with the modifications that constant variations of $\me$ introduce, suggesting that a similarly orchestrated change in the ionization history may be at work behind the scenes.

A complimentary study using a joint PCA for recombination and reionisation seems highly motivated by these findings. Specifically for the VFC model and the time-dependent eigenmodes, the inclusion of reionisation effects are beyond the scope of this paper. Due to the logarithmic relationship between conformal time and redshift ($\delta\ln\eta\simeq-\delta\ln z$), the implementation of basis functions into the reionisation era is more complicated for perturbations at redshifts $15\leq z_i\leq300$. Variable basis functions across the same grid could help but have been shown to create significant correlations between eigenmodes when recasting the Fisher elements back into the $\xe$-basis (see PCA20 for more details).
Returning to $\xe$-modes for both recombination and reionisation may be beneficial, combining the methods of PCA20 and \citet{Mortonson2008}, for instance.
These explorations are left for a future study but most likely are at the core of the issues seen here.

\section{Forecasting eigenmodes with Simons Observatory noise curves}\label{sec:so}
%----------------------------------------------
To conclude our study, we turn to one of the interesting future projects that will involve CMB observables: The Simons Observatory (SO) \citep{SOWP2018}. For this analysis, we make use of the publicly available {\tt so\_noise} models code to generate an added noise term in the Fisher matrices for our analytic model. In this section, we add in the Simons noise curves with a $40\%$ sky coverage according to their preliminary forecasts as well as an adjusted $\ell$ range for their Large Aperture Telescope (LAT) where $40\leq\ell\leq8000$. For simplicity, we will be considering the standard-ILC noise that emerges from the SO forecasts and not any of the deprojection effects from foregrounds (e.g., dust and synchrotron constrained-ILCs). The noise curves for the LAT agree with the forecasting paper for SO \citep{SOWP2018}. The machinery is modified such that $C_\ell^{X}\rightarrow C_\ell^{X}+N_\ell^{X}$ within the covariance matrix which changes the effective signal-to-noise of certain responses in the Fisher matrix \citep[see][for more details]{Tegmark1997}. 

\begin{figure}
    \centering
    \includegraphics[width=\linewidth]{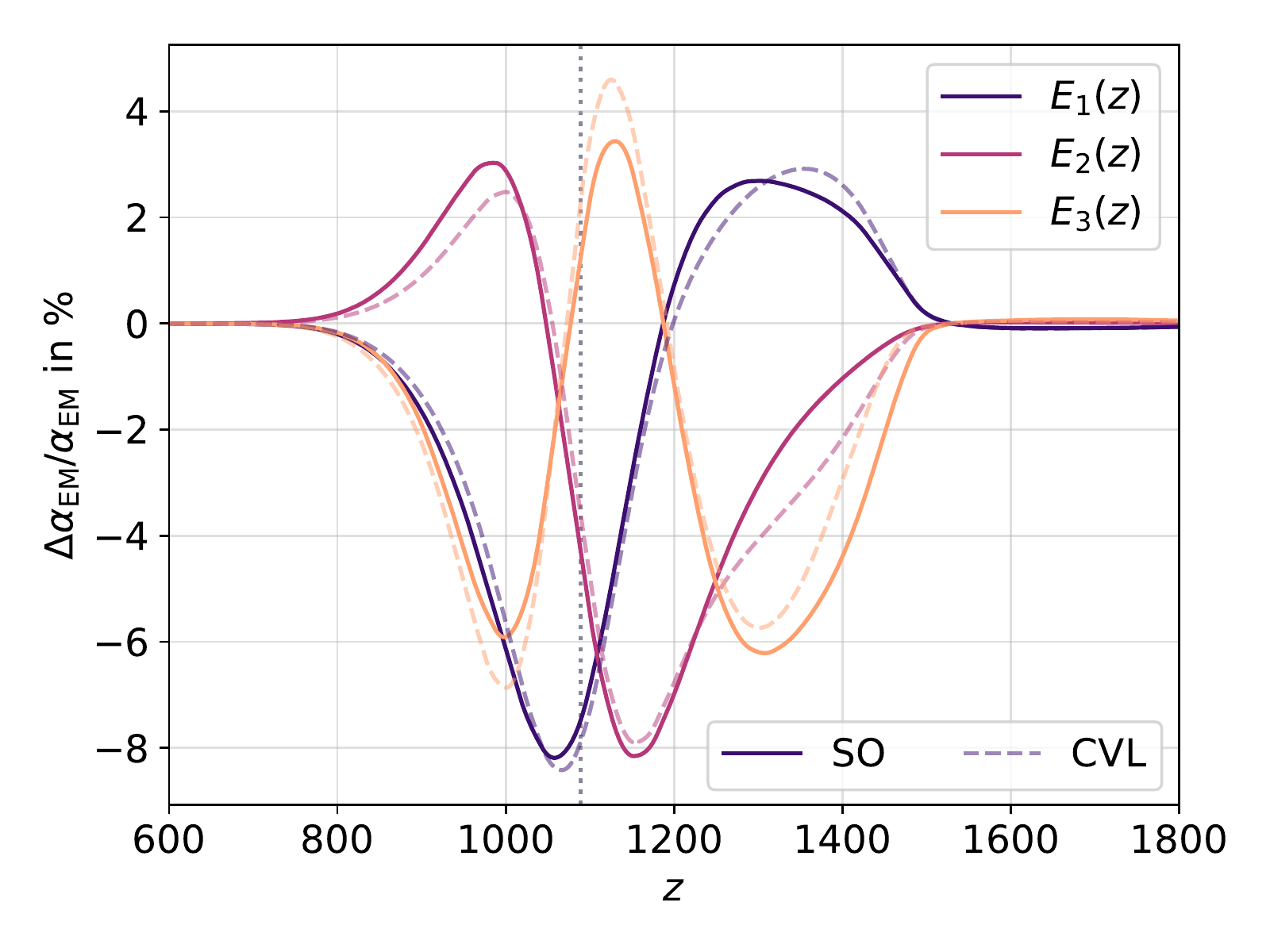}
    \includegraphics[width=\linewidth]{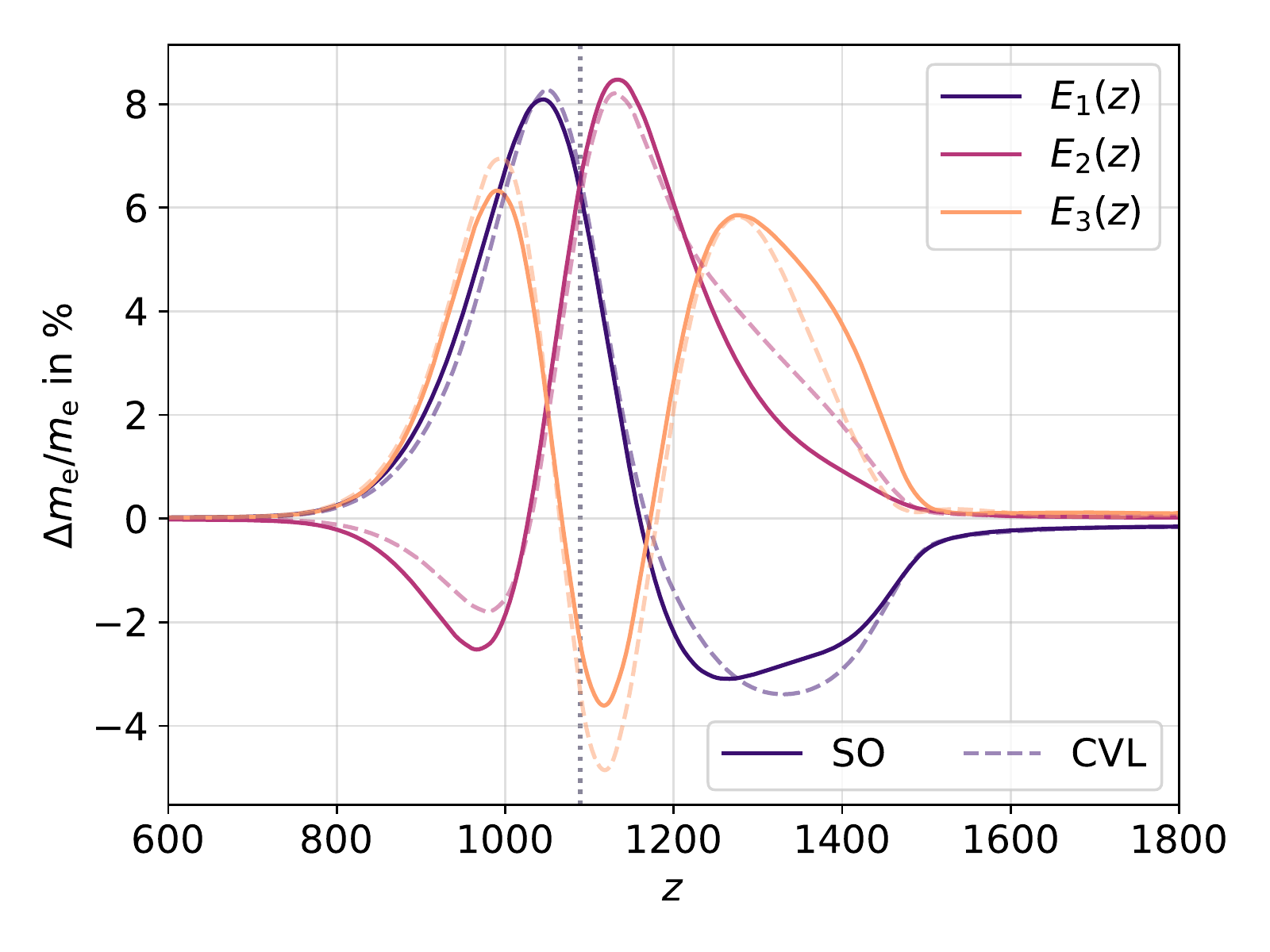}
    \caption{Principal components predicted using the analytic method with the Simons Observatory noise curves. For this particular experiment, it has been configured for the Large Aperture Telescope (LAT) with $f_{\rm sky} = 0.4$ and $l_{\rm max} = 8000$.}
    \label{fig:modesSO}
\end{figure}

In Fig.~\ref{fig:modesSO}, the SO modes are shown together with the CVL modes from Sect.~\ref{sec:cvl}. For both $\aEM$ (\emph{top}, Fig.~\ref{fig:modesSO}) and $\me$ (\emph{bottom}, Fig.~\ref{fig:modesSO}), the biggest impact lies in the second and third eigenmodes. The kink that was present in $E_2$ for the CVL case (Sect.~\ref{sec:cvl}) at $z\sim1300$ has been removed for the SO modes. Specifically, whatever relic in the $\Dell$ power spectra that caused the modes to quickly truncate to 0 around $z\sim1500$ has been removed for a smoother exponential-like decline. 
The rapid drop in the first mode, $E_1$, has also been subtly changed due to the introduction of the SO noise. The third mode in both cases also exhibits an amplitude reduction for the peaks where $z<1200$ however a larger amount of mode information (larger area) for the final peak at $z\sim 1300$. The trading of feature information in the modes could explain the removal of the kink in $E_2$ as well. 

The forecasted errors for the SO modes are shown in Table~\ref{tab:fisherErrors} alongside the previously discussed CVL and \planck results. The predicted errors for a PCA with SO parameters sit nicely between the marginalised \planck components and the idealised CVL setup.
However, the reduction of these errors with data from SO heavily relies on careful treatment of the likelihood, covariances and the data in general. Furthermore, the wider implications of the SO forecasted modes are that recombination eigenmodes (such as the fundamental constant eigenmodes) are approaching a critical constraining limit. Applications of the PCA method to \planck data have made huge strides in constraining these eigenmodes shown in previous studies where these were forecasted \citep{Farhang2011}. However, since $\sigma(E_i^{\rm P18})\simeq 4\sigma(E_i^{\rm SO})$, we are very close to the CVL floor of constraining power available for this kind of analysis.

Feeding the estimated errors for SO into the projection machinery discussed in Sect.~\ref{sec:proj}, the predicted error for a constant measurement of fine structure constant is $\sigma_{\rm SO}\left(\aEM\right)\simeq0.0001$. Similarly the predicted error for the electron mass is $\sigma_{\rm SO}\left(\me\right)\simeq0.0003$. Although in both cases this is $\simeq 20$ times smaller than the \planck projection result, this neglects the marginalisation over standard parameters. The \emph{CORE} collaboration forecasted for $\aEM$ detectability for several experimental configurations and their baseline, CORE-M5, was similar to SO for high-$\ell$ noise \citep[see][for more details]{CORE2016}. The constrained error they found for this setup was $\simeq0.0007$ which is $\sim 5$ times larger than our expected error; however, they anticipate the degeneracy between $\aEM$ and $\ho$ to start being a limiting factor.

The projection error for SO can be refined with more detailed forecasting models in the future (including foregrounds and other experimental effects). However, our estimates are already promising and cement the idea that SO could be approaching the limit of exceptional constraints for $\aEM$ and $\me$ in upcoming analyses. In particular, sensitivity to time-dependent variations may be possible, and a VFC PCA provides a robust framework for mapping various VFC model parameters to direct observables, separating the model-dependent interpretation step from the data analysis.

\subsection{Responses in the CMB power spectra with added noise suppression}\label{sec:residualsSO}
\begin{figure}
    \centering
    \includegraphics[width=\linewidth]{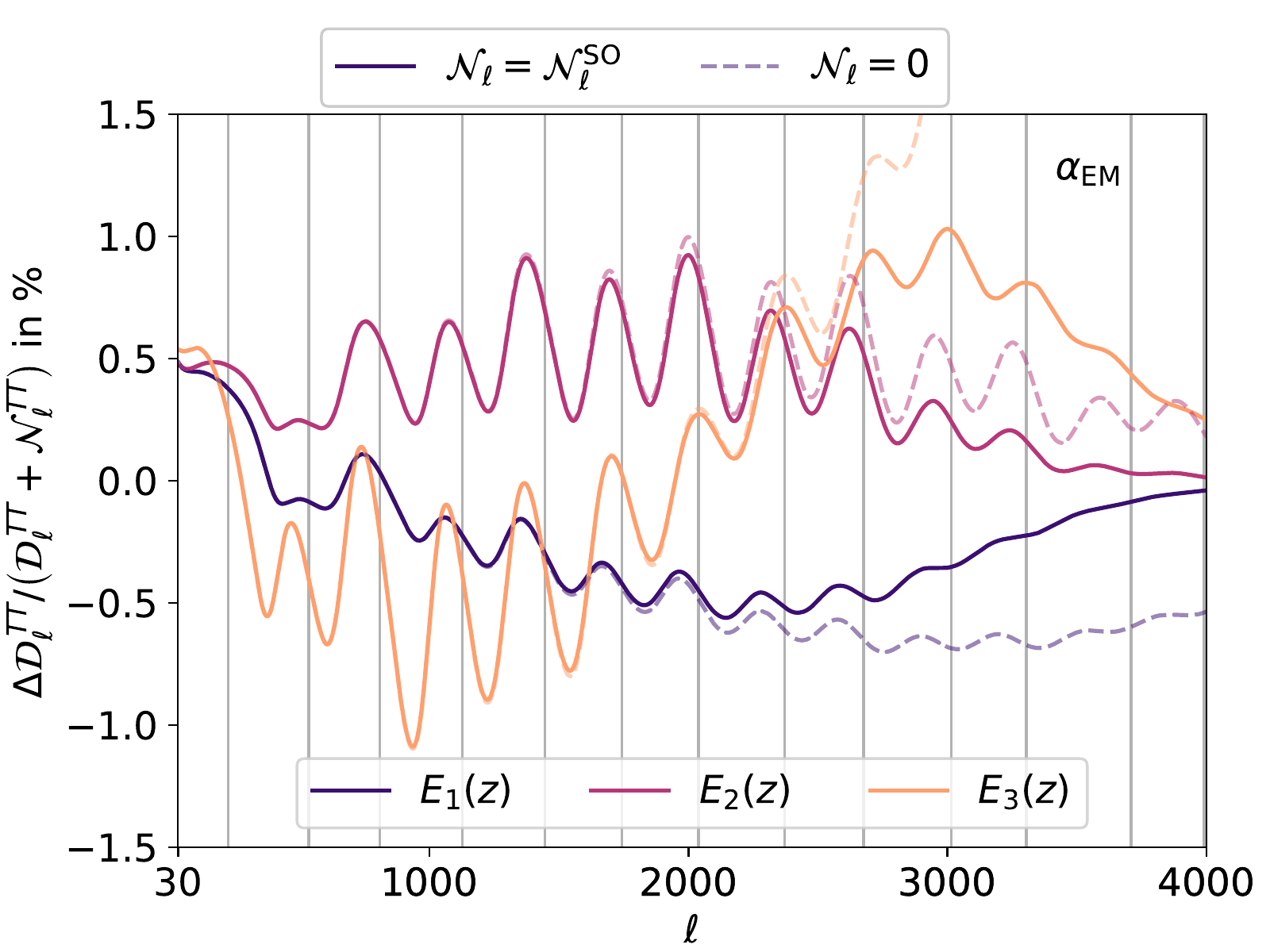}
    \includegraphics[width=\linewidth]{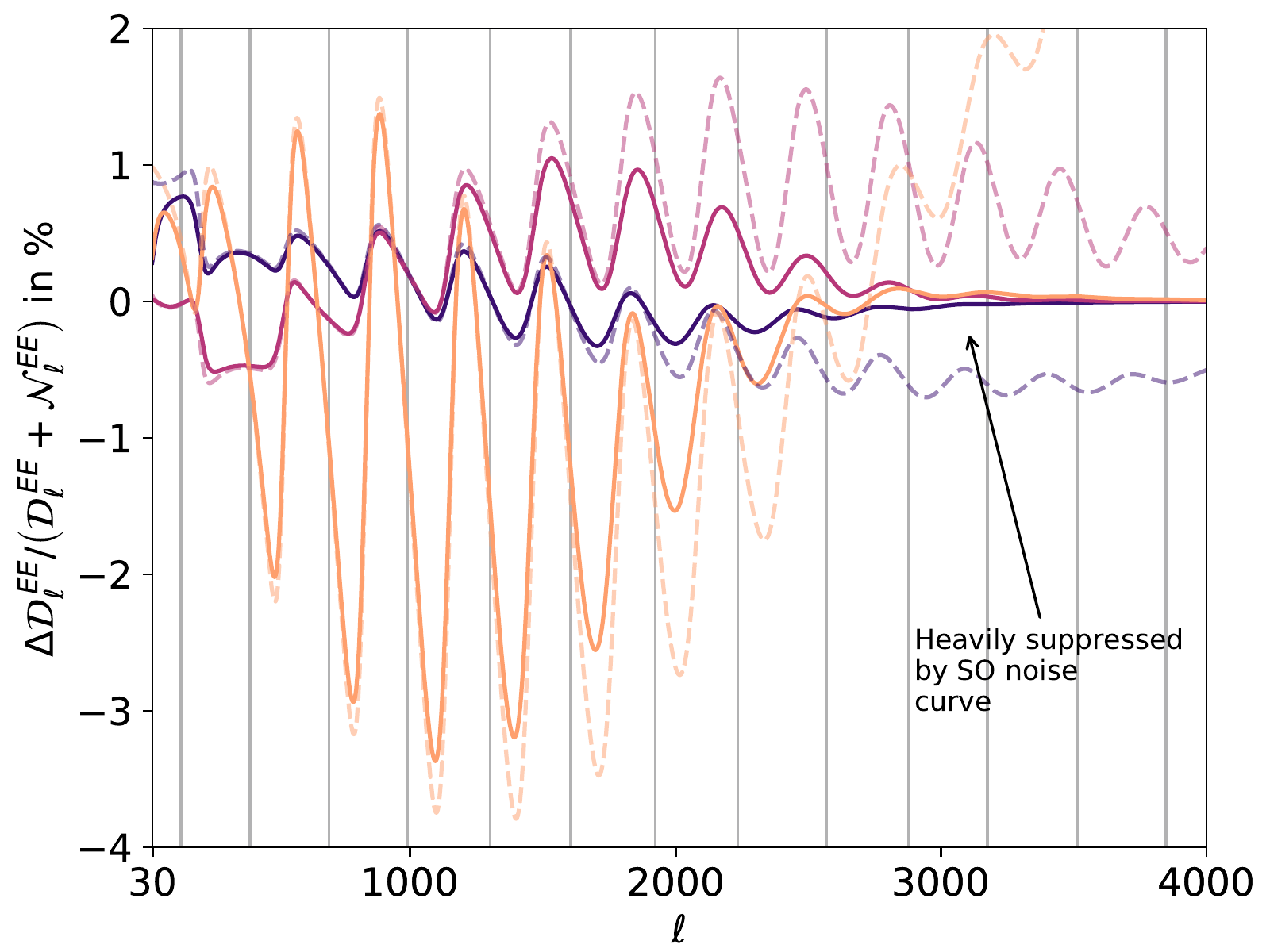}
    \caption{Responses from the eigenmodes generates with $\mathcal{N}_\ell$ noise curves from the SO forecasts compared against the simplest CVL alternative ($N_\ell = 0$). The differences in the CMB power spectra, $\Delta\mathcal{D}_\ell$ are shown as a ratio with the noiseless case (\emph{dashed}) and the SO noise case with $f_{\rm sky} = 0.4$ and $\ell_{\rm max} = 8000$ (\emph{solid}). Grey bands indicate the peaks of the fiducial \LCDM CMB spectra according to \planck 2018.}
    \label{fig:cmbSO}
\end{figure}

%----------------------------------------------
As an additional illustration, the differences from the responses for these eigenmodes are shown in Fig.~\ref{fig:cmbSO} with and without SO noise weighting. For this example, we are only focused on $\aEM$ principal components and there is a truncation of $\ell\gtrsim4000$ due to the forecasted noise from SO. This suppression happens much lower at $\ell\simeq 3000$ for the polarisation spectra, suggesting the noise has a sharper cutoff than the temperature spectra. Interestingly, the third eigenmode $E_3$ exhibits an exponentially large response in the both panels at high-$\ell$ (\emph{orange, dashed}), however the noise helps to damp these residuals away. 

For the other two eigenmodes, the noise changes a non-zero floor in $\Delta\mathcal{D}_\ell^{\rm TT}$ into an exponential decay at higher-$\ell$, showing the influence of the SO noise. This is particularly noticeable for the $\DellE$ residual of $E_2$, where the  out of phase responses in the CMB $EE$ peaks are much smaller in the SO case. The discrepancies in the modes between CVL and SO arise from the added $\ell$ modes before the noise kicks in at $\ell\lsim 3000$. Even a few hundred extra modes compared to the CVL case can make the difference seen. Furthermore, the interplay of the different suppression levels between temperature and polarisation will influence the Fisher matrix as well. Adding the SAT noise curves may sharply change the eigenmode shapes in the large scales as well, however these noise curves target the larger scale features associated with BB error constraints thus far (i.e., tensor spectral index $r$), which are not affected by the ionisation history \citep{SOWP2018}.

%----------------------------------------------
%----------------------------------------------

%----------------------------------------------

\section{Conclusion}
\label{sec:conclusion}
%----------------------------------------------
In this work, we have performed the first PCA for fundamental constant variations across recombination. 
Fundamental constant variations modelled in \citet{Hart2017} can now be broken into a set of basis functions which provoke responses in the CMB spectra through the changes to the ionisation history and the opacity scaling (via the Thomson cross section) using {\tt FEARec++}. 
Generalizing the methodology of PCA20, we have constrained the principal components for a number of experimental setups using an analytical method (i.e., CVL experiments and Simons Observatory) and a direct likelihood method (i.e., \planck). The obtained principal components all cut off smoothly above $z\simeq 1500$, as the deviation become negligible in the ionisation history and the CMB spectra in the tails. The modes given here have been constructed with the same rigour as PCA20 including stability analysis, minimisation of non-orthogonalities and parameter marginalisation. 

We have shown that the majority of our component analysis does not point to deviations from \LCDM; however, the marginalised \planck 2018 VFC modes for both constants hint at a $\simeq 1-1.5\sigma$ deviation in the $\mu_3$ amplitudes. This mode is not strongly excited by time-independent $\aEM$ and $\me$ variations (see Table~\ref{tab:rho}), which have been thoroughly studied in the past, hence suggesting that the story could be more complicated. 

Given the principal components for $\aEM$ and $\me$ have now been constrained, these can be easily applied to complex models of these variations using rudimentary linear algebra. For constant variations and our phenomenological power law first discussed in \citet{Hart2017}, the results from our projections method are consistent with the direct MCMC runs (Table~\ref{tab:projectionResults}). For example, constant $\aEM$ variations agree with the MCMC result to the level of $\simeq0.08\sigma$ whilst the power law is consistent to $\simeq0.25\sigma$ for both $\aEM$ and $\me$. Similar studies could be used to constrain physically-motivated models such as the runaway dilaton and BSBM model variations during recombination \citep[see][for BSBM model]{Sandvik2001}. 

An inconsistency appears for our $\me$ modes; however, this also delivers one of the interesting contentions of this work. Specifically, the constant $\me$ projection results differ radically from the MCMC results, which strongly exploit a $\ho$ geometric degeneracy (see VFC20). Since here the basis functions are created at $z>300$ and there are non-negligible variations in the reionisation visibility function arising from $\me$ variations missing (see Fig.~\ref{fig:visiMe}), we suggest that a more detailed analysis with reionisation modes explicitly included could recreate the degeneracy. More importantly, since the reionisation physics interplay with recombination is important to these components, we posit the idea that certain aspects of this interplay are integral to full solutions to the Hubble tension (see Sect.~\ref{sec:troubleshoot} for discussion). This could help rectify current issues with other promising solutions for the Hubble tension such as those induced by primordial magnetic fields \citep{Jedamzik2020}, as we speculate here.

We also forecast how well VFC may be constrained with The Simons Observatory (see Sect.~\ref{sec:so}). Here, we see a distinct improvement in the detectability over the idealised \planck eigenmodes, with some of the structural features from a CVL experiment. Differences arise due to the carefully computed noise curves lifted from the SO forecast data release \citep{SOWP2018}. While more detailed forecast including instrumental and foreground effects will be needed, our estimates highlight the immense potential of future ground-based observations in this respect.

We close by remarking that variations in $\aEM$ and $\me$ are often motivated by scalar fields (e.g., the BSBM model) which could conceivably stem from the same variations that give rise to \emph{early dark energy} effects \citep[e.g.,][]{Poulin2019}. Developing more realistic models on $\aEM$ and $\me$ could have an impact on future constraints for the early dark energy mechanisms such as the ultra-light axion model. The extended variations of these constants have not been forecasted here, however, there is lots of room to continue this in the future with The Simons Observatory. Since the forecasted eigenmode errors for SO are significantly improved when compared to \planck, this could put important constraints on these models and add more information to the current picture surrounding these more complex  formulations for changes to $\aEM$ and $\me$.

\section*{Acknowledgements}
%----------------------------------------------
This work was supported by the ERC Consolidator Grant {\it CMBSPEC} (No.~725456) as part of the European Union's Horizon 2020 research and innovation program.
JC was also supported as a Royal Society URF at the University of Manchester.

\section*{Data Availability}
%----------------------------------------------
The current {\tt FEARec++} software package will be available at \url{https://github.com/cosmologyluke/FEARec} for solving PCAs during recombination. The \planck 2018 likelihood files are available at \url{http://pla.esac.esa.int/pla/}. Forecasted data for the different Simons Observatory specifications are given at \url{https://github.com/simonsobs/so_noise_models}.

\appendix
%----------------------------------------------
\section{Stability analysis of the \planck 2018 likelihood}\label{app:planck}
%----------------------------------------------
In this section, we corroborate the analysis from PCA20 by following the same \emph{direct likelihood} methodology, mentioned in Sect.~\ref{sec:howtoPCA}. Numerical derivatives requires an appropriate step size and stable minima position for the \planck 2018 likelihood with respect to all the standard cosmological parameters $\left(\left\{\omb,\omc,\thetaMC,\tau,\ns,\As\right\}\right)$. In our analysis we have also included derivatives about the minima for nuisance parameters from the \planck 2018 likelihood.

%----------------------------------------------
\begin{figure}
    \centering
    \includegraphics[width=\linewidth]{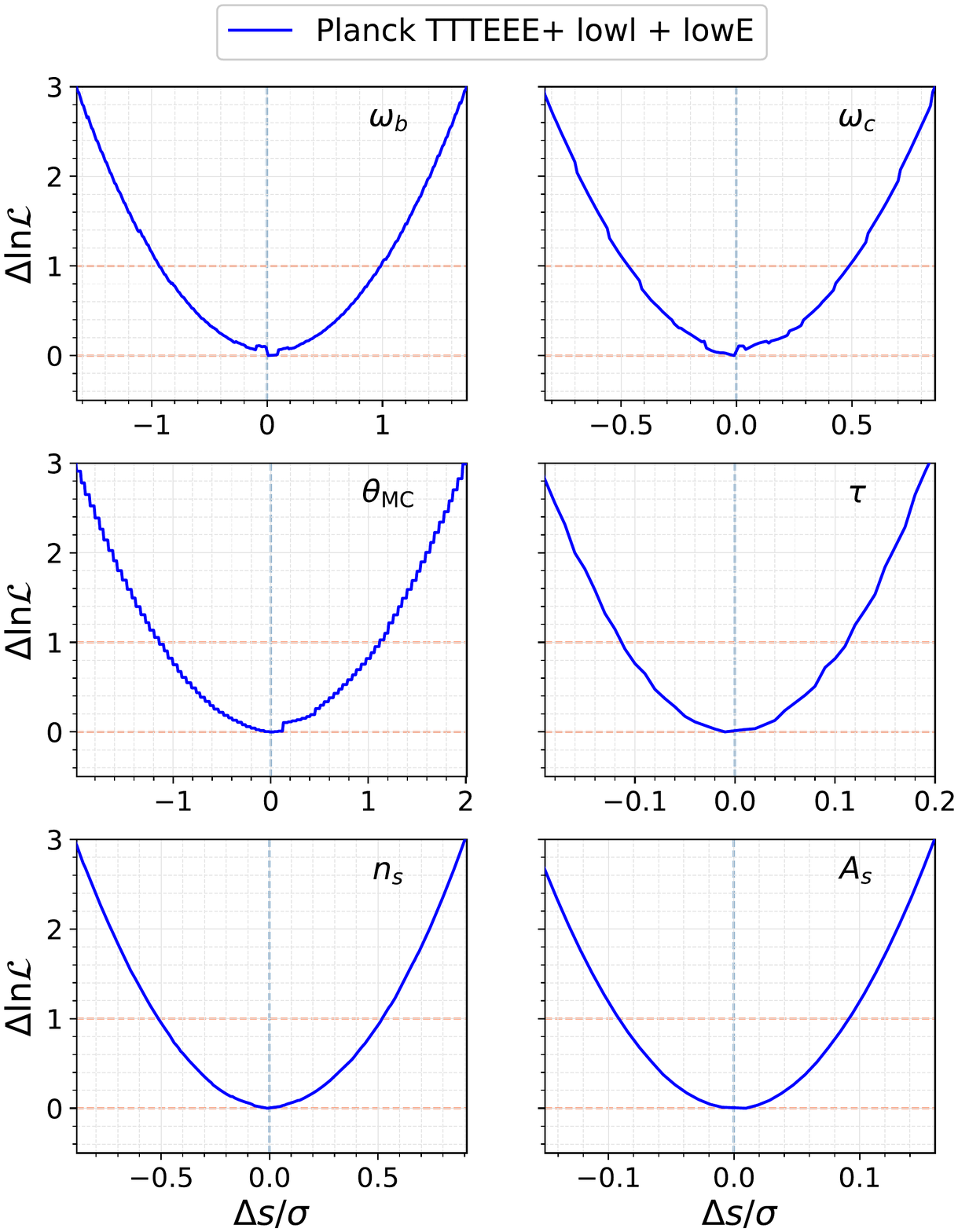}
    \caption{Log-likelihood functions for the 6 typical standard cosmological parameters: $[\omb,\omc,\thetaMC,\tau,\ns,\As]$. Here the steps, $\Delta s/\sigma$ are in units of their respective \planck 2018 marginalised errors and the likelihood used is \planck 2018 TTTEEE + low-$\ell$ + low-E data. This is the recommended likelihood for the standard \planck 2018 analysis \citep{Planck2018over}. Blue dashed lines refer to the zero minima of the likelihood residuals, $\DelL$, whereas the red lines show the position of $\DelL=1$ for each parameter respectively.}
    \label{fig:planck_curves}
\end{figure}
%----------------------------------------------

\subsection{Optimisation of 1D likelihood curves}\label{sec:curves}
%----------------------------------------------
The curves were optimised using the likelihood minimisation routines found in {\tt CosmoMC}. The minimisation mode was run for several starting points to check that the location of the maximum likelihood value (from hereon, MLV) was attained correctly with respect to all the parameters. The chosen likelihood configuration was \planck 2018 TTTEEE + low-$\ell$ + low-E dataset, given that this was the baseline for the 2018 papers \citep{Planck2018over, Planck2018like}.
The likelihood function was then perturbed from the MLV, $L_{\rm m}\,\equiv\, L(\vec{p}_{\rm m})$ and the residual was calculated: $\Delta L_i = L\left(\vec{p}_i\right)-L_{\rm m}$. 
For the purposes of this paper, we use the log-likelihood $L\equiv \ln\mathcal{L}$, where $\mathcal{L}$ is the actual likelihood function and $\vec{p}_{\rm m}$ refers to the fiducial set of parameters that define the location of the MLV. The variations in the likelihood parametrised by a fractional standard deviation (according to \planck 2018 fiducial results), $\Delta s /\sigma$, are shown in Fig.~\ref{fig:planck_curves}. The blue dashed lines show that the likelihood minima have been offset to $\Delta s=0$, whilst the red dashed lines show the $\Delta \ln\mathcal{L} = 1$ limits above the minima. 

The noisy structure of the likelihood around the minima for $\omb$, $\omc$ and $\thetaMC$ has not disappeared from the previous analysis in PCA20, where the \planck 2015 data was used \citep{Planck2015like}. However $\ns$, $\tau$ and $\As$ all look relatively smooth for very small changes in the parameters $\Delta s/\sigma$ (with respect to their standard deviations). Furthermore, the likelihood variations in the Fisher matrix are insensitive to small changes in the nuisance parameters. Both these details are consistent with the conclusions from our previous paper, where even the functional form of the noisiness around $\Delta s \sim 0$ has similar structure. Given the clear parabolic shapes of the log-normal distributions in Fig.~\ref{fig:planck_curves} around $\Delta s = 0$, the minimised likelihood here is an ideal configuration for the Fisher method outlined in Sect.~\ref{sec:howtoPCA}.

\subsection{Stability of step sizes for cosmological parameters}\label{sec:stepsize}
\begin{figure}
    \centering
    \includegraphics[width=\linewidth]{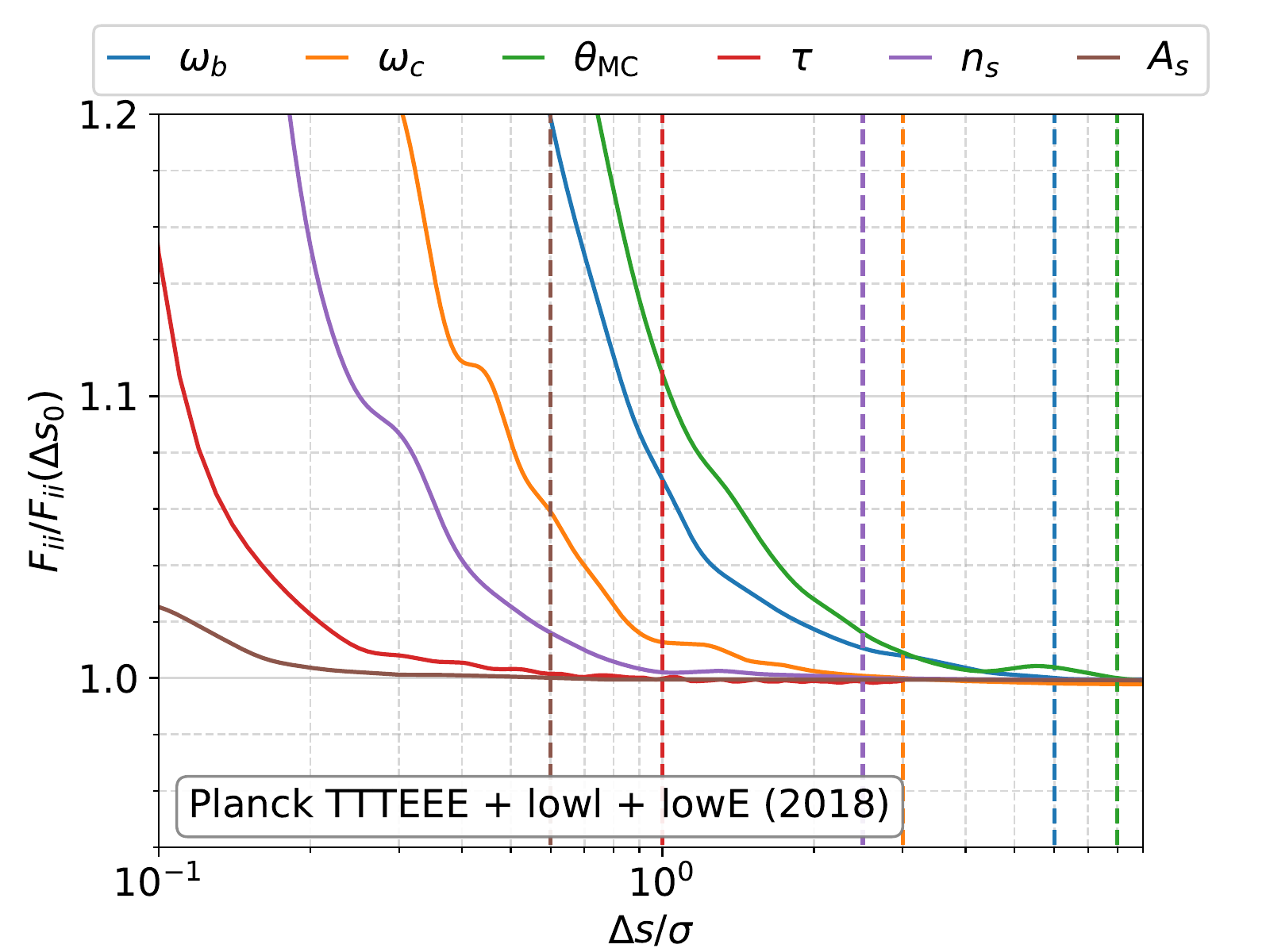}
    \caption{The diagonal Fisher matrix elements for the standard cosmological parameters, $F_{ii}$ referenced in Appendix~\ref{app:planck} and PCA20. These have been weighted by $F_{ii}\left(\Delta s_0\right)$, where $\Delta s_0$ is the location of the dashed lines, which have been chosen for the appropriate step sizes in the direct likelihood method.} 
    \label{fig:planck_fisher}
\end{figure}

%----------------------------------------------
Once the minimum value of the N-D log-likelihood distribution is \emph{approximately} found, one can start to optimise the Fisher for the direct likelihood method. In this case, we repeat the methodology from Appendix~C of the previous paper, by finding the parameter step size that allows for a stable evaluation of the Fisher elements $F_{ij}$. Here we focused on the most sensitive, non-component parameters which were again the standard six cosmological parameters analysed by \emph{Planck}. The results from evaluating the diagonal Fisher elements are shown in Fig.~\ref{fig:planck_fisher}.
In this figure, the lines correspond to the diagonal Fisher element at some step size for a given parameter. These are weighted by the value of the given diagonal Fisher element at the x-value of the dotted lines. For example. for $\As$, the weighted value of $\Delta s_0 = 0.6\sigma$. 
The step-sizes that represent an adequate level of numerical stability are represented by dashed lines for each parameter in Fig.~\ref{fig:planck_fisher}. For $\omb,\omc,\thetaMC$ and $\ns$, these lines are complemented by the curves in Fig.~\ref{fig:planck_curves}. The pivot values shown in Fig.~\ref{fig:planck_fisher} were chosen as they tend to some constant value, implying the derivatives are stable. On the LHS of Fig~\ref{fig:planck_fisher}, the Fisher elements are affected by propagating noise within the Boltzmann code and the likelihood function; whereas on the RHS these responses become non-linear and parabolic (since the Taylor expansion of the likelihood no longer works in this regime). For comparison, see Fig.~C2 of PCA20. In a similar vein to the previous paper, we are using a \emph{five-point stencil} \footnote{Five-point is misleading when you are using the 2D finite difference method. In reality, for two unique parameters, this scheme requires evaluating 16 points (full calculation in  PCA20).} for derivatives for the same reasons as before: the higher order scheme allows for derivatives at step-sizes that do not incur noisy likelihood responses.

The increments used in the direct likelihood method for the modes shown in Sect.~\ref{sec:planckModes} are given in Table~\ref{tab:steps}. For $\omb$, $\omc$ and $\ns$, the step sizes given are roughly the same size as in PCA20, however slightly modified according to the shifts in the best-fit values between the 2015 and 2018 \planck likelihood. The value of $\Delta s$ for $\tau$ is $\sim 70\%$ smaller for the case of the 2018 likelihood; however this is intrinsically linked to the improvements in the polarisation data and the much smaller value of $\tau$ in the \planck 2018 parameters \citep{Planck2018params}. The only anomaly is the shift for $\As$ is much larger; however as shown in Fig.~\ref{fig:planck_fisher}, the amplitude of $\As$ is much more forgiving for the Fisher method. 
The chosen step-size is well outside the ranges of these curves where small-scale noise dominates the likelihood. However, one can notice that the step-sizes for $\tau$ and $\As$ are noticeably larger than their likelihood curves would imply. This is due to the huge degeneracy arising from these parameters in the analysis of the CMB power spectra. The amplitude of the matter power spectrum $\As$ contributes to an overall change in the normalisation of the CMB anisotropies, similar to the net damping effect caused by an increase in the reionisation optical depth $\tau$. As such, their stability is far weaker than their 1D curves would suggest. There are similar correlations across the standard 6 parameters, but the degeneracy between $\tau$ and $\As$ is by far the largest \citep[see][for more details]{Planck2018params}.

%----------------------------------------------
\begin{table}
     \centering
     \begin{tabular*}{\linewidth}{@{\extracolsep{\fill}} l c c c }
        \hline\hline
        Parameter ($p$) & $\bar{\mu}_p$ & $\sigma_p$ & $\Delta s_p/\sigma_p$ \\
        \hline
        $\omb$ & 0.02237 & 0.00015 & 6.0 \\ 
        $\omc$ & 0.1201 & 0.0014 & $3.0$ \\ 
        $\thetaxMC$ & 1.04085 & 0.00031& $8.0$ \\
        $\tau$ & 0.0533 & 0.0074 & $1.0$\\
        $\ns$ & 0.9658 & 0.0044 & $2.5$\\
        $\logA$ & 3.043 & 0.016 & $0.6$\\
        \hline
     \end{tabular*}
     \caption{The standard 6 cosmological parameters used in the direct method analysis with \planck 2018 TTTEEE + low-$\ell$ + low-E data along with their best-fit values ($\bar{\mu}_p$), their standard deviation ($\sigma_p$) and the choice of $\Delta s/\sigma$ for the calculation of stable derivatives (shown in Fig.~\ref{fig:planck_fisher}). The best-fit values come from iterated minimisation and the standard deviations come from MCMC, both obtained using the {\tt CosmoMC} software package.}
     \label{tab:steps}
 \end{table}
 %----------------------------------------------

%----------------------------------------------
%----------------------------------------------

\section{Revisiting free electron fraction PCA with 2018 data}\label{app:xe18}
%----------------------------------------------

\begin{figure}
    \centering
    \includegraphics[width=\linewidth]{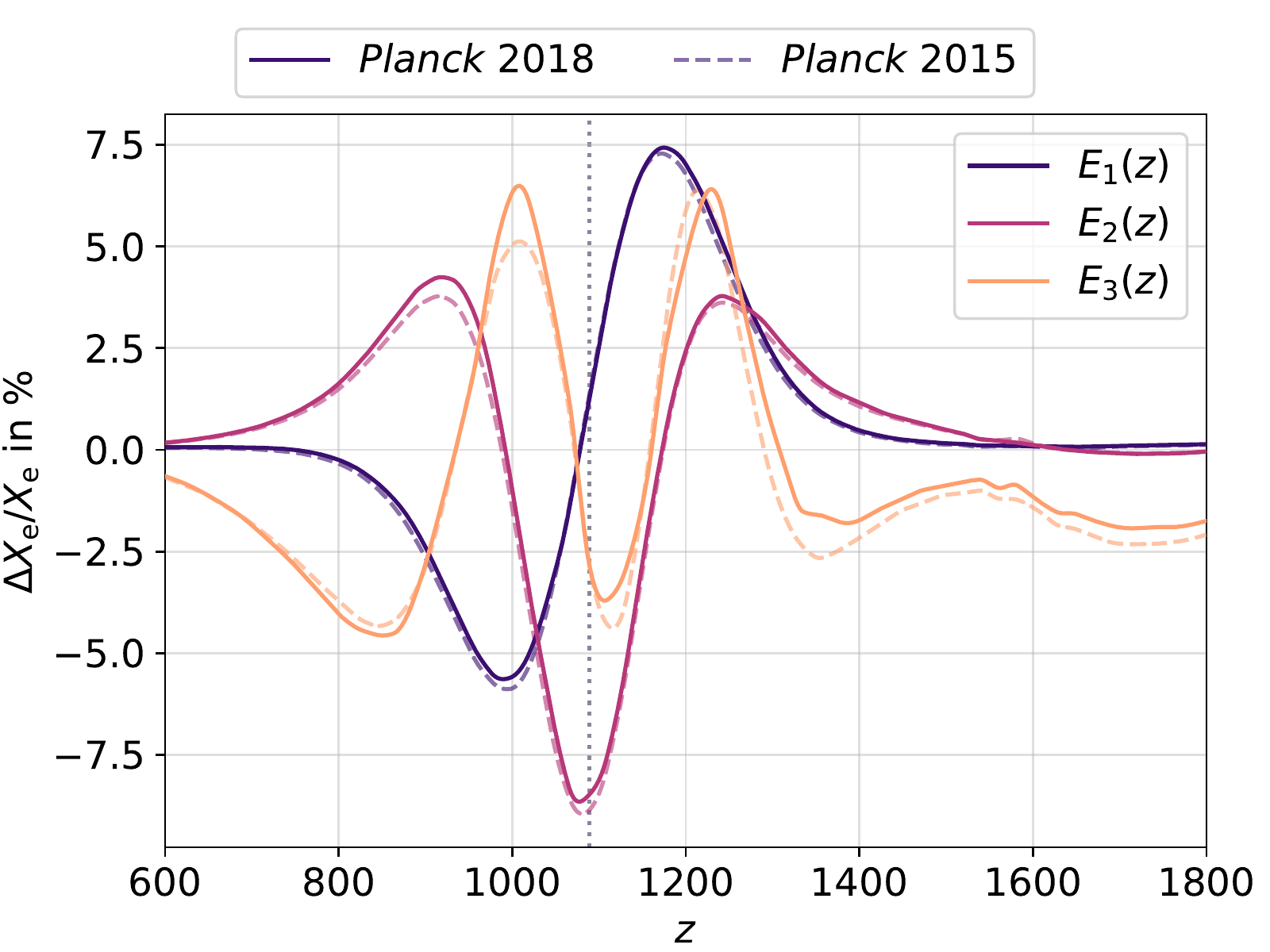}
    \caption{Eigenmodes from the $\xe$ PCA with the \planck 2018 likelihood. The reference lines on the plot (\emph{dashed}) are the 2015 converged eigenmodes found in PCA20. The \planck 2018  $\xe$ results use the same stability analysis explained in Appendix~\ref{app:planck}. The dotted line represents the most probable last scattering redshift, $z_*$ under \LCDM.}
    \label{fig:xePlanck}
\end{figure}

\begin{figure}
    \centering
    \includegraphics[width=\linewidth]{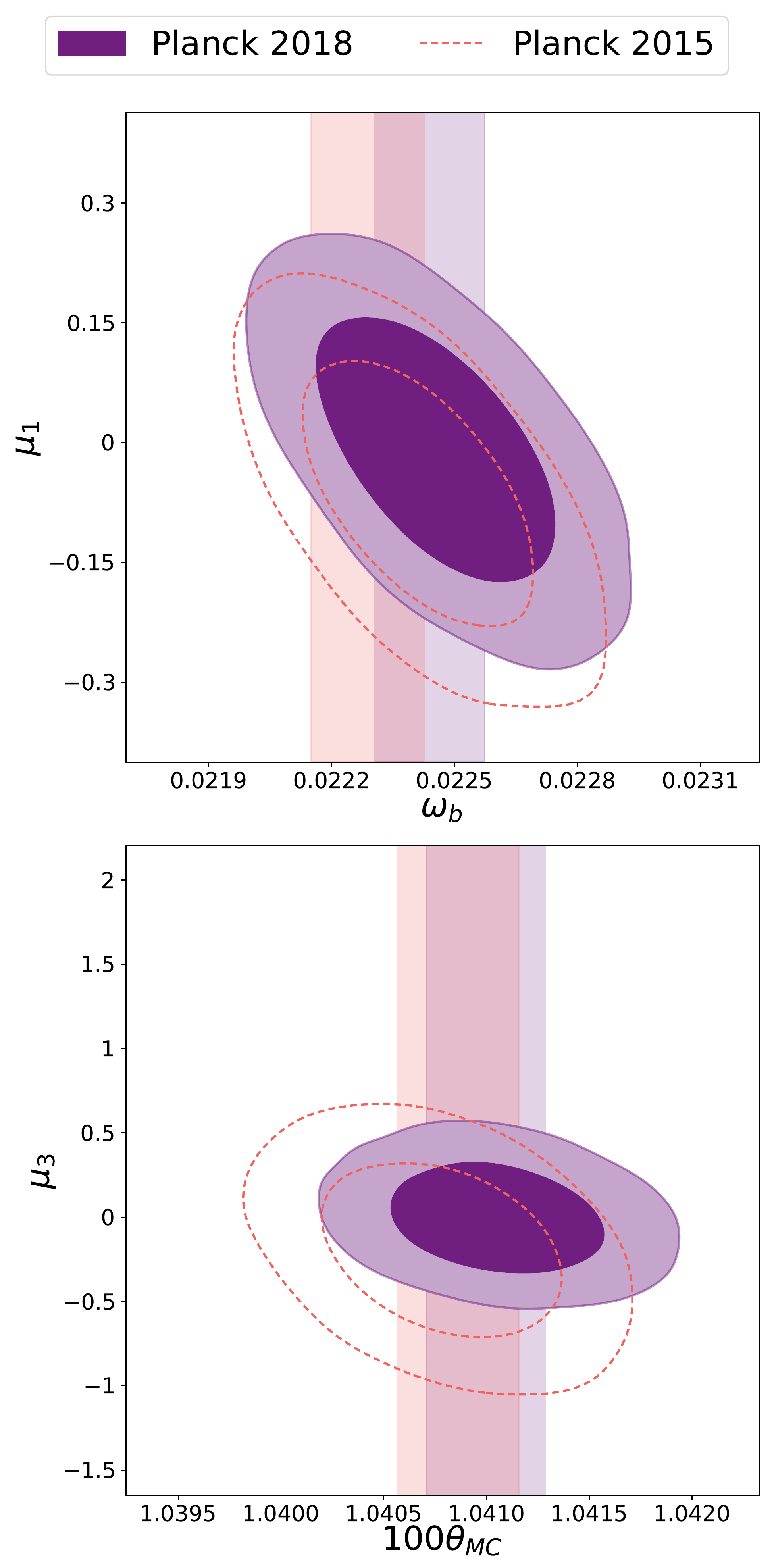}
    \caption{Posterior contours for the free electron fraction $\xe$ for the largest correlations from PCA20: $\mu_1$ vs. $\omb$ (\emph{top}) and $\mu_3$ vs. $\thetaMC$ (\emph{bottom}). The two contours shown are the \planck 2015 modes (\emph{orange, dashed}) and the \planck 2018 modes (\emph{purple, solid}) with the fiducial \LCDM bands included as well. Here we have included CMB lensing as well as BAO data from each respective data release.}
    \label{fig:xeShorts}
\end{figure}

In this appendix, we show the results for the $\xe$ eigenmodes, similar to the method in PCA20, as a consistency check for the PCA methodology. The ionisation history eigenmodes generated with the most recent \planck data are shown in Fig.~\ref{fig:xePlanck}. With the exception of the small changes in peak height at $z\lsim 1200$ for all modes, the shapes across all 3 eigenmodes are congruous with the 2015 eigenmodes. Given that these were produced via the same optimisation and stability protocols as defined in Appendix~\ref{app:planck}, we are confident that the PCA methodology is robust for the production of fundamental constant eigenmodes using the likelihood as we have in our previous work. The stability is also shown in the marginalised results in Table~\ref{tab:xeTable}. Here we have applied the marginalised 2015 and 2018 modes to \planck likelihood data (2015 and 2018 respectively) with CMB lensing and BAO data included also. This gave us another check against the previously held \planck 2018 constraints for the recombination eigenmodes. Compared to the \planck 2018 paper, the errors are agreeable with a $\sim70\%$ decrease in the error on $\mu_3$ and a distinct lack of residual degeneracies across the cosmological parameters. As shown in PCA20, the data combination has little impact on the final MCMC values from the $\xe$ eigenmodes save a few small fluctuations due to the shifts in the best fit values when lensing and galaxy clustering data in the analysis \citep[see][for BAO results]{SDSSDR12}. These conclusions are shown clearly in the posterior contours in Fig.~\ref{fig:xeShorts}. The 2018 (\emph{solid}) contour is slightly compressed in the $\mu_3$ dimension compared to the previous work (\emph{dashed}). There is also a drift in the likelihood contours due to the baseline changes in the \planck 2018 parameters.  

%----------------------------------------------
%----------------------------------------------

%----------------------------------------------
\begin{table}
    \centering
    \begin{tabular*}{\linewidth}{@{\extracolsep{\fill} } l c c}
    \hline\hline
        Parameter & \emph{Planck} 2015  & \emph{Planck} 2018\\
        & + lensing + BAO & + lensing + BAO \\
        \hline
        $\omega_b  $ &  $0.02241\pm 0.00019  $ &  $0.02246\pm 0.00019  $\\
        $\omega_c  $ &  $0.1183\pm 0.0011  $ &  $0.1189^{+0.0011}_{-0.00096}  $\\
        $100\theta_{MC}  $ &  $1.04079\pm 0.00038  $ &  $1.04104\pm 0.00036  $\\
        $\tau  $ &  $0.070\pm 0.013  $ &  $0.0572\pm 0.0075  $\\
        ${\rm{ln}}(10^{10} A_s)  $ &  $3.071\pm 0.024  $ &  $3.049\pm 0.015  $\\
        $n_s  $ &  $0.9686\pm 0.0055  $ &  $0.9681\pm 0.0053  $\\
        \hline
        $\mu_1 \;\left(\xe\right)  $ &  $-0.06\pm 0.11  $ &  $-0.01\pm 0.11  $\\
        $\mu_2 \;\left(\xe\right)  $ &  $-0.16\pm 0.19  $ &  $0.06\pm 0.18  $\\
        $\mu_3 \;\left(\xe\right)  $ &  $-0.19\pm 0.35  $ &  $0.02^{+0.19}_{-0.24}  $\\[.5mm]
        \hline
        $H_0  $ &  $67.93\pm 0.49  $ &  $67.85^{+0.46}_{-0.51}  $\\
        $\sigma_8  $ &  $0.8174\pm 0.0091  $ &  $0.8100\pm 0.0065  $\\
        \hline\hline
    \end{tabular*}
    \caption{Marginalised $68\%$ limit results for the $\xe$ eigenmodes generated with \planck 2018 compared against the results generated with the \planck 2015 likelihood. The MCMC has been carried out with \planck + lensing + BAO to easily compare against the previous \planck papers \citep{Planck2018params}.}
    \label{tab:xeTable}
\end{table}
%----------------------------------------------

%----------------------------------------------
%----------------------------------------------

\small 
\bibliographystyle{mn2e}
\bibliography{Lit}

\begin{thebibliography}{92}
\expandafter\ifx\csname natexlab\endcsname\relax\def\natexlab#1{#1}\fi

\bibitem[{{Abazajian} {et~al}\mbox{.}(2016){Abazajian}, {Adshead}, {Ahmed},
  {Allen}, {Alonso}, {Arnold}, {Baccigalupi}, {Bartlett}, {Battaglia},
  {Benson}, {Bischoff}, {Borrill}, {Buza}, {Calabrese}, {Caldwell},
  {Carlstrom}, {Chang}, {Crawford}, {Cyr-Racine}, {De Bernardis}, {de Haan},
  {di Serego Alighieri}, {Dunkley}, {Dvorkin}, {Errard}, {Fabbian}, {Feeney},
  {Ferraro}, {Filippini}, {Flauger}, {Fuller}, {Gluscevic}, {Green}, {Grin},
  {Grohs}, {Henning}, {Hill}, {Hlozek}, {Holder}, {Holzapfel}, {Hu},
  {Huffenberger}, {Keskitalo}, {Knox}, {Kosowsky}, {Kovac}, {Kovetz}, {Kuo},
  {Kusaka}, {Le Jeune}, {Lee}, {Lilley}, {Loverde}, {Madhavacheril}, {Mantz},
  {Marsh}, {McMahon}, {Meerburg}, {Meyers}, {Miller}, {Munoz}, {Nguyen},
  {Niemack}, {Peloso}, {Peloton}, {Pogosian}, {Pryke}, {Raveri}, {Reichardt},
  {Rocha}, {Rotti}, {Schaan}, {Schmittfull}, {Scott}, {Sehgal}, {Shandera},
  {Sherwin}, {Smith}, {Sorbo}, {Starkman}, {Story}, {van Engelen}, {Vieira},
  {Watson}, {Whitehorn}, \& {Kimmy Wu}}]{CMBS42016}
{Abazajian} K.~N. {et~al.}, 2016, ArXiv:1610.0274

\bibitem[{{Abazajian} {et~al}\mbox{.}(2015){Abazajian}, {Arnold}, {Austermann},
  {Benson}, {Bischoff}, {Bock}, {Bond}, {Borrill}, {Calabrese}, {Carlstrom},
  {Carvalho}, {Chang}, {Chiang}, {Church}, {Cooray}, {Crawford}, {Dawson},
  {Das}, {Devlin}, {Dobbs}, {Dodelson}, {Dor{\'e}}, {Dunkley}, {Errard},
  {Fraisse}, {Gallicchio}, {Halverson}, {Hanany}, {Hildebrandt}, {Hincks},
  {Hlozek}, {Holder}, {Holzapfel}, {Honscheid}, {Hu}, {Hubmayr}, {Irwin},
  {Jones}, {Kamionkowski}, {Keating}, {Keisler}, {Knox}, {Komatsu}, {Kovac},
  {Kuo}, {Lawrence}, {Lee}, {Leitch}, {Linder}, {Lubin}, {McMahon}, {Miller},
  {Newburgh}, {Niemack}, {Nguyen}, {Nguyen}, {Page}, {Pryke}, {Reichardt},
  {Ruhl}, {Sehgal}, {Seljak}, {Sievers}, {Silverstein}, {Slosar}, {Smith},
  {Spergel}, {Staggs}, {Stark}, {Stompor}, {Vieregg}, {Wang}, {Watson},
  {Wollack}, {Wu}, {Yoon}, \& {Zahn}}]{Abazajian2015}
{Abazajian} K.~N. {et~al.}, 2015, Astroparticle Physics, 63, 66

\bibitem[{{Addison}(2021)}]{Addison2021}
{Addison} G.~E., 2021, \apjl, 912, L1

\bibitem[{{Ade} {et~al}\mbox{.}(2019){Ade}, {Aguirre}, {Ahmed}, {Aiola}, {Ali},
  {Alonso}, {Alvarez}, {Arnold}, {Ashton}, {Austermann}, {Awan}, {Baccigalupi},
  {Baildon}, {Barron}, {Battaglia}, {Battye}, {Baxter}, {Bazarko}, {Beall},
  {Bean}, {Beck}, {Beckman}, {Beringue}, {Bianchini}, {Boada}, {Boettger},
  {Bond}, {Borrill}, {Brown}, {Bruno}, {Bryan}, {Calabrese}, {Calafut},
  {Calisse}, {Carron}, {Challinor}, {Chesmore}, {Chinone}, {Chluba}, {Cho},
  {Choi}, {Coppi}, {Cothard}, {Coughlin}, {Crichton}, {Crowley}, {Crowley},
  {Cukierman}, {D'Ewart}, {D{\"u}nner}, {de Haan}, {Devlin}, {Dicker},
  {Didier}, {Dobbs}, {Dober}, {Duell}, {Duff}, {Duivenvoorden}, {Dunkley},
  {Dusatko}, {Errard}, {Fabbian}, {Feeney}, {Ferraro}, {Flux{\`a}}, {Freese},
  {Frisch}, {Frolov}, {Fuller}, {Fuzia}, {Galitzki}, {Gallardo}, {Tomas Galvez
  Ghersi}, {Gao}, {Gawiser}, {Gerbino}, {Gluscevic}, {Goeckner-Wald}, {Golec},
  {Gordon}, {Gralla}, {Green}, {Grigorian}, {Groh}, {Groppi}, {Guan},
  {Gudmundsson}, {Han}, {Hargrave}, {Hasegawa}, {Hasselfield}, {Hattori},
  {Haynes}, {Hazumi}, {He}, {Healy}, {Henderson}, {Hervias-Caimapo}, {Hill},
  {Hill}, {Hilton}, {Hilton}, {Hincks}, {Hinshaw}, {Hlo{\v{z}}ek}, {Ho}, {Ho},
  {Howe}, {Huang}, {Hubmayr}, {Huffenberger}, {Hughes}, {Ijjas}, {Ikape},
  {Irwin}, {Jaffe}, {Jain}, {Jeong}, {Kaneko}, {Karpel}, {Katayama}, {Keating},
  {Kernasovskiy}, {Keskitalo}, {Kisner}, {Kiuchi}, {Klein}, {Knowles},
  {Koopman}, {Kosowsky}, {Krachmalnicoff}, {Kuenstner}, {Kuo}, {Kusaka},
  {Lashner}, {Lee}, {Lee}, {Leon}, {Leung}, {Lewis}, {Li}, {Li}, {Limon},
  {Linder}, {Lopez-Caraballo}, {Louis}, {Lowry}, {Lungu}, {Madhavacheril},
  {Mak}, {Maldonado}, {Mani}, {Mates}, {Matsuda}, {Maurin}, {Mauskopf}, {May},
  {McCallum}, {McKenney}, {McMahon}, {Meerburg}, {Meyers}, {Miller},
  {Mirmelstein}, {Moodley}, {Munchmeyer}, {Munson}, {Naess}, {Nati},
  {Navaroli}, {Newburgh}, {Nguyen}, {Niemack}, {Nishino}, {Orlowski-Scherer},
  {Page}, {Partridge}, {Peloton}, {Perrotta}, {Piccirillo}, {Pisano},
  {Poletti}, {Puddu}, {Puglisi}, {Raum}, {Reichardt}, {Remazeilles},
  {Rephaeli}, {Riechers}, {Rojas}, {Roy}, {Sadeh}, {Sakurai}, {Salatino},
  {Sathyanarayana Rao}, {Schaan}, {Schmittfull}, {Sehgal}, {Seibert}, {Seljak},
  {Sherwin}, {Shimon}, {Sierra}, {Sievers}, {Sikhosana}, {Silva-Feaver},
  {Simon}, {Sinclair}, {Siritanasak}, {Smith}, {Smith}, {Spergel}, {Staggs},
  {Stein}, {Stevens}, {Stompor}, {Suzuki}, {Tajima}, {Takakura}, {Teply},
  {Thomas}, {Thorne}, {Thornton}, {Trac}, {Tsai}, {Tucker}, {Ullom},
  {Vagnozzi}, {van Engelen}, {Van Lanen}, {Van Winkle}, {Vavagiakis},
  {Verg{\`e}s}, {Vissers}, {Wagoner}, {Walker}, {Ward}, {Westbrook},
  {Whitehorn}, {Williams}, {Williams}, {Wollack}, {Xu}, {Yu}, {Yu}, {Zago},
  {Zhang}, {Zhu}, \& {Simons Observatory Collaboration}}]{SOWP2018}
{Ade} P. {et~al.}, 2019, \jcap, 2019, 056

\bibitem[{{Ade} {et~al}\mbox{.}(2014){Ade}, {Akiba}, {Anthony}, {Arnold},
  {Atlas}, {Barron}, {Boettger}, {Borrill}, {Chapman}, {Chinone}, {Dobbs},
  {Elleflot}, {Errard}, {Fabbian}, {Feng}, {Flanigan}, {Gilbert}, {Grainger},
  {Halverson}, {Hasegawa}, {Hattori}, {Hazumi}, {Holzapfel}, {Hori}, {Howard},
  {Hyland}, {Inoue}, {Jaehnig}, {Jaffe}, {Keating}, {Kermish}, {Keskitalo},
  {Kisner}, {Le Jeune}, {Lee}, {Linder}, {Leitch}, {Lungu}, {Matsuda},
  {Matsumura}, {Meng}, {Miller}, {Morii}, {Moyerman}, {Myers}, {Navaroli},
  {Nishino}, {Paar}, {Peloton}, {Quealy}, {Rebeiz}, {Reichardt}, {Richards},
  {Ross}, {Schanning}, {Schenck}, {Sherwin}, {Shimizu}, {Shimmin}, {Shimon},
  {Siritanasak}, {Smecher}, {Spieler}, {Stebor}, {Steinbach}, {Stompor},
  {Suzuki}, {Takakura}, {Tomaru}, {Wilson}, {Yadav}, {Zahn}, \& {Polarbear
  Collaboration}}]{polarbear_results}
{Ade} P.~A.~R. {et~al.}, 2014, Physical Review Letters, 113, 021301

\bibitem[{Alam {et~al}\mbox{.}(2015)Alam, Albareti, Prieto, Anders, Anderson,
  Anderton, Andrews, Armengaud, Aubourg, Bailey, \& et~al.}]{SDSSDR12}
Alam S. {et~al.}, 2015, The Astrophysical Journal Supplement Series, 219, 12

\bibitem[{{Avelino} {et~al}\mbox{.}(2001){Avelino}, {Esposito}, {Mangano},
  {Martins}, {Melchiorri}, {Miele}, {Pisanti}, {Rocha}, \&
  {Viana}}]{Avelino2001}
{Avelino} P.~P. {et~al.}, 2001, \prd, 64, 103505

\bibitem[{{Barrow} \& {Graham}(2013)}]{Barrow2013}
{Barrow} J.~D., {Graham} A.~A.~H., 2013, \prd, 88, 103513

\bibitem[{{Battye} {et~al}\mbox{.}(2001){Battye}, {Crittenden}, \&
  {Weller}}]{Battye2001}
{Battye} R.~A., {Crittenden} R., {Weller} J., 2001, \prd, 63, 043505

\bibitem[{{Battye} \& {Moss}(2014)}]{BattyeNeutrinos}
{Battye} R.~A., {Moss} A., 2014, Physical Review Letters, 112, 051303

\bibitem[{Bekenstein(1982)}]{Bekenstein1982}
Bekenstein J.~D., 1982, Phys. Rev. D, 25, 1527

\bibitem[{{Bennett} {et~al}\mbox{.}(1996){Bennett}, {Banday}, {Gorski},
  {Hinshaw}, {Jackson}, {Keegstra}, {Kogut}, {Smoot}, {Wilkinson}, \&
  {Wright}}]{COBE4yr}
{Bennett} C.~L. {et~al.}, 1996, \apjl, 464, L1

\bibitem[{{Bennett} {et~al}\mbox{.}(2013){Bennett}, {Larson}, {Weiland},
  {Jarosik}, {Hinshaw}, {Odegard}, {Smith}, {Hill}, {Gold}, {Halpern},
  {Komatsu}, {Nolta}, {Page}, {Spergel}, {Wollack}, {Dunkley}, {Kogut},
  {Limon}, {Meyer}, {Tucker}, \& {Wright}}]{wmap9results}
{Bennett} C.~L. {et~al.}, 2013, \apjs, 208, 20

\bibitem[{{Bolliet} {et~al}\mbox{.}(2020){Bolliet}, {Chluba}, \&
  {Battye}}]{Bolliet2020}
{Bolliet} B., {Chluba} J., {Battye} R., 2020, arXiv e-prints, arXiv:2012.07292

\bibitem[{{Bonifacio} {et~al}\mbox{.}(2014){Bonifacio}, {Rahmani}, {Whitmore},
  {Wendt}, {Centurion}, {Molaro}, {Srianand}, {Murphy}, {Petitjean},
  {Agafonova}, {D'Odorico}, {Evans}, {Levshakov}, {Lopez}, {Martins},
  {Reimers}, \& {Vladilo}}]{Bonifacio2014}
{Bonifacio} P. {et~al.}, 2014, Astronomische Nachrichten, 335, 83

\bibitem[{{Campeti} {et~al}\mbox{.}(2019){Campeti}, {Poletti}, \&
  {Baccigalupi}}]{Campeti2019}
{Campeti} P., {Poletti} D., {Baccigalupi} C., 2019, arXiv e-prints,
  arXiv:1905.08200

\bibitem[{{Carlstrom} {et~al}\mbox{.}(2019){Carlstrom}, {Abazajian}, {Addison},
  {Adshead}, {Ahmed}, {Allen}, {Alonso}, {Alvarez}, {Anderson}, {Arnold},
  {Baccigalupi}, {Bailey}, {Barkats}, {Barron}, {Barry}, {Bartlett}, {Basu
  Thakur}, {Battaglia}, {Baxter}, {Bean}, {Bebek}, {Bender}, {Benson},
  {Berger}, {Bhimani}, {Bischoff}, {Bleem}, {Bocquet}, {Boddy}, {Bonato},
  {Bond}, {Borrill}, {Bouchet}, {Brown}, {Bryan}, {Burkhart}, {Buza}, {Byrum},
  {Calabrese}, {Calafut}, {Caldwell}, {Carlstrom}, {Carron}, {Cecil},
  {Challinor}, {Chang}, {Chinone}, {Cho}, {Cooray}, {Crawford}, {Crites},
  {Cukierman}, {Cyr-Racine}, {de Haan}, {de Zotti}, {Delabrouille},
  {Demarteau}, {Devlin}, {Di Valentino}, {Dobbs}, {Duff}, {Duivenvoorden},
  {Dvorkin}, {Edwards}, {Eimer}, {Errard}, {Essinger-Hileman}, {Fabbian},
  {Feng}, {Ferraro}, {Filippini}, {Flauger}, {Flaugher}, {Fraisse}, {Frolov},
  {Galitzki}, {Galli}, {Ganga}, {Gerbino}, {Gilchriese}, {Gluscevic}, {Green},
  {Grin}, {Grohs}, {Gualtieri}, {Guarino}, {Gudmundsson}, {Habib}, {Haller},
  {Halpern}, {Halverson}, {Hanany}, {Harrington}, {Hasegawa}, {Hasselfield},
  {Hazumi}, {Heitmann}, {Henderson}, {Henning}, {Hill}, {Hlo{\v{z}}ek},
  {Holder}, {Holzapfel}, {Hubmayr}, {Huffenberger}, {Huffer}, {Hui}, {Irwin},
  {Johnson}, {Johnstone}, {Jones}, {Karkare}, {Katayama}, {Kerby}, {Kernovsky},
  {Keskitalo}, {Kisner}, {Knox}, {Kosowsky}, {Kovac}, {Kovetz}, {Kuhlmann},
  {Kuo}, {Kurita}, {Kusaka}, {Lahteenmaki}, {Lawrence}, {Lee}, {Lewis}, {Li},
  {Linder}, {Loverde}, {Lowitz}, {Madhavacheril}, {Mantz}, {Matsuda},
  {Mauskopf}, {McMahon}, {Meerburg}, {Melin}, {Meyers}, {Millea}, {Mohr},
  {Moncelsi}, {Mroczkowski}, {Mukherjee}, {Munchmeyer}, {Nagai}, {Nagy},
  {Namikawa}, {Nati}, {Natoli}, {Negrello}, {Newburgh}, {Niemack}, {Nishino},
  {Nordby}, {Novosad}, {O'Connor}, {Obied}, {Padin}, {Pandey}, {Partridge},
  {Pierpaoli}, {Pogosian}, {Pryke}, {Puglisi}, {Racine}, {Raghunathan},
  {Rahlin}, {Rajagopalan}, {Raveri}, {Reichanadter}, {Reichardt},
  {Remazeilles}, {Rocha}, {Roe}, {Roy}, {Ruhl}, {Salatino}, {Saliwanchik},
  {Schaan}, {Schillaci}, {Schmittfull}, {Scott}, {Sehgal}, {Shandera},
  {Sheehy}, {Sherwin}, {Shirokoff}, {Simon}, {Slosar}, {Somerville}, {Staggs},
  {Stark}, {Stompor}, {Story}, {Stoughton}, {Suzuki}, {Tajima}, {Teply},
  {Thompson}, {Timbie}, {Tomasi}, {Treu}, {Tristram}, {Tucker}, {Umilta}, {van
  Engelen}, {Vieira}, {Vieregg}, {Vogelsberger}, {Wang}, {Watson}, {White},
  {Whitehorn}, {Wollack}, {Wu}, {Xu}, {Yasini}, {Yeck}, {Yoon}, {Young}, \&
  {Zonca}}]{CMBS4WP}
{Carlstrom} J. {et~al.}, 2019, in Bulletin of the American Astronomical
  Society, Vol.~51, p. 209

\bibitem[{Chen \& Wang(2021)}]{Chen2021}
Chen L., Wang K., 2021, Does the reionization model influence the constraints
  on dark matter decay or annihilation?

\bibitem[{{Chen} \& {Kamionkowski}(2004)}]{Chen2004}
{Chen} X., {Kamionkowski} M., 2004, \prd, 70, 043502

\bibitem[{{Chluba}(2010)}]{Chluba2010a}
{Chluba} J., 2010, \mnras, 402, 1195

\bibitem[{{Chluba} {et~al}\mbox{.}(2015){Chluba}, {Paoletti}, {Finelli}, \&
  {Rubino-Martin}}]{Chluba2015PMF}
{Chluba} J., {Paoletti} D., {Finelli} F., {Rubino-Martin} J.-A., 2015,
  ArXiv:1503.04827

\bibitem[{{Chluba} \& {Thomas}(2011)}]{Chluba2010b}
{Chluba} J., {Thomas} R.~M., 2011, \mnras, 412, 748

\bibitem[{Dai {et~al}\mbox{.}(2018)Dai, Ma, Guo, \& Cai}]{Dai2018}
Dai W.-M., Ma Y.-Z., Guo Z.-K., Cai R.-G., 2018, Phys. Rev., astro-ph.CO

\bibitem[{{Di Valentino} {et~al}\mbox{.}(2016){Di Valentino}, {Brinckmann},
  {Gerbino}, {Poulin}, {Bouchet}, {Lesgourgues}, {Melchiorri}, {Chluba},
  {Clesse}, {Delabrouille}, {Dvorkin}, {Forastieri}, {Galli}, {Hooper},
  {Lattanzi}, {Martins}, {Salvati}, {Cabass}, {Caputo}, {Giusarma}, {Hivon},
  {Natoli}, {Pagano}, {Paradiso}, {Rubino-Martin}, {Achucarro}, {Ade},
  {Allison}, {Arroja}, {Ashdown}, {Ballardini}, {Banday}, {Banerji}, {Bartolo},
  {Bartlett}, {Basak}, {Baselmans}, {Baumann}, {de Bernardis}, {Bersanelli},
  {Bonaldi}, {Bonato}, {Borrill}, {Boulanger}, {Bucher}, {Burigana},
  {Buzzelli}, {Cai}, {Calvo}, {Carvalho}, {Castellano}, {Challinor}, {Charles},
  {Colantoni}, {Coppolecchia}, {Crook}, {D'Alessandro}, {De Petris}, {De
  Zotti}, {Diego}, {Errard}, {Feeney}, {Fernandez-Cobos}, {Ferraro}, {Finelli},
  {de Gasperis}, {G{\'e}nova-Santos}, {Gonz{\'a}lez-Nuevo}, {Grandis},
  {Greenslade}, {Hagstotz}, {Hanany}, {Handley}, {Hazra},
  {Hern{\'a}ndez-Monteagudo}, {Hervias-Caimapo}, {Hills}, {Kiiveri}, {Kisner},
  {Kitching}, {Kunz}, {Kurki-Suonio}, {Lamagna}, {Lasenby}, {Lewis}, {Liguori},
  {Lindholm}, {Lopez-Caniego}, {Luzzi}, {Maffei}, {Martin},
  {Martinez-Gonzalez}, {Masi}, {McCarthy}, {Melin}, {Mohr}, {Molinari},
  {Monfardini}, {Negrello}, {Notari}, {Paiella}, {Paoletti}, {Patanchon},
  {Piacentini}, {Piat}, {Pisano}, {Polastri}, {Polenta}, {Pollo}, {Quartin},
  {Remazeilles}, {Roman}, {Ringeval}, {Tartari}, {Tomasi}, {Tramonte},
  {Trappe}, {Trombetti}, {Tucker}, {V{\"a}liviita}, {van de Weygaert}, {Van
  Tent}, {Vennin}, {Vermeulen}, {Vielva}, {Vittorio}, {Young}, {Zannoni}, \&
  {for the CORE collaboration}}]{CORE2016}
{Di Valentino} E. {et~al.}, 2016, ArXiv:1612.00021

\bibitem[{{Di Valentino} {et~al}\mbox{.}(2017){Di Valentino}, {Melchiorri}, \&
  {Mena}}]{DiValentino2017}
{Di Valentino} E., {Melchiorri} A., {Mena} O., 2017, \prd, 96, 043503

\bibitem[{Di~Valentino {et~al}\mbox{.}(2019)Di~Valentino, Melchiorri, \&
  Silk}]{DiValentino2019}
Di~Valentino E., Melchiorri A., Silk J., 2019, Nature Astronomy

\bibitem[{{Farhang} {et~al}\mbox{.}(2012){Farhang}, {Bond}, \&
  {Chluba}}]{Farhang2011}
{Farhang} M., {Bond} J.~R., {Chluba} J., 2012, \apj, 752, 88

\bibitem[{{Farhang} {et~al}\mbox{.}(2013){Farhang}, {Bond}, {Chluba}, \&
  {Switzer}}]{Farhang2013}
{Farhang} M., {Bond} J.~R., {Chluba} J., {Switzer} E.~R., 2013, \apj, 764, 137

\bibitem[{{Finkbeiner} {et~al}\mbox{.}(2012){Finkbeiner}, {Galli}, {Lin}, \&
  {Slatyer}}]{Finkbeiner2012}
{Finkbeiner} D.~P., {Galli} S., {Lin} T., {Slatyer} T.~R., 2012, \prd, 85,
  043522

\bibitem[{{Galli} {et~al}\mbox{.}(2009){Galli}, {Iocco}, {Bertone}, \&
  {Melchiorri}}]{Galli2009}
{Galli} S., {Iocco} F., {Bertone} G., {Melchiorri} A., 2009, \prd, 80, 023505

\bibitem[{{Gratton} {et~al}\mbox{.}(2008){Gratton}, {Lewis}, \&
  {Efstathiou}}]{PlanckNeutrino}
{Gratton} S., {Lewis} A., {Efstathiou} G., 2008, \prd, 77, 083507

\bibitem[{Guennebaud {et~al}\mbox{.}(2010)Guennebaud, Jacob,
  {et~al.}}]{EigenCode}
Guennebaud G., Jacob B., {et~al.}, 2010, Eigen v3. http://eigen.tuxfamily.org

\bibitem[{Hart \& Chluba(2018)}]{Hart2017}
Hart L., Chluba J., 2018, MNRAS, 474, 1850

\bibitem[{Hart \& Chluba(2020{\natexlab{a}})}]{Hart2020b}
Hart L., Chluba J., 2020{\natexlab{a}}, \mnras, 495, 4210

\bibitem[{Hart \& Chluba(2020{\natexlab{b}})}]{Hart2020a}
Hart L., Chluba J., 2020{\natexlab{b}}, \mnras, 493, 3255

\bibitem[{Hees {et~al}\mbox{.}(2020)Hees, Do, Roberts, Ghez, Nishiyama,
  Bentley, Gautam, Jia, Kara, Lu, \& et~al.}]{Hees2020VFC}
Hees A. {et~al.}, 2020, Physical Review Letters, 124

\bibitem[{{Henderson} {et~al}\mbox{.}(2016){Henderson}, {Allison},
  {Austermann}, {Baildon}, {Battaglia}, {Beall}, {Becker}, {De Bernardis},
  {Bond}, {Calabrese}, {Choi}, {Coughlin}, {Crowley}, {Datta}, {Devlin},
  {Duff}, {Dunkley}, {D{\"u}nner}, {van Engelen}, {Gallardo}, {Grace},
  {Hasselfield}, {Hills}, {Hilton}, {Hincks}, {Hloẑek}, {Ho}, {Hubmayr},
  {Huffenberger}, {Hughes}, {Irwin}, {Koopman}, {Kosowsky}, {Li}, {McMahon},
  {Munson}, {Nati}, {Newburgh}, {Niemack}, {Niraula}, {Page}, {Pappas},
  {Salatino}, {Schillaci}, {Schmitt}, {Sehgal}, {Sherwin}, {Sievers}, {Simon},
  {Spergel}, {Staggs}, {Stevens}, {Thornton}, {Van Lanen}, {Vavagiakis},
  {Ward}, \& {Wollack}}]{AdvancedACTPol}
{Henderson} S.~W. {et~al.}, 2016, Journal of Low Temperature Physics, 184, 772

\bibitem[{Hu {et~al}\mbox{.}(2020)Hu, Webb, Ayres, Bainbridge, Barrow, Barstow,
  Berengut, Carswell, Dumont, Dzuba, \& et~al.}]{Hu2020VFC}
Hu J. {et~al.}, 2020, Monthly Notices of the Royal Astronomical Society

\bibitem[{{H{\"u}tsi} {et~al}\mbox{.}(2009){H{\"u}tsi}, {Hektor}, \&
  {Raidal}}]{Huetsi2009}
{H{\"u}tsi} G., {Hektor} A., {Raidal} M., 2009, \aap, 505, 999

\bibitem[{{Ishida} \& {de Souza}(2011)}]{Ishida2011}
{Ishida} E.~E.~O., {de Souza} R.~S., 2011, \aap, 527, A49

\bibitem[{Jedamzik \& Pogosian(2020)}]{Jedamzik2020}
Jedamzik K., Pogosian L., 2020, Physical Review Letters, 125

\bibitem[{Jedamzik \& Saveliev(2019)}]{Jedamzik2019}
Jedamzik K., Saveliev A., 2019, Physical Review Letters, 123

\bibitem[{{Kaplinghat} {et~al}\mbox{.}(1999){Kaplinghat}, {Scherrer}, \&
  {Turner}}]{Kaplinghat1999}
{Kaplinghat} M., {Scherrer} R.~J., {Turner} M.~S., 1999, \prd, 60, 023516

\bibitem[{{Keisler} {et~al}\mbox{.}(2015){Keisler}, {Hoover}, {Harrington},
  {Henning}, {Ade}, {Aird}, {Austermann}, {Beall}, {Bender}, {Benson}, {Bleem},
  {Carlstrom}, {Chang}, {Chiang}, {Cho}, {Citron}, {Crawford}, {Crites}, {de
  Haan}, {Dobbs}, {Everett}, {Gallicchio}, {Gao}, {George}, {Gilbert},
  {Halverson}, {Hanson}, {Hilton}, {Holder}, {Holzapfel}, {Hou}, {Hrubes},
  {Huang}, {Hubmayr}, {Irwin}, {Knox}, {Lee}, {Leitch}, {Li}, {Luong-Van},
  {Marrone}, {McMahon}, {Mehl}, {Meyer}, {Mocanu}, {Natoli}, {Nibarger},
  {Novosad}, {Padin}, {Pryke}, {Reichardt}, {Ruhl}, {Saliwanchik}, {Sayre},
  {Schaffer}, {Shirokoff}, {Smecher}, {Stark}, {Story}, {Tucker},
  {Vanderlinde}, {Vieira}, {Wang}, {Whitehorn}, {Yefremenko}, \&
  {Zahn}}]{sptpol_results}
{Keisler} R. {et~al.}, 2015, \apj, 807, 151

\bibitem[{Knox \& Millea(2020)}]{Knox2020}
Knox L., Millea M., 2020, Physical Review D, 101

\bibitem[{Kotu\v{s} {et~al}\mbox{.}(2017)Kotu\v{s}, Murphy, \&
  Carswell}]{Kotus2017}
Kotu\v{s} S.~M., Murphy M.~T., Carswell R.~F., 2017, \mnras, 464, 3679

\bibitem[{{Kunze} \& {Komatsu}(2014)}]{Kunze2014}
{Kunze} K.~E., {Komatsu} E., 2014, \jcap, 1, 9

\bibitem[{Levshakov {et~al}\mbox{.}(2020)Levshakov, Kozlov, \&
  Agafonova}]{Levshakov2020_me}
Levshakov S.~A., Kozlov M.~G., Agafonova I.~I., 2020, Monthly Notices of the
  Royal Astronomical Society, 498, 3624–3632

\bibitem[{{Levshakov} {et~al}\mbox{.}(2019){Levshakov}, {Ng}, {Henkel},
  {Mookerjea}, {Agafonova}, {Liu}, \& {Wang}}]{Levshakov2019}
{Levshakov} S.~A., {Ng} K.~W., {Henkel} C., {Mookerjea} B., {Agafonova} I.~I.,
  {Liu} S.~Y., {Wang} W.~H., 2019, \mnras, 487, 5175

\bibitem[{Lewis(2013)}]{Lewis2013}
Lewis A., 2013, \prd, 87

\bibitem[{Lewis \& Bridle(2002)}]{COSMOMC}
Lewis A., Bridle S., 2002, Phys. Rev., D66, 103511

\bibitem[{{Lewis} {et~al}\mbox{.}(2000){Lewis}, {Challinor}, \&
  {Lasenby}}]{CAMB}
{Lewis} A., {Challinor} A., {Lasenby} A., 2000, \apj, 538, 473

\bibitem[{Lin {et~al}\mbox{.}(2020)Lin, Hu, \& Raveri}]{Lin2020}
Lin M.-X., Hu W., Raveri M., 2020, Physical Review D, 102

\bibitem[{Martins(2017)}]{Martins2017review}
Martins C. J. A.~P., 2017, Reports on Progress in Physics, 80, 126902

\bibitem[{{Menegoni} {et~al}\mbox{.}(2012){Menegoni}, {Archidiacono},
  {Calabrese}, {Galli}, {Martins}, \& {Melchiorri}}]{Menegoni2012}
{Menegoni} E., {Archidiacono} M., {Calabrese} E., {Galli} S., {Martins}
  C.~J.~A.~P., {Melchiorri} A., 2012, \prd, 85, 107301

\bibitem[{{Menegoni} {et~al}\mbox{.}(2009){Menegoni}, {Galli}, {Bartlett},
  {Martins}, \& {Melchiorri}}]{Menegoni2009}
{Menegoni} E., {Galli} S., {Bartlett} J.~G., {Martins} C.~J.~A.~P.,
  {Melchiorri} A., 2009, \prd, 80, 087302

\bibitem[{{Mortonson} \& {Hu}(2008)}]{Mortonson2008}
{Mortonson} M.~J., {Hu} W., 2008, \apj, 672, 737

\bibitem[{Mota \& Barrow(2004)}]{Mota2004}
Mota D.~F., Barrow J.~D., 2004, \mnras, 349, 291

\bibitem[{Murphy \& Cooksey(2017)}]{Murphy2017}
Murphy M.~T., Cooksey K.~L., 2017, Mon. Not. Roy. Astron. Soc., 471, 4930

\bibitem[{{Naess} {et~al}\mbox{.}(2014){Naess}, {Hasselfield}, {McMahon},
  {Niemack}, {Addison}, {Ade}, {Allison}, {Amiri}, {Battaglia}, {Beall}, {de
  Bernardis}, {Bond}, {Britton}, {Calabrese}, {Cho}, {Coughlin}, {Crichton},
  {Das}, {Datta}, {Devlin}, {Dicker}, {Dunkley}, {D{\"u}nner}, {Fowler}, {Fox},
  {Gallardo}, {Grace}, {Gralla}, {Hajian}, {Halpern}, {Henderson}, {Hill},
  {Hilton}, {Hilton}, {Hincks}, {Hlozek}, {Ho}, {Hubmayr}, {Huffenberger},
  {Hughes}, {Infante}, {Irwin}, {Jackson}, {Muya Kasanda}, {Klein}, {Koopman},
  {Kosowsky}, {Li}, {Louis}, {Lungu}, {Madhavacheril}, {Marriage}, {Maurin},
  {Menanteau}, {Moodley}, {Munson}, {Newburgh}, {Nibarger}, {Nolta}, {Page},
  {Pappas}, {Partridge}, {Rojas}, {Schmitt}, {Sehgal}, {Sherwin}, {Sievers},
  {Simon}, {Spergel}, {Staggs}, {Switzer}, {Thornton}, {Trac}, {Tucker},
  {Uehara}, {Van Engelen}, {Ward}, \& {Wollack}}]{actpol_polresults}
{Naess} S. {et~al.}, 2014, \jcap, 10, 007

\bibitem[{Negrelli {et~al}\mbox{.}(2018)Negrelli, Kraiselburd, Landau, \&
  García-Berro}]{Negrelli2018}
Negrelli C., Kraiselburd L., Landau S., García-Berro E., 2018, International
  Journal of Modern Physics D, 27, 1850099

\bibitem[{{Netterfield} {et~al}\mbox{.}(2002){Netterfield}, {Ade}, {Bock},
  {Bond}, {Borrill}, {Boscaleri}, {Coble}, {Contaldi}, {Crill}, {de Bernardis},
  {Farese}, {Ganga}, {Giacometti}, {Hivon}, {Hristov}, {Iacoangeli}, \&
  {Jaffe}}]{Netterfield2002}
{Netterfield} C.~B. {et~al.}, 2002, \apj, 571, 604

\bibitem[{{Pace} {et~al}\mbox{.}(2019){Pace}, {Battye}, {Bolliet}, \&
  {Trinh}}]{Pace2019}
{Pace} F., {Battye} R.~A., {Bolliet} B., {Trinh} D., 2019, \jcap, 2019, 018

\bibitem[{{Padmanabhan} \& {Finkbeiner}(2005)}]{Padmanabhan2005}
{Padmanabhan} N., {Finkbeiner} D.~P., 2005, \prd, 72, 023508

\bibitem[{Paoletti {et~al}\mbox{.}(2019)Paoletti, Chluba, Finelli, \&
  Rubiño-Martín}]{Paoletti2019}
Paoletti D., Chluba J., Finelli F., Rubiño-Martín J.~A., 2019, Monthly
  Notices of the Royal Astronomical Society, 484, 185–195

\bibitem[{{Pearson} {et~al}\mbox{.}(2003){Pearson}, {Mason}, {Readhead},
  {Shepherd}, {Sievers}, {Udomprasert}, {Cartwright}, {Farmer}, {Padin},
  {Myers}, {Bond}, {Contaldi}, {Pen}, {Prunet}, {Pogosyan}, {Carlstrom},
  {Kovac}, {Leitch}, {Pryke}, {Halverson}, {Holzapfel}, {Altamirano},
  {Bronfman}, {Casassus}, {May}, \& {Joy}}]{CBI03}
{Pearson} T.~J. {et~al.}, 2003, \apj, 591, 556

\bibitem[{{Planck Collaboration et al.}(2015{\natexlab{a}})}]{Planck2015params}
{Planck Collaboration et al.}, 2015{\natexlab{a}}, ArXiv:1502.01589

\bibitem[{{Planck Collaboration et
  al.}(2015{\natexlab{b}})}]{Planck2015var_alp}
{Planck Collaboration et al.}, 2015{\natexlab{b}}, \aap, 580, A22

\bibitem[{{Planck Collaboration et al.}(2016)}]{Planck2015like}
{Planck Collaboration et al.}, 2016, A\&A, 594, A11

\bibitem[{{Planck Collaboration et al.}(2018{\natexlab{a}})}]{Planck2018over}
{Planck Collaboration et al.}, 2018{\natexlab{a}}, arXiv e-prints,
  arXiv:1807.06205

\bibitem[{{Planck Collaboration et al.}(2018{\natexlab{b}})}]{Planck2018params}
{Planck Collaboration et al.}, 2018{\natexlab{b}}, ArXiv:1807.06209

\bibitem[{{Planck Collaboration et al.}(2019)}]{Planck2018like}
{Planck Collaboration et al.}, 2019, arXiv e-prints, arXiv:1907.12875

\bibitem[{Poulin {et~al}\mbox{.}(2018)Poulin, Smith, Grin, Karwal, \&
  Kamionkowski}]{Poulin2018}
Poulin V., Smith T.~L., Grin D., Karwal T., Kamionkowski M., 2018, Physical
  Review D, 98

\bibitem[{{Poulin} {et~al}\mbox{.}(2019){Poulin}, {Smith}, {Karwal}, \&
  {Kamionkowski}}]{Poulin2019}
{Poulin} V., {Smith} T.~L., {Karwal} T., {Kamionkowski} M., 2019, \prl, 122,
  221301

\bibitem[{Riess {et~al}\mbox{.}(2019)Riess, Casertano, Yuan, Macri, \&
  Scolnic}]{Riess2019}
Riess A.~G., Casertano S., Yuan W., Macri L.~M., Scolnic D., 2019, Astrophys.
  J., 876, 85

\bibitem[{{Rubi{\~n}o-Martin} {et~al}\mbox{.}(2003){Rubi{\~n}o-Martin},
  {Rebolo}, {Carreira}, {Cleary}, {Davies}, {Davis}, {Dickinson}, {Grainge},
  {Guti{\'e}rrez}, {Hobson}, {Jones}, {Kneissl}, {Lasenby}, \&
  {Maisinger}}]{VSA2003}
{Rubi{\~n}o-Martin} J.~A. {et~al.}, 2003, \mnras, 341, 1084

\bibitem[{Sandvik {et~al}\mbox{.}(2002)Sandvik, Barrow, \&
  Magueijo}]{Sandvik2001}
Sandvik H.~B., Barrow J.~D., Magueijo J., 2002, \prl, 88, 031302

\bibitem[{{Sch{\"o}neberg} {et~al}\mbox{.}(2021){Sch{\"o}neberg},
  {Abell{\'a}n}, {P{\'e}rez S{\'a}nchez}, {Witte}, {Poulin}, \&
  {Lesgourgues}}]{Nils2021}
{Sch{\"o}neberg} N., {Abell{\'a}n} G.~F., {P{\'e}rez S{\'a}nchez} A., {Witte}
  S.~J., {Poulin} c.~V., {Lesgourgues} J., 2021, arXiv e-prints,
  arXiv:2107.10291

\bibitem[{{Sc{\'o}ccola} {et~al}\mbox{.}(2009){Sc{\'o}ccola}, {Landau}, \&
  {Vucetich}}]{Scoccola2009}
{Sc{\'o}ccola} C.~G., {Landau} S.~J., {Vucetich} H., 2009, Memorie della Societ
  Astronomica Italiana, 80, 814

\bibitem[{{Sethi} \& {Subramanian}(2005)}]{Sethi2005}
{Sethi} S.~K., {Subramanian} K., 2005, \mnras, 356, 778

\bibitem[{{Sharma} {et~al}\mbox{.}(2020){Sharma}, {Mukherjee}, \&
  {Jassal}}]{Sharma2020}
{Sharma} R., {Mukherjee} A., {Jassal} H.~K., 2020, arXiv e-prints,
  arXiv:2004.01393

\bibitem[{{Shaw} \& {Chluba}(2011)}]{Shaw2011}
{Shaw} J.~R., {Chluba} J., 2011, \mnras, 415, 1343

\bibitem[{{Shaw} \& {Lewis}(2010)}]{Shaw2010PMF}
{Shaw} J.~R., {Lewis} A., 2010, \prd, 81, 043517

\bibitem[{{Silvestri} \& {Trodden}(2009)}]{Silvestri2009}
{Silvestri} A., {Trodden} M., 2009, Reports on Progress in Physics, 72, 096901

\bibitem[{Slatyer {et~al}\mbox{.}(2009)Slatyer, Padmanabhan, \&
  Finkbeiner}]{Slatyer2009}
Slatyer T.~R., Padmanabhan N., Finkbeiner D.~P., 2009, Physical Review D
  (Particles, Fields, Gravitation, and Cosmology), 80, 043526

\bibitem[{Slatyer \& Wu(2017)}]{Slatyer2016}
Slatyer T.~R., Wu C.-L., 2017, Phys. Rev., D95, 023010

\bibitem[{Tegmark {et~al}\mbox{.}(1997)Tegmark, Taylor, \&
  Heavens}]{Tegmark1997}
Tegmark M., Taylor A.~N., Heavens A.~F., 1997, The Astrophysical Journal, 480,
  22

\bibitem[{{Thiele} {et~al}\mbox{.}(2021){Thiele}, {Guan}, {Hill}, {Kosowsky},
  \& {Spergel}}]{Thiele2021ACT}
{Thiele} L., {Guan} Y., {Hill} J.~C., {Kosowsky} A., {Spergel} D.~N., 2021,
  arXiv e-prints, arXiv:2105.03003

\bibitem[{{Uzan}(2003)}]{Uzan2003}
{Uzan} J.-P., 2003, Reviews of Modern Physics, 75, 403

\bibitem[{{Uzan}(2011)}]{Uzan2011}
{Uzan} J.-P., 2011, Living Reviews in Relativity, 14, 2

\bibitem[{{Verde}(2010)}]{Verde2009}
{Verde} L., 2010, {Statistical Methods in Cosmology}, Vol. 800, Berlin Springer
  Verlag, pp. 147--177

\bibitem[{{Wilczynska} {et~al}\mbox{.}(2020){Wilczynska}, {Webb}, {Bainbridge},
  {Barrow}, {Bosman}, {Carswell}, {D{\k{a}}browski}, {Dumont}, {Lee}, {Leite},
  {Leszczy{\'n}ska}, {Liske}, {Marosek}, {Martins}, {Milakovi{\'c}}, {Molaro},
  \& {Pasquini}}]{Wilczynska2020VFC}
{Wilczynska} M.~R. {et~al.}, 2020, Science Advances, 6, eaay9672

\end{thebibliography}

% Don't change these lines
\bsp	% typesetting comment
\label{lastpage}
\end{document}